\documentclass[longauth]{aaEC}

\usepackage{graphicx}
\usepackage{natbib}
\usepackage{scalerel}
\usepackage{comment}

\usepackage[table]{xcolor}

\usepackage{ulem}
\usepackage{txfonts}
\usepackage[pdfencoding=auto,psdextra]{hyperref}
\hypersetup{
    colorlinks=true,
    linkcolor=blue,
    filecolor=magenta,      
    urlcolor=blue,
    citecolor=blue
}
\makeatletter
\renewcommand*\aa@pageof{, page \thepage{} of \pageref*{LastPage}}
\makeatother

\usepackage[utf8]{inputenc}

\usepackage[switch, modulo]{lineno}

\usepackage{euclid}
\usepackage{ISTL_macros}

\usepackage[english]{babel}
\usepackage{amsmath}
\usepackage{color, colortbl}
\usepackage{amsfonts}
\usepackage{amssymb}
\usepackage{bm}
\usepackage[capitalise,nameinlink]{cleveref}
\usepackage{glossaries}
\glsdisablehyper
\usepackage{xspace}
\usepackage{comment}

\newacronym{lss}{LSS}{large-scale structure}
\newacronym{eft}{EFTofLSS}{effective field theory of large-scale structure}
\newacronym{rsd}{RSD}{Redshift space distortions}
\newacronym{gcsp}{GCsp}{spectroscopic galaxy clustering}
\newacronym{2pcf}{2PCF}{two-point correlation function}
\newacronym{mcmc}{MCMC}{Monte Carlo Markov chain}
\newacronym{gcph}{GCph}{photometric galaxy clustering}
\newacronym{wl}{WL}{weak lensing}
\newacronym{ggl}{GGL}{galaxy-galaxy lensing}

\defcitealias{Paper1}{Paper 1}
\defcitealias{Paper2}{Paper 2}
\defcitealias{Paper3}{Paper 3}

\begin{document}

\title{\Euclid\/ preparation.}
\subtitle{Cosmology Likelihood for Observables in \Euclid (CLOE). 1. Theoretical recipe}

   
\newcommand{\orcid}[1]{} 


\author{Euclid Collaboration: V.~F.~Cardone\thanks{\email{vincenzo.cardone@inaf.it}}\inst{\ref{aff1},\ref{aff2}}
\and S.~Joudaki\orcid{0000-0001-8820-673X}\inst{\ref{aff3},\ref{aff4},\ref{aff5},\ref{aff6}}
\and L.~Blot\orcid{0000-0002-9622-7167}\inst{\ref{aff7},\ref{aff8}}
\and M.~Bonici\orcid{0000-0002-8430-126X}\inst{\ref{aff5},\ref{aff9}}
\and S.~Camera\orcid{0000-0003-3399-3574}\inst{\ref{aff10},\ref{aff11},\ref{aff12}}
\and G.~Ca\~nas-Herrera\orcid{0000-0003-2796-2149}\inst{\ref{aff13},\ref{aff14},\ref{aff15}}
\and P.~Carrilho\orcid{0000-0003-1339-0194}\inst{\ref{aff16}}
\and S.~Casas\orcid{0000-0002-4751-5138}\inst{\ref{aff17}}
\and S.~Davini\orcid{0000-0003-3269-1718}\inst{\ref{aff18}}
\and S.~Di~Domizio\orcid{0000-0003-2863-5895}\inst{\ref{aff19},\ref{aff18}}
\and S.~Farrens\orcid{0000-0002-9594-9387}\inst{\ref{aff20}}
\and L.~W.~K.~Goh\orcid{0000-0002-0104-8132}\inst{\ref{aff20}}
\and S.~Gouyou~Beauchamps\inst{\ref{aff21},\ref{aff22}}
\and S.~Ili\'c\orcid{0000-0003-4285-9086}\inst{\ref{aff23},\ref{aff24}}
\and F.~Keil\orcid{0000-0002-8108-1679}\inst{\ref{aff24}}
\and A.~M.~C.~Le~Brun\orcid{0000-0002-0936-4594}\inst{\ref{aff8}}
\and M.~Martinelli\orcid{0000-0002-6943-7732}\inst{\ref{aff1},\ref{aff2}}
\and C.~Moretti\orcid{0000-0003-3314-8936}\inst{\ref{aff25},\ref{aff26},\ref{aff27},\ref{aff28},\ref{aff29}}
\and V.~Pettorino\inst{\ref{aff13}}
\and A.~Pezzotta\orcid{0000-0003-0726-2268}\inst{\ref{aff30},\ref{aff31}}
\and A.~G.~S\'anchez\orcid{0000-0003-1198-831X}\inst{\ref{aff31}}
\and Z.~Sakr\orcid{0000-0002-4823-3757}\inst{\ref{aff32},\ref{aff24},\ref{aff33}}
\and D.~Sciotti\orcid{0009-0008-4519-2620}\inst{\ref{aff1},\ref{aff2}}
\and K.~Tanidis\orcid{0000-0001-9843-5130}\inst{\ref{aff34}}
\and I.~Tutusaus\orcid{0000-0002-3199-0399}\inst{\ref{aff24}}
\and V.~Ajani\orcid{0000-0001-9442-2527}\inst{\ref{aff20},\ref{aff35},\ref{aff36}}
\and M.~Crocce\orcid{0000-0002-9745-6228}\inst{\ref{aff22},\ref{aff21}}
\and C.~Giocoli\orcid{0000-0002-9590-7961}\inst{\ref{aff37},\ref{aff38}}
\and L.~Legrand\orcid{0000-0003-0610-5252}\inst{\ref{aff39},\ref{aff40}}
\and M.~Lembo\orcid{0000-0002-5271-5070}\inst{\ref{aff41},\ref{aff42}}
\and G.~F.~Lesci\orcid{0000-0002-4607-2830}\inst{\ref{aff43},\ref{aff37}}
\and D.~Navarro~Girones\orcid{0000-0003-0507-372X}\inst{\ref{aff15}}
\and A.~Nouri-Zonoz\orcid{0009-0006-6164-8670}\inst{\ref{aff44}}
\and S.~Pamuk\orcid{0009-0004-0852-8624}\inst{\ref{aff45}}
\and M.~Tsedrik\orcid{0000-0002-0020-5343}\inst{\ref{aff16},\ref{aff46}}
\and J.~Bel\inst{\ref{aff47}}
\and C.~Carbone\orcid{0000-0003-0125-3563}\inst{\ref{aff9}}
\and C.~A.~J.~Duncan\orcid{0009-0003-3573-0791}\inst{\ref{aff16},\ref{aff48}}
\and M.~Kilbinger\orcid{0000-0001-9513-7138}\inst{\ref{aff20}}
\and F.~Lacasa\orcid{0000-0002-7268-3440}\inst{\ref{aff49},\ref{aff50}}
\and M.~Lattanzi\orcid{0000-0003-1059-2532}\inst{\ref{aff42}}
\and D.~Sapone\orcid{0000-0001-7089-4503}\inst{\ref{aff51}}
\and E.~Sellentin\inst{\ref{aff52},\ref{aff15}}
\and P.~L.~Taylor\orcid{0000-0001-6999-4718}\inst{\ref{aff53},\ref{aff54}}
\and N.~Aghanim\orcid{0000-0002-6688-8992}\inst{\ref{aff50}}
\and B.~Altieri\orcid{0000-0003-3936-0284}\inst{\ref{aff55}}
\and L.~Amendola\orcid{0000-0002-0835-233X}\inst{\ref{aff32}}
\and S.~Andreon\orcid{0000-0002-2041-8784}\inst{\ref{aff56}}
\and N.~Auricchio\orcid{0000-0003-4444-8651}\inst{\ref{aff37}}
\and H.~Aussel\orcid{0000-0002-1371-5705}\inst{\ref{aff20}}
\and C.~Baccigalupi\orcid{0000-0002-8211-1630}\inst{\ref{aff28},\ref{aff27},\ref{aff29},\ref{aff25}}
\and M.~Baldi\orcid{0000-0003-4145-1943}\inst{\ref{aff57},\ref{aff37},\ref{aff38}}
\and S.~Bardelli\orcid{0000-0002-8900-0298}\inst{\ref{aff37}}
\and P.~Battaglia\orcid{0000-0002-7337-5909}\inst{\ref{aff37}}
\and A.~Biviano\orcid{0000-0002-0857-0732}\inst{\ref{aff27},\ref{aff28}}
\and E.~Branchini\orcid{0000-0002-0808-6908}\inst{\ref{aff19},\ref{aff18},\ref{aff56}}
\and M.~Brescia\orcid{0000-0001-9506-5680}\inst{\ref{aff58},\ref{aff59}}
\and J.~Brinchmann\orcid{0000-0003-4359-8797}\inst{\ref{aff60},\ref{aff61}}
\and V.~Capobianco\orcid{0000-0002-3309-7692}\inst{\ref{aff12}}
\and J.~Carretero\orcid{0000-0002-3130-0204}\inst{\ref{aff3},\ref{aff62}}
\and M.~Castellano\orcid{0000-0001-9875-8263}\inst{\ref{aff1}}
\and G.~Castignani\orcid{0000-0001-6831-0687}\inst{\ref{aff37}}
\and S.~Cavuoti\orcid{0000-0002-3787-4196}\inst{\ref{aff59},\ref{aff63}}
\and K.~C.~Chambers\orcid{0000-0001-6965-7789}\inst{\ref{aff64}}
\and A.~Cimatti\inst{\ref{aff65}}
\and C.~Colodro-Conde\inst{\ref{aff66}}
\and G.~Congedo\orcid{0000-0003-2508-0046}\inst{\ref{aff16}}
\and C.~J.~Conselice\orcid{0000-0003-1949-7638}\inst{\ref{aff48}}
\and L.~Conversi\orcid{0000-0002-6710-8476}\inst{\ref{aff67},\ref{aff55}}
\and Y.~Copin\orcid{0000-0002-5317-7518}\inst{\ref{aff68}}
\and F.~Courbin\orcid{0000-0003-0758-6510}\inst{\ref{aff69},\ref{aff70}}
\and H.~M.~Courtois\orcid{0000-0003-0509-1776}\inst{\ref{aff71}}
\and M.~Cropper\orcid{0000-0003-4571-9468}\inst{\ref{aff72}}
\and A.~Da~Silva\orcid{0000-0002-6385-1609}\inst{\ref{aff73},\ref{aff74}}
\and H.~Degaudenzi\orcid{0000-0002-5887-6799}\inst{\ref{aff75}}
\and G.~De~Lucia\orcid{0000-0002-6220-9104}\inst{\ref{aff27}}
\and A.~M.~Di~Giorgio\orcid{0000-0002-4767-2360}\inst{\ref{aff76}}
\and M.~Douspis\orcid{0000-0003-4203-3954}\inst{\ref{aff50}}
\and F.~Dubath\orcid{0000-0002-6533-2810}\inst{\ref{aff75}}
\and X.~Dupac\inst{\ref{aff55}}
\and S.~Dusini\orcid{0000-0002-1128-0664}\inst{\ref{aff77}}
\and A.~Ealet\orcid{0000-0003-3070-014X}\inst{\ref{aff68}}
\and S.~Escoffier\orcid{0000-0002-2847-7498}\inst{\ref{aff78}}
\and M.~Farina\orcid{0000-0002-3089-7846}\inst{\ref{aff76}}
\and R.~Farinelli\inst{\ref{aff37}}
\and F.~Faustini\orcid{0000-0001-6274-5145}\inst{\ref{aff1},\ref{aff79}}
\and S.~Ferriol\inst{\ref{aff68}}
\and F.~Finelli\orcid{0000-0002-6694-3269}\inst{\ref{aff37},\ref{aff80}}
\and P.~Fosalba\orcid{0000-0002-1510-5214}\inst{\ref{aff21},\ref{aff22}}
\and S.~Fotopoulou\orcid{0000-0002-9686-254X}\inst{\ref{aff81}}
\and M.~Frailis\orcid{0000-0002-7400-2135}\inst{\ref{aff27}}
\and E.~Franceschi\orcid{0000-0002-0585-6591}\inst{\ref{aff37}}
\and M.~Fumana\orcid{0000-0001-6787-5950}\inst{\ref{aff9}}
\and S.~Galeotta\orcid{0000-0002-3748-5115}\inst{\ref{aff27}}
\and B.~Gillis\orcid{0000-0002-4478-1270}\inst{\ref{aff16}}
\and P.~G\'omez-Alvarez\orcid{0000-0002-8594-5358}\inst{\ref{aff82},\ref{aff55}}
\and J.~Gracia-Carpio\inst{\ref{aff31}}
\and B.~R.~Granett\orcid{0000-0003-2694-9284}\inst{\ref{aff56}}
\and A.~Grazian\orcid{0000-0002-5688-0663}\inst{\ref{aff83}}
\and F.~Grupp\inst{\ref{aff31},\ref{aff84}}
\and L.~Guzzo\orcid{0000-0001-8264-5192}\inst{\ref{aff85},\ref{aff56},\ref{aff86}}
\and S.~V.~H.~Haugan\orcid{0000-0001-9648-7260}\inst{\ref{aff87}}
\and H.~Hoekstra\orcid{0000-0002-0641-3231}\inst{\ref{aff15}}
\and W.~Holmes\inst{\ref{aff88}}
\and I.~M.~Hook\orcid{0000-0002-2960-978X}\inst{\ref{aff89}}
\and F.~Hormuth\inst{\ref{aff90}}
\and A.~Hornstrup\orcid{0000-0002-3363-0936}\inst{\ref{aff91},\ref{aff92}}
\and K.~Jahnke\orcid{0000-0003-3804-2137}\inst{\ref{aff93}}
\and M.~Jhabvala\inst{\ref{aff94}}
\and E.~Keih\"anen\orcid{0000-0003-1804-7715}\inst{\ref{aff95}}
\and S.~Kermiche\orcid{0000-0002-0302-5735}\inst{\ref{aff78}}
\and A.~Kiessling\orcid{0000-0002-2590-1273}\inst{\ref{aff88}}
\and B.~Kubik\orcid{0009-0006-5823-4880}\inst{\ref{aff68}}
\and M.~K\"ummel\orcid{0000-0003-2791-2117}\inst{\ref{aff84}}
\and M.~Kunz\orcid{0000-0002-3052-7394}\inst{\ref{aff44}}
\and H.~Kurki-Suonio\orcid{0000-0002-4618-3063}\inst{\ref{aff96},\ref{aff97}}
\and O.~Lahav\orcid{0000-0002-1134-9035}\inst{\ref{aff98}}
\and P.~Liebing\inst{\ref{aff72}}
\and P.~B.~Lilje\orcid{0000-0003-4324-7794}\inst{\ref{aff87}}
\and V.~Lindholm\orcid{0000-0003-2317-5471}\inst{\ref{aff96},\ref{aff97}}
\and I.~Lloro\orcid{0000-0001-5966-1434}\inst{\ref{aff99}}
\and G.~Mainetti\orcid{0000-0003-2384-2377}\inst{\ref{aff100}}
\and D.~Maino\inst{\ref{aff85},\ref{aff9},\ref{aff86}}
\and E.~Maiorano\orcid{0000-0003-2593-4355}\inst{\ref{aff37}}
\and O.~Mansutti\orcid{0000-0001-5758-4658}\inst{\ref{aff27}}
\and S.~Marcin\inst{\ref{aff101}}
\and O.~Marggraf\orcid{0000-0001-7242-3852}\inst{\ref{aff102}}
\and N.~Martinet\orcid{0000-0003-2786-7790}\inst{\ref{aff103}}
\and F.~Marulli\orcid{0000-0002-8850-0303}\inst{\ref{aff43},\ref{aff37},\ref{aff38}}
\and R.~Massey\orcid{0000-0002-6085-3780}\inst{\ref{aff104}}
\and S.~Maurogordato\inst{\ref{aff105}}
\and E.~Medinaceli\orcid{0000-0002-4040-7783}\inst{\ref{aff37}}
\and S.~Mei\orcid{0000-0002-2849-559X}\inst{\ref{aff106},\ref{aff107}}
\and Y.~Mellier\inst{\ref{aff108},\ref{aff109}}
\and M.~Meneghetti\orcid{0000-0003-1225-7084}\inst{\ref{aff37},\ref{aff38}}
\and E.~Merlin\orcid{0000-0001-6870-8900}\inst{\ref{aff1}}
\and G.~Meylan\inst{\ref{aff110}}
\and A.~Mora\orcid{0000-0002-1922-8529}\inst{\ref{aff111}}
\and M.~Moresco\orcid{0000-0002-7616-7136}\inst{\ref{aff43},\ref{aff37}}
\and L.~Moscardini\orcid{0000-0002-3473-6716}\inst{\ref{aff43},\ref{aff37},\ref{aff38}}
\and R.~Nakajima\orcid{0009-0009-1213-7040}\inst{\ref{aff102}}
\and C.~Neissner\orcid{0000-0001-8524-4968}\inst{\ref{aff112},\ref{aff62}}
\and S.-M.~Niemi\orcid{0009-0005-0247-0086}\inst{\ref{aff13}}
\and C.~Padilla\orcid{0000-0001-7951-0166}\inst{\ref{aff112}}
\and S.~Paltani\orcid{0000-0002-8108-9179}\inst{\ref{aff75}}
\and F.~Pasian\orcid{0000-0002-4869-3227}\inst{\ref{aff27}}
\and K.~Pedersen\inst{\ref{aff113}}
\and W.~J.~Percival\orcid{0000-0002-0644-5727}\inst{\ref{aff5},\ref{aff6},\ref{aff114}}
\and S.~Pires\orcid{0000-0002-0249-2104}\inst{\ref{aff20}}
\and G.~Polenta\orcid{0000-0003-4067-9196}\inst{\ref{aff79}}
\and M.~Poncet\inst{\ref{aff115}}
\and L.~A.~Popa\inst{\ref{aff116}}
\and L.~Pozzetti\orcid{0000-0001-7085-0412}\inst{\ref{aff37}}
\and G.~D.~Racca\orcid{0000-0002-9883-8981}\inst{\ref{aff13},\ref{aff15}}
\and F.~Raison\orcid{0000-0002-7819-6918}\inst{\ref{aff31}}
\and R.~Rebolo\orcid{0000-0003-3767-7085}\inst{\ref{aff66},\ref{aff117},\ref{aff118}}
\and A.~Renzi\orcid{0000-0001-9856-1970}\inst{\ref{aff119},\ref{aff77}}
\and J.~Rhodes\orcid{0000-0002-4485-8549}\inst{\ref{aff88}}
\and G.~Riccio\inst{\ref{aff59}}
\and E.~Romelli\orcid{0000-0003-3069-9222}\inst{\ref{aff27}}
\and M.~Roncarelli\orcid{0000-0001-9587-7822}\inst{\ref{aff37}}
\and R.~Saglia\orcid{0000-0003-0378-7032}\inst{\ref{aff84},\ref{aff31}}
\and B.~Sartoris\orcid{0000-0003-1337-5269}\inst{\ref{aff84},\ref{aff27}}
\and R.~Scaramella\orcid{0000-0003-2229-193X}\inst{\ref{aff1},\ref{aff2}}
\and J.~A.~Schewtschenko\orcid{0000-0002-4913-6393}\inst{\ref{aff16}}
\and P.~Schneider\orcid{0000-0001-8561-2679}\inst{\ref{aff102}}
\and T.~Schrabback\orcid{0000-0002-6987-7834}\inst{\ref{aff120}}
\and A.~Secroun\orcid{0000-0003-0505-3710}\inst{\ref{aff78}}
\and E.~Sefusatti\orcid{0000-0003-0473-1567}\inst{\ref{aff27},\ref{aff28},\ref{aff29}}
\and G.~Seidel\orcid{0000-0003-2907-353X}\inst{\ref{aff93}}
\and S.~Serrano\orcid{0000-0002-0211-2861}\inst{\ref{aff21},\ref{aff121},\ref{aff22}}
\and P.~Simon\inst{\ref{aff102}}
\and C.~Sirignano\orcid{0000-0002-0995-7146}\inst{\ref{aff119},\ref{aff77}}
\and G.~Sirri\orcid{0000-0003-2626-2853}\inst{\ref{aff38}}
\and L.~Stanco\orcid{0000-0002-9706-5104}\inst{\ref{aff77}}
\and J.~Steinwagner\orcid{0000-0001-7443-1047}\inst{\ref{aff31}}
\and P.~Tallada-Cresp\'{i}\orcid{0000-0002-1336-8328}\inst{\ref{aff3},\ref{aff62}}
\and A.~N.~Taylor\inst{\ref{aff16}}
\and I.~Tereno\orcid{0000-0002-4537-6218}\inst{\ref{aff73},\ref{aff122}}
\and S.~Toft\orcid{0000-0003-3631-7176}\inst{\ref{aff123},\ref{aff124}}
\and R.~Toledo-Moreo\orcid{0000-0002-2997-4859}\inst{\ref{aff125}}
\and F.~Torradeflot\orcid{0000-0003-1160-1517}\inst{\ref{aff62},\ref{aff3}}
\and L.~Valenziano\orcid{0000-0002-1170-0104}\inst{\ref{aff37},\ref{aff80}}
\and J.~Valiviita\orcid{0000-0001-6225-3693}\inst{\ref{aff96},\ref{aff97}}
\and T.~Vassallo\orcid{0000-0001-6512-6358}\inst{\ref{aff84},\ref{aff27}}
\and G.~Verdoes~Kleijn\orcid{0000-0001-5803-2580}\inst{\ref{aff126}}
\and A.~Veropalumbo\orcid{0000-0003-2387-1194}\inst{\ref{aff56},\ref{aff18},\ref{aff19}}
\and Y.~Wang\orcid{0000-0002-4749-2984}\inst{\ref{aff127}}
\and J.~Weller\orcid{0000-0002-8282-2010}\inst{\ref{aff84},\ref{aff31}}
\and A.~Zacchei\orcid{0000-0003-0396-1192}\inst{\ref{aff27},\ref{aff28}}
\and G.~Zamorani\orcid{0000-0002-2318-301X}\inst{\ref{aff37}}
\and F.~M.~Zerbi\inst{\ref{aff56}}
\and E.~Zucca\orcid{0000-0002-5845-8132}\inst{\ref{aff37}}
\and V.~Allevato\orcid{0000-0001-7232-5152}\inst{\ref{aff59}}
\and M.~Ballardini\orcid{0000-0003-4481-3559}\inst{\ref{aff41},\ref{aff42},\ref{aff37}}
\and M.~Bolzonella\orcid{0000-0003-3278-4607}\inst{\ref{aff37}}
\and E.~Bozzo\orcid{0000-0002-8201-1525}\inst{\ref{aff75}}
\and C.~Burigana\orcid{0000-0002-3005-5796}\inst{\ref{aff128},\ref{aff80}}
\and R.~Cabanac\orcid{0000-0001-6679-2600}\inst{\ref{aff24}}
\and M.~Calabrese\orcid{0000-0002-2637-2422}\inst{\ref{aff129},\ref{aff9}}
\and A.~Cappi\inst{\ref{aff37},\ref{aff105}}
\and D.~Di~Ferdinando\inst{\ref{aff38}}
\and J.~A.~Escartin~Vigo\inst{\ref{aff31}}
\and L.~Gabarra\orcid{0000-0002-8486-8856}\inst{\ref{aff34}}
\and W.~G.~Hartley\inst{\ref{aff75}}
\and J.~Mart\'{i}n-Fleitas\orcid{0000-0002-8594-569X}\inst{\ref{aff111}}
\and S.~Matthew\orcid{0000-0001-8448-1697}\inst{\ref{aff16}}
\and M.~Maturi\orcid{0000-0002-3517-2422}\inst{\ref{aff32},\ref{aff130}}
\and N.~Mauri\orcid{0000-0001-8196-1548}\inst{\ref{aff65},\ref{aff38}}
\and R.~B.~Metcalf\orcid{0000-0003-3167-2574}\inst{\ref{aff43},\ref{aff37}}
\and M.~P\"ontinen\orcid{0000-0001-5442-2530}\inst{\ref{aff96}}
\and C.~Porciani\orcid{0000-0002-7797-2508}\inst{\ref{aff102}}
\and I.~Risso\orcid{0000-0003-2525-7761}\inst{\ref{aff131}}
\and V.~Scottez\inst{\ref{aff108},\ref{aff132}}
\and M.~Sereno\orcid{0000-0003-0302-0325}\inst{\ref{aff37},\ref{aff38}}
\and M.~Tenti\orcid{0000-0002-4254-5901}\inst{\ref{aff38}}
\and M.~Viel\orcid{0000-0002-2642-5707}\inst{\ref{aff28},\ref{aff27},\ref{aff25},\ref{aff29},\ref{aff26}}
\and M.~Wiesmann\orcid{0009-0000-8199-5860}\inst{\ref{aff87}}
\and Y.~Akrami\orcid{0000-0002-2407-7956}\inst{\ref{aff133},\ref{aff134}}
\and S.~Alvi\orcid{0000-0001-5779-8568}\inst{\ref{aff41}}
\and I.~T.~Andika\orcid{0000-0001-6102-9526}\inst{\ref{aff135},\ref{aff136}}
\and S.~Anselmi\orcid{0000-0002-3579-9583}\inst{\ref{aff77},\ref{aff119},\ref{aff137}}
\and M.~Archidiacono\orcid{0000-0003-4952-9012}\inst{\ref{aff85},\ref{aff86}}
\and F.~Atrio-Barandela\orcid{0000-0002-2130-2513}\inst{\ref{aff138}}
\and A.~Balaguera-Antolinez\orcid{0000-0001-5028-3035}\inst{\ref{aff66},\ref{aff139}}
\and M.~Bethermin\orcid{0000-0002-3915-2015}\inst{\ref{aff140}}
\and S.~Borgani\orcid{0000-0001-6151-6439}\inst{\ref{aff141},\ref{aff28},\ref{aff27},\ref{aff29},\ref{aff26}}
\and M.~L.~Brown\orcid{0000-0002-0370-8077}\inst{\ref{aff48}}
\and S.~Bruton\orcid{0000-0002-6503-5218}\inst{\ref{aff142}}
\and A.~Calabro\orcid{0000-0003-2536-1614}\inst{\ref{aff1}}
\and B.~Camacho~Quevedo\orcid{0000-0002-8789-4232}\inst{\ref{aff21},\ref{aff22}}
\and F.~Caro\inst{\ref{aff1}}
\and C.~S.~Carvalho\inst{\ref{aff122}}
\and T.~Castro\orcid{0000-0002-6292-3228}\inst{\ref{aff27},\ref{aff29},\ref{aff28},\ref{aff26}}
\and F.~Cogato\orcid{0000-0003-4632-6113}\inst{\ref{aff43},\ref{aff37}}
\and S.~Conseil\orcid{0000-0002-3657-4191}\inst{\ref{aff68}}
\and S.~Contarini\orcid{0000-0002-9843-723X}\inst{\ref{aff31}}
\and A.~R.~Cooray\orcid{0000-0002-3892-0190}\inst{\ref{aff143}}
\and O.~Cucciati\orcid{0000-0002-9336-7551}\inst{\ref{aff37}}
\and F.~De~Paolis\orcid{0000-0001-6460-7563}\inst{\ref{aff144},\ref{aff145},\ref{aff146}}
\and G.~Desprez\orcid{0000-0001-8325-1742}\inst{\ref{aff126}}
\and A.~D\'iaz-S\'anchez\orcid{0000-0003-0748-4768}\inst{\ref{aff147}}
\and J.~J.~Diaz\orcid{0000-0003-2101-1078}\inst{\ref{aff66}}
\and J.~M.~Diego\orcid{0000-0001-9065-3926}\inst{\ref{aff45}}
\and P.~Dimauro\orcid{0000-0001-7399-2854}\inst{\ref{aff1},\ref{aff148}}
\and A.~Enia\orcid{0000-0002-0200-2857}\inst{\ref{aff57},\ref{aff37}}
\and Y.~Fang\inst{\ref{aff84}}
\and A.~G.~Ferrari\orcid{0009-0005-5266-4110}\inst{\ref{aff38}}
\and P.~G.~Ferreira\orcid{0000-0002-3021-2851}\inst{\ref{aff34}}
\and A.~Finoguenov\orcid{0000-0002-4606-5403}\inst{\ref{aff96}}
\and A.~Fontana\orcid{0000-0003-3820-2823}\inst{\ref{aff1}}
\and A.~Franco\orcid{0000-0002-4761-366X}\inst{\ref{aff145},\ref{aff144},\ref{aff146}}
\and K.~Ganga\orcid{0000-0001-8159-8208}\inst{\ref{aff106}}
\and J.~Garc\'ia-Bellido\orcid{0000-0002-9370-8360}\inst{\ref{aff133}}
\and T.~Gasparetto\orcid{0000-0002-7913-4866}\inst{\ref{aff27}}
\and V.~Gautard\inst{\ref{aff149}}
\and E.~Gaztanaga\orcid{0000-0001-9632-0815}\inst{\ref{aff22},\ref{aff21},\ref{aff4}}
\and F.~Giacomini\orcid{0000-0002-3129-2814}\inst{\ref{aff38}}
\and F.~Gianotti\orcid{0000-0003-4666-119X}\inst{\ref{aff37}}
\and G.~Gozaliasl\orcid{0000-0002-0236-919X}\inst{\ref{aff150},\ref{aff96}}
\and A.~Gruppuso\orcid{0000-0001-9272-5292}\inst{\ref{aff37},\ref{aff38}}
\and M.~Guidi\orcid{0000-0001-9408-1101}\inst{\ref{aff57},\ref{aff37}}
\and C.~M.~Gutierrez\orcid{0000-0001-7854-783X}\inst{\ref{aff151}}
\and C.~Hern\'andez-Monteagudo\orcid{0000-0001-5471-9166}\inst{\ref{aff118},\ref{aff66}}
\and H.~Hildebrandt\orcid{0000-0002-9814-3338}\inst{\ref{aff152}}
\and J.~Hjorth\orcid{0000-0002-4571-2306}\inst{\ref{aff113}}
\and J.~J.~E.~Kajava\orcid{0000-0002-3010-8333}\inst{\ref{aff153},\ref{aff154}}
\and Y.~Kang\orcid{0009-0000-8588-7250}\inst{\ref{aff75}}
\and V.~Kansal\orcid{0000-0002-4008-6078}\inst{\ref{aff155},\ref{aff156}}
\and D.~Karagiannis\orcid{0000-0002-4927-0816}\inst{\ref{aff41},\ref{aff157}}
\and K.~Kiiveri\inst{\ref{aff95}}
\and C.~C.~Kirkpatrick\inst{\ref{aff95}}
\and S.~Kruk\orcid{0000-0001-8010-8879}\inst{\ref{aff55}}
\and F.~Lepori\orcid{0009-0000-5061-7138}\inst{\ref{aff158}}
\and G.~Leroy\orcid{0009-0004-2523-4425}\inst{\ref{aff159},\ref{aff104}}
\and J.~Lesgourgues\orcid{0000-0001-7627-353X}\inst{\ref{aff17}}
\and L.~Leuzzi\orcid{0009-0006-4479-7017}\inst{\ref{aff43},\ref{aff37}}
\and T.~I.~Liaudat\orcid{0000-0002-9104-314X}\inst{\ref{aff160}}
\and S.~J.~Liu\orcid{0000-0001-7680-2139}\inst{\ref{aff76}}
\and A.~Loureiro\orcid{0000-0002-4371-0876}\inst{\ref{aff161},\ref{aff162}}
\and J.~Macias-Perez\orcid{0000-0002-5385-2763}\inst{\ref{aff163}}
\and G.~Maggio\orcid{0000-0003-4020-4836}\inst{\ref{aff27}}
\and M.~Magliocchetti\orcid{0000-0001-9158-4838}\inst{\ref{aff76}}
\and F.~Mannucci\orcid{0000-0002-4803-2381}\inst{\ref{aff164}}
\and R.~Maoli\orcid{0000-0002-6065-3025}\inst{\ref{aff165},\ref{aff1}}
\and C.~J.~A.~P.~Martins\orcid{0000-0002-4886-9261}\inst{\ref{aff166},\ref{aff60}}
\and L.~Maurin\orcid{0000-0002-8406-0857}\inst{\ref{aff50}}
\and M.~Migliaccio\inst{\ref{aff167},\ref{aff168}}
\and M.~Miluzio\inst{\ref{aff55},\ref{aff169}}
\and P.~Monaco\orcid{0000-0003-2083-7564}\inst{\ref{aff141},\ref{aff27},\ref{aff29},\ref{aff28}}
\and G.~Morgante\inst{\ref{aff37}}
\and S.~Nadathur\orcid{0000-0001-9070-3102}\inst{\ref{aff4}}
\and K.~Naidoo\orcid{0000-0002-9182-1802}\inst{\ref{aff4}}
\and A.~Navarro-Alsina\orcid{0000-0002-3173-2592}\inst{\ref{aff102}}
\and S.~Nesseris\orcid{0000-0002-0567-0324}\inst{\ref{aff133}}
\and L.~Pagano\orcid{0000-0003-1820-5998}\inst{\ref{aff41},\ref{aff42}}
\and F.~Passalacqua\orcid{0000-0002-8606-4093}\inst{\ref{aff119},\ref{aff77}}
\and K.~Paterson\orcid{0000-0001-8340-3486}\inst{\ref{aff93}}
\and L.~Patrizii\inst{\ref{aff38}}
\and A.~Pisani\orcid{0000-0002-6146-4437}\inst{\ref{aff78}}
\and D.~Potter\orcid{0000-0002-0757-5195}\inst{\ref{aff158}}
\and S.~Quai\orcid{0000-0002-0449-8163}\inst{\ref{aff43},\ref{aff37}}
\and M.~Radovich\orcid{0000-0002-3585-866X}\inst{\ref{aff83}}
\and P.~Reimberg\orcid{0000-0003-3410-0280}\inst{\ref{aff108}}
\and S.~Sacquegna\orcid{0000-0002-8433-6630}\inst{\ref{aff144},\ref{aff145},\ref{aff146}}
\and M.~Sahl\'en\orcid{0000-0003-0973-4804}\inst{\ref{aff170}}
\and D.~B.~Sanders\orcid{0000-0002-1233-9998}\inst{\ref{aff64}}
\and E.~Sarpa\orcid{0000-0002-1256-655X}\inst{\ref{aff25},\ref{aff26},\ref{aff29}}
\and J.~Schaye\orcid{0000-0002-0668-5560}\inst{\ref{aff15}}
\and A.~Schneider\orcid{0000-0001-7055-8104}\inst{\ref{aff158}}
\and M.~Schultheis\inst{\ref{aff105}}
\and A.~Silvestri\orcid{0000-0001-6904-5061}\inst{\ref{aff14}}
\and L.~C.~Smith\orcid{0000-0002-3259-2771}\inst{\ref{aff171}}
\and C.~Tao\orcid{0000-0001-7961-8177}\inst{\ref{aff78}}
\and G.~Testera\inst{\ref{aff18}}
\and R.~Teyssier\orcid{0000-0001-7689-0933}\inst{\ref{aff172}}
\and S.~Tosi\orcid{0000-0002-7275-9193}\inst{\ref{aff19},\ref{aff18},\ref{aff56}}
\and A.~Troja\orcid{0000-0003-0239-4595}\inst{\ref{aff119},\ref{aff77}}
\and M.~Tucci\inst{\ref{aff75}}
\and C.~Valieri\inst{\ref{aff38}}
\and A.~Venhola\orcid{0000-0001-6071-4564}\inst{\ref{aff173}}
\and D.~Vergani\orcid{0000-0003-0898-2216}\inst{\ref{aff37}}
\and F.~Vernizzi\orcid{0000-0003-3426-2802}\inst{\ref{aff174}}
\and G.~Verza\orcid{0000-0002-1886-8348}\inst{\ref{aff175}}
\and N.~A.~Walton\orcid{0000-0003-3983-8778}\inst{\ref{aff171}}}
										   
\institute{INAF-Osservatorio Astronomico di Roma, Via Frascati 33, 00078 Monteporzio Catone, Italy\label{aff1}
\and
INFN-Sezione di Roma, Piazzale Aldo Moro, 2 - c/o Dipartimento di Fisica, Edificio G. Marconi, 00185 Roma, Italy\label{aff2}
\and
Centro de Investigaciones Energ\'eticas, Medioambientales y Tecnol\'ogicas (CIEMAT), Avenida Complutense 40, 28040 Madrid, Spain\label{aff3}
\and
Institute of Cosmology and Gravitation, University of Portsmouth, Portsmouth PO1 3FX, UK\label{aff4}
\and
Waterloo Centre for Astrophysics, University of Waterloo, Waterloo, Ontario N2L 3G1, Canada\label{aff5}
\and
Department of Physics and Astronomy, University of Waterloo, Waterloo, Ontario N2L 3G1, Canada\label{aff6}
\and
Center for Data-Driven Discovery, Kavli IPMU (WPI), UTIAS, The University of Tokyo, Kashiwa, Chiba 277-8583, Japan\label{aff7}
\and
Laboratoire d'etude de l'Univers et des phenomenes eXtremes, Observatoire de Paris, Universit\'e PSL, Sorbonne Universit\'e, CNRS, 92190 Meudon, France\label{aff8}
\and
INAF-IASF Milano, Via Alfonso Corti 12, 20133 Milano, Italy\label{aff9}
\and
Dipartimento di Fisica, Universit\`a degli Studi di Torino, Via P. Giuria 1, 10125 Torino, Italy\label{aff10}
\and
INFN-Sezione di Torino, Via P. Giuria 1, 10125 Torino, Italy\label{aff11}
\and
INAF-Osservatorio Astrofisico di Torino, Via Osservatorio 20, 10025 Pino Torinese (TO), Italy\label{aff12}
\and
European Space Agency/ESTEC, Keplerlaan 1, 2201 AZ Noordwijk, The Netherlands\label{aff13}
\and
Institute Lorentz, Leiden University, Niels Bohrweg 2, 2333 CA Leiden, The Netherlands\label{aff14}
\and
Leiden Observatory, Leiden University, Einsteinweg 55, 2333 CC Leiden, The Netherlands\label{aff15}
\and
Institute for Astronomy, University of Edinburgh, Royal Observatory, Blackford Hill, Edinburgh EH9 3HJ, UK\label{aff16}
\and
Institute for Theoretical Particle Physics and Cosmology (TTK), RWTH Aachen University, 52056 Aachen, Germany\label{aff17}
\and
INFN-Sezione di Genova, Via Dodecaneso 33, 16146, Genova, Italy\label{aff18}
\and
Dipartimento di Fisica, Universit\`a di Genova, Via Dodecaneso 33, 16146, Genova, Italy\label{aff19}
\and
Universit\'e Paris-Saclay, Universit\'e Paris Cit\'e, CEA, CNRS, AIM, 91191, Gif-sur-Yvette, France\label{aff20}
\and
Institut d'Estudis Espacials de Catalunya (IEEC),  Edifici RDIT, Campus UPC, 08860 Castelldefels, Barcelona, Spain\label{aff21}
\and
Institute of Space Sciences (ICE, CSIC), Campus UAB, Carrer de Can Magrans, s/n, 08193 Barcelona, Spain\label{aff22}
\and
Universit\'e Paris-Saclay, CNRS/IN2P3, IJCLab, 91405 Orsay, France\label{aff23}
\and
Institut de Recherche en Astrophysique et Plan\'etologie (IRAP), Universit\'e de Toulouse, CNRS, UPS, CNES, 14 Av. Edouard Belin, 31400 Toulouse, France\label{aff24}
\and
SISSA, International School for Advanced Studies, Via Bonomea 265, 34136 Trieste TS, Italy\label{aff25}
\and
ICSC - Centro Nazionale di Ricerca in High Performance Computing, Big Data e Quantum Computing, Via Magnanelli 2, Bologna, Italy\label{aff26}
\and
INAF-Osservatorio Astronomico di Trieste, Via G. B. Tiepolo 11, 34143 Trieste, Italy\label{aff27}
\and
IFPU, Institute for Fundamental Physics of the Universe, via Beirut 2, 34151 Trieste, Italy\label{aff28}
\and
INFN, Sezione di Trieste, Via Valerio 2, 34127 Trieste TS, Italy\label{aff29}
\and
INAF - Osservatorio Astronomico di Brera, via Emilio Bianchi 46, 23807 Merate, Italy\label{aff30}
\and
Max Planck Institute for Extraterrestrial Physics, Giessenbachstr. 1, 85748 Garching, Germany\label{aff31}
\and
Institut f\"ur Theoretische Physik, University of Heidelberg, Philosophenweg 16, 69120 Heidelberg, Germany\label{aff32}
\and
Universit\'e St Joseph; Faculty of Sciences, Beirut, Lebanon\label{aff33}
\and
Department of Physics, Oxford University, Keble Road, Oxford OX1 3RH, UK\label{aff34}
\and
LINKS Foundation, Via Pier Carlo Boggio, 61 10138 Torino, Italy\label{aff35}
\and
Institute for Particle Physics and Astrophysics, Dept. of Physics, ETH Zurich, Wolfgang-Pauli-Strasse 27, 8093 Zurich, Switzerland\label{aff36}
\and
INAF-Osservatorio di Astrofisica e Scienza dello Spazio di Bologna, Via Piero Gobetti 93/3, 40129 Bologna, Italy\label{aff37}
\and
INFN-Sezione di Bologna, Viale Berti Pichat 6/2, 40127 Bologna, Italy\label{aff38}
\and
DAMTP, Centre for Mathematical Sciences, Wilberforce Road, Cambridge CB3 0WA, UK\label{aff39}
\and
Kavli Institute for Cosmology Cambridge, Madingley Road, Cambridge, CB3 0HA, UK\label{aff40}
\and
Dipartimento di Fisica e Scienze della Terra, Universit\`a degli Studi di Ferrara, Via Giuseppe Saragat 1, 44122 Ferrara, Italy\label{aff41}
\and
Istituto Nazionale di Fisica Nucleare, Sezione di Ferrara, Via Giuseppe Saragat 1, 44122 Ferrara, Italy\label{aff42}
\and
Dipartimento di Fisica e Astronomia "Augusto Righi" - Alma Mater Studiorum Universit\`a di Bologna, via Piero Gobetti 93/2, 40129 Bologna, Italy\label{aff43}
\and
Universit\'e de Gen\`eve, D\'epartement de Physique Th\'eorique and Centre for Astroparticle Physics, 24 quai Ernest-Ansermet, CH-1211 Gen\`eve 4, Switzerland\label{aff44}
\and
Instituto de F\'isica de Cantabria, Edificio Juan Jord\'a, Avenida de los Castros, 39005 Santander, Spain\label{aff45}
\and
Higgs Centre for Theoretical Physics, School of Physics and Astronomy, The University of Edinburgh, Edinburgh EH9 3FD, UK\label{aff46}
\and
Aix-Marseille Universit\'e, Universit\'e de Toulon, CNRS, CPT, Marseille, France\label{aff47}
\and
Jodrell Bank Centre for Astrophysics, Department of Physics and Astronomy, University of Manchester, Oxford Road, Manchester M13 9PL, UK\label{aff48}
\and
Universit\'e Libre de Bruxelles (ULB), Service de Physique Th\'eorique CP225, Boulevard du Triophe, 1050 Bruxelles, Belgium\label{aff49}
\and
Universit\'e Paris-Saclay, CNRS, Institut d'astrophysique spatiale, 91405, Orsay, France\label{aff50}
\and
Departamento de F\'isica, FCFM, Universidad de Chile, Blanco Encalada 2008, Santiago, Chile\label{aff51}
\and
Mathematical Institute, University of Leiden, Einsteinweg 55, 2333 CA Leiden, The Netherlands\label{aff52}
\and
Center for Cosmology and AstroParticle Physics, The Ohio State University, 191 West Woodruff Avenue, Columbus, OH 43210, USA\label{aff53}
\and
Department of Physics, The Ohio State University, Columbus, OH 43210, USA\label{aff54}
\and
ESAC/ESA, Camino Bajo del Castillo, s/n., Urb. Villafranca del Castillo, 28692 Villanueva de la Ca\~nada, Madrid, Spain\label{aff55}
\and
INAF-Osservatorio Astronomico di Brera, Via Brera 28, 20122 Milano, Italy\label{aff56}
\and
Dipartimento di Fisica e Astronomia, Universit\`a di Bologna, Via Gobetti 93/2, 40129 Bologna, Italy\label{aff57}
\and
Department of Physics "E. Pancini", University Federico II, Via Cinthia 6, 80126, Napoli, Italy\label{aff58}
\and
INAF-Osservatorio Astronomico di Capodimonte, Via Moiariello 16, 80131 Napoli, Italy\label{aff59}
\and
Instituto de Astrof\'isica e Ci\^encias do Espa\c{c}o, Universidade do Porto, CAUP, Rua das Estrelas, PT4150-762 Porto, Portugal\label{aff60}
\and
Faculdade de Ci\^encias da Universidade do Porto, Rua do Campo de Alegre, 4150-007 Porto, Portugal\label{aff61}
\and
Port d'Informaci\'{o} Cient\'{i}fica, Campus UAB, C. Albareda s/n, 08193 Bellaterra (Barcelona), Spain\label{aff62}
\and
INFN section of Naples, Via Cinthia 6, 80126, Napoli, Italy\label{aff63}
\and
Institute for Astronomy, University of Hawaii, 2680 Woodlawn Drive, Honolulu, HI 96822, USA\label{aff64}
\and
Dipartimento di Fisica e Astronomia "Augusto Righi" - Alma Mater Studiorum Universit\`a di Bologna, Viale Berti Pichat 6/2, 40127 Bologna, Italy\label{aff65}
\and
Instituto de Astrof\'{\i}sica de Canarias, V\'{\i}a L\'actea, 38205 La Laguna, Tenerife, Spain\label{aff66}
\and
European Space Agency/ESRIN, Largo Galileo Galilei 1, 00044 Frascati, Roma, Italy\label{aff67}
\and
Universit\'e Claude Bernard Lyon 1, CNRS/IN2P3, IP2I Lyon, UMR 5822, Villeurbanne, F-69100, France\label{aff68}
\and
Institut de Ci\`{e}ncies del Cosmos (ICCUB), Universitat de Barcelona (IEEC-UB), Mart\'{i} i Franqu\`{e}s 1, 08028 Barcelona, Spain\label{aff69}
\and
Instituci\'o Catalana de Recerca i Estudis Avan\c{c}ats (ICREA), Passeig de Llu\'{\i}s Companys 23, 08010 Barcelona, Spain\label{aff70}
\and
UCB Lyon 1, CNRS/IN2P3, IUF, IP2I Lyon, 4 rue Enrico Fermi, 69622 Villeurbanne, France\label{aff71}
\and
Mullard Space Science Laboratory, University College London, Holmbury St Mary, Dorking, Surrey RH5 6NT, UK\label{aff72}
\and
Departamento de F\'isica, Faculdade de Ci\^encias, Universidade de Lisboa, Edif\'icio C8, Campo Grande, PT1749-016 Lisboa, Portugal\label{aff73}
\and
Instituto de Astrof\'isica e Ci\^encias do Espa\c{c}o, Faculdade de Ci\^encias, Universidade de Lisboa, Campo Grande, 1749-016 Lisboa, Portugal\label{aff74}
\and
Department of Astronomy, University of Geneva, ch. d'Ecogia 16, 1290 Versoix, Switzerland\label{aff75}
\and
INAF-Istituto di Astrofisica e Planetologia Spaziali, via del Fosso del Cavaliere, 100, 00100 Roma, Italy\label{aff76}
\and
INFN-Padova, Via Marzolo 8, 35131 Padova, Italy\label{aff77}
\and
Aix-Marseille Universit\'e, CNRS/IN2P3, CPPM, Marseille, France\label{aff78}
\and
Space Science Data Center, Italian Space Agency, via del Politecnico snc, 00133 Roma, Italy\label{aff79}
\and
INFN-Bologna, Via Irnerio 46, 40126 Bologna, Italy\label{aff80}
\and
School of Physics, HH Wills Physics Laboratory, University of Bristol, Tyndall Avenue, Bristol, BS8 1TL, UK\label{aff81}
\and
FRACTAL S.L.N.E., calle Tulip\'an 2, Portal 13 1A, 28231, Las Rozas de Madrid, Spain\label{aff82}
\and
INAF-Osservatorio Astronomico di Padova, Via dell'Osservatorio 5, 35122 Padova, Italy\label{aff83}
\and
Universit\"ats-Sternwarte M\"unchen, Fakult\"at f\"ur Physik, Ludwig-Maximilians-Universit\"at M\"unchen, Scheinerstrasse 1, 81679 M\"unchen, Germany\label{aff84}
\and
Dipartimento di Fisica "Aldo Pontremoli", Universit\`a degli Studi di Milano, Via Celoria 16, 20133 Milano, Italy\label{aff85}
\and
INFN-Sezione di Milano, Via Celoria 16, 20133 Milano, Italy\label{aff86}
\and
Institute of Theoretical Astrophysics, University of Oslo, P.O. Box 1029 Blindern, 0315 Oslo, Norway\label{aff87}
\and
Jet Propulsion Laboratory, California Institute of Technology, 4800 Oak Grove Drive, Pasadena, CA, 91109, USA\label{aff88}
\and
Department of Physics, Lancaster University, Lancaster, LA1 4YB, UK\label{aff89}
\and
Felix Hormuth Engineering, Goethestr. 17, 69181 Leimen, Germany\label{aff90}
\and
Technical University of Denmark, Elektrovej 327, 2800 Kgs. Lyngby, Denmark\label{aff91}
\and
Cosmic Dawn Center (DAWN), Denmark\label{aff92}
\and
Max-Planck-Institut f\"ur Astronomie, K\"onigstuhl 17, 69117 Heidelberg, Germany\label{aff93}
\and
NASA Goddard Space Flight Center, Greenbelt, MD 20771, USA\label{aff94}
\and
Department of Physics and Helsinki Institute of Physics, Gustaf H\"allstr\"omin katu 2, 00014 University of Helsinki, Finland\label{aff95}
\and
Department of Physics, P.O. Box 64, 00014 University of Helsinki, Finland\label{aff96}
\and
Helsinki Institute of Physics, Gustaf H{\"a}llstr{\"o}min katu 2, University of Helsinki, Helsinki, Finland\label{aff97}
\and
Department of Physics and Astronomy, University College London, Gower Street, London WC1E 6BT, UK\label{aff98}
\and
SKA Observatory, Jodrell Bank, Lower Withington, Macclesfield, Cheshire SK11 9FT, UK\label{aff99}
\and
Centre de Calcul de l'IN2P3/CNRS, 21 avenue Pierre de Coubertin 69627 Villeurbanne Cedex, France\label{aff100}
\and
University of Applied Sciences and Arts of Northwestern Switzerland, School of Computer Science, 5210 Windisch, Switzerland\label{aff101}
\and
Universit\"at Bonn, Argelander-Institut f\"ur Astronomie, Auf dem H\"ugel 71, 53121 Bonn, Germany\label{aff102}
\and
Aix-Marseille Universit\'e, CNRS, CNES, LAM, Marseille, France\label{aff103}
\and
Department of Physics, Institute for Computational Cosmology, Durham University, South Road, Durham, DH1 3LE, UK\label{aff104}
\and
Universit\'e C\^{o}te d'Azur, Observatoire de la C\^{o}te d'Azur, CNRS, Laboratoire Lagrange, Bd de l'Observatoire, CS 34229, 06304 Nice cedex 4, France\label{aff105}
\and
Universit\'e Paris Cit\'e, CNRS, Astroparticule et Cosmologie, 75013 Paris, France\label{aff106}
\and
CNRS-UCB International Research Laboratory, Centre Pierre Bin\'etruy, IRL2007, CPB-IN2P3, Berkeley, USA\label{aff107}
\and
Institut d'Astrophysique de Paris, 98bis Boulevard Arago, 75014, Paris, France\label{aff108}
\and
Institut d'Astrophysique de Paris, UMR 7095, CNRS, and Sorbonne Universit\'e, 98 bis boulevard Arago, 75014 Paris, France\label{aff109}
\and
Institute of Physics, Laboratory of Astrophysics, Ecole Polytechnique F\'ed\'erale de Lausanne (EPFL), Observatoire de Sauverny, 1290 Versoix, Switzerland\label{aff110}
\and
Aurora Technology for European Space Agency (ESA), Camino bajo del Castillo, s/n, Urbanizacion Villafranca del Castillo, Villanueva de la Ca\~nada, 28692 Madrid, Spain\label{aff111}
\and
Institut de F\'{i}sica d'Altes Energies (IFAE), The Barcelona Institute of Science and Technology, Campus UAB, 08193 Bellaterra (Barcelona), Spain\label{aff112}
\and
DARK, Niels Bohr Institute, University of Copenhagen, Jagtvej 155, 2200 Copenhagen, Denmark\label{aff113}
\and
Perimeter Institute for Theoretical Physics, Waterloo, Ontario N2L 2Y5, Canada\label{aff114}
\and
Centre National d'Etudes Spatiales -- Centre spatial de Toulouse, 18 avenue Edouard Belin, 31401 Toulouse Cedex 9, France\label{aff115}
\and
Institute of Space Science, Str. Atomistilor, nr. 409 M\u{a}gurele, Ilfov, 077125, Romania\label{aff116}
\and
Consejo Superior de Investigaciones Cientificas, Calle Serrano 117, 28006 Madrid, Spain\label{aff117}
\and
Universidad de La Laguna, Departamento de Astrof\'{\i}sica, 38206 La Laguna, Tenerife, Spain\label{aff118}
\and
Dipartimento di Fisica e Astronomia "G. Galilei", Universit\`a di Padova, Via Marzolo 8, 35131 Padova, Italy\label{aff119}
\and
Universit\"at Innsbruck, Institut f\"ur Astro- und Teilchenphysik, Technikerstr. 25/8, 6020 Innsbruck, Austria\label{aff120}
\and
Satlantis, University Science Park, Sede Bld 48940, Leioa-Bilbao, Spain\label{aff121}
\and
Instituto de Astrof\'isica e Ci\^encias do Espa\c{c}o, Faculdade de Ci\^encias, Universidade de Lisboa, Tapada da Ajuda, 1349-018 Lisboa, Portugal\label{aff122}
\and
Cosmic Dawn Center (DAWN)\label{aff123}
\and
Niels Bohr Institute, University of Copenhagen, Jagtvej 128, 2200 Copenhagen, Denmark\label{aff124}
\and
Universidad Polit\'ecnica de Cartagena, Departamento de Electr\'onica y Tecnolog\'ia de Computadoras,  Plaza del Hospital 1, 30202 Cartagena, Spain\label{aff125}
\and
Kapteyn Astronomical Institute, University of Groningen, PO Box 800, 9700 AV Groningen, The Netherlands\label{aff126}
\and
Infrared Processing and Analysis Center, California Institute of Technology, Pasadena, CA 91125, USA\label{aff127}
\and
INAF, Istituto di Radioastronomia, Via Piero Gobetti 101, 40129 Bologna, Italy\label{aff128}
\and
Astronomical Observatory of the Autonomous Region of the Aosta Valley (OAVdA), Loc. Lignan 39, I-11020, Nus (Aosta Valley), Italy\label{aff129}
\and
Zentrum f\"ur Astronomie, Universit\"at Heidelberg, Philosophenweg 12, 69120 Heidelberg, Germany\label{aff130}
\and
INAF-Osservatorio Astronomico di Brera, Via Brera 28, 20122 Milano, Italy, and INFN-Sezione di Genova, Via Dodecaneso 33, 16146, Genova, Italy\label{aff131}
\and
ICL, Junia, Universit\'e Catholique de Lille, LITL, 59000 Lille, France\label{aff132}
\and
Instituto de F\'isica Te\'orica UAM-CSIC, Campus de Cantoblanco, 28049 Madrid, Spain\label{aff133}
\and
CERCA/ISO, Department of Physics, Case Western Reserve University, 10900 Euclid Avenue, Cleveland, OH 44106, USA\label{aff134}
\and
Technical University of Munich, TUM School of Natural Sciences, Physics Department, James-Franck-Str.~1, 85748 Garching, Germany\label{aff135}
\and
Max-Planck-Institut f\"ur Astrophysik, Karl-Schwarzschild-Str.~1, 85748 Garching, Germany\label{aff136}
\and
Laboratoire Univers et Th\'eorie, Observatoire de Paris, Universit\'e PSL, Universit\'e Paris Cit\'e, CNRS, 92190 Meudon, France\label{aff137}
\and
Departamento de F{\'\i}sica Fundamental. Universidad de Salamanca. Plaza de la Merced s/n. 37008 Salamanca, Spain\label{aff138}
\and
Instituto de Astrof\'isica de Canarias (IAC); Departamento de Astrof\'isica, Universidad de La Laguna (ULL), 38200, La Laguna, Tenerife, Spain\label{aff139}
\and
Universit\'e de Strasbourg, CNRS, Observatoire astronomique de Strasbourg, UMR 7550, 67000 Strasbourg, France\label{aff140}
\and
Dipartimento di Fisica - Sezione di Astronomia, Universit\`a di Trieste, Via Tiepolo 11, 34131 Trieste, Italy\label{aff141}
\and
California Institute of Technology, 1200 E California Blvd, Pasadena, CA 91125, USA\label{aff142}
\and
Department of Physics \& Astronomy, University of California Irvine, Irvine CA 92697, USA\label{aff143}
\and
Department of Mathematics and Physics E. De Giorgi, University of Salento, Via per Arnesano, CP-I93, 73100, Lecce, Italy\label{aff144}
\and
INFN, Sezione di Lecce, Via per Arnesano, CP-193, 73100, Lecce, Italy\label{aff145}
\and
INAF-Sezione di Lecce, c/o Dipartimento Matematica e Fisica, Via per Arnesano, 73100, Lecce, Italy\label{aff146}
\and
Departamento F\'isica Aplicada, Universidad Polit\'ecnica de Cartagena, Campus Muralla del Mar, 30202 Cartagena, Murcia, Spain\label{aff147}
\and
Observatorio Nacional, Rua General Jose Cristino, 77-Bairro Imperial de Sao Cristovao, Rio de Janeiro, 20921-400, Brazil\label{aff148}
\and
CEA Saclay, DFR/IRFU, Service d'Astrophysique, Bat. 709, 91191 Gif-sur-Yvette, France\label{aff149}
\and
Department of Computer Science, Aalto University, PO Box 15400, Espoo, FI-00 076, Finland\label{aff150}
\and
Instituto de Astrof\'\i sica de Canarias, c/ Via Lactea s/n, La Laguna 38200, Spain. Departamento de Astrof\'\i sica de la Universidad de La Laguna, Avda. Francisco Sanchez, La Laguna, 38200, Spain\label{aff151}
\and
Ruhr University Bochum, Faculty of Physics and Astronomy, Astronomical Institute (AIRUB), German Centre for Cosmological Lensing (GCCL), 44780 Bochum, Germany\label{aff152}
\and
Department of Physics and Astronomy, Vesilinnantie 5, 20014 University of Turku, Finland\label{aff153}
\and
Serco for European Space Agency (ESA), Camino bajo del Castillo, s/n, Urbanizacion Villafranca del Castillo, Villanueva de la Ca\~nada, 28692 Madrid, Spain\label{aff154}
\and
ARC Centre of Excellence for Dark Matter Particle Physics, Melbourne, Australia\label{aff155}
\and
Centre for Astrophysics \& Supercomputing, Swinburne University of Technology,  Hawthorn, Victoria 3122, Australia\label{aff156}
\and
Department of Physics and Astronomy, University of the Western Cape, Bellville, Cape Town, 7535, South Africa\label{aff157}
\and
Department of Astrophysics, University of Zurich, Winterthurerstrasse 190, 8057 Zurich, Switzerland\label{aff158}
\and
Department of Physics, Centre for Extragalactic Astronomy, Durham University, South Road, Durham, DH1 3LE, UK\label{aff159}
\and
IRFU, CEA, Universit\'e Paris-Saclay 91191 Gif-sur-Yvette Cedex, France\label{aff160}
\and
Oskar Klein Centre for Cosmoparticle Physics, Department of Physics, Stockholm University, Stockholm, SE-106 91, Sweden\label{aff161}
\and
Astrophysics Group, Blackett Laboratory, Imperial College London, London SW7 2AZ, UK\label{aff162}
\and
Univ. Grenoble Alpes, CNRS, Grenoble INP, LPSC-IN2P3, 53, Avenue des Martyrs, 38000, Grenoble, France\label{aff163}
\and
INAF-Osservatorio Astrofisico di Arcetri, Largo E. Fermi 5, 50125, Firenze, Italy\label{aff164}
\and
Dipartimento di Fisica, Sapienza Universit\`a di Roma, Piazzale Aldo Moro 2, 00185 Roma, Italy\label{aff165}
\and
Centro de Astrof\'{\i}sica da Universidade do Porto, Rua das Estrelas, 4150-762 Porto, Portugal\label{aff166}
\and
Dipartimento di Fisica, Universit\`a di Roma Tor Vergata, Via della Ricerca Scientifica 1, Roma, Italy\label{aff167}
\and
INFN, Sezione di Roma 2, Via della Ricerca Scientifica 1, Roma, Italy\label{aff168}
\and
HE Space for European Space Agency (ESA), Camino bajo del Castillo, s/n, Urbanizacion Villafranca del Castillo, Villanueva de la Ca\~nada, 28692 Madrid, Spain\label{aff169}
\and
Theoretical astrophysics, Department of Physics and Astronomy, Uppsala University, Box 516, 751 37 Uppsala, Sweden\label{aff170}
\and
Institute of Astronomy, University of Cambridge, Madingley Road, Cambridge CB3 0HA, UK\label{aff171}
\and
Department of Astrophysical Sciences, Peyton Hall, Princeton University, Princeton, NJ 08544, USA\label{aff172}
\and
Space physics and astronomy research unit, University of Oulu, Pentti Kaiteran katu 1, FI-90014 Oulu, Finland\label{aff173}
\and
Institut de Physique Th\'eorique, CEA, CNRS, Universit\'e Paris-Saclay 91191 Gif-sur-Yvette Cedex, France\label{aff174}
\and
Center for Computational Astrophysics, Flatiron Institute, 162 5th Avenue, 10010, New York, NY, USA\label{aff175}} 


%
%
\abstract{As the statistical precision of cosmological measurements increases, the accuracy of the theoretical description of these measurements needs to increase correspondingly in order to infer the underlying cosmology that governs the Universe. 
To this end, we have created the Cosmology Likelihood for Observables in \Euclid (CLOE), which is a novel cosmological parameter inference pipeline developed within the Euclid Consortium to translate measurements and covariances into cosmological parameter constraints. In this first in a series of five papers, we describe the theoretical recipe of this code for the \Euclid primary probes. These probes are composed of the photometric 3$\times$2pt observables of cosmic shear, galaxy-galaxy lensing, and galaxy clustering, along with spectroscopic galaxy clustering. We provide this description in both Fourier and configuration space for standard and extended summary statistics, including the wide range of systematic uncertainties that affect them. This includes systematic uncertainties such as intrinsic galaxy alignments, baryonic feedback, photometric and spectroscopic redshift uncertainties, shear calibration uncertainties, sample impurities, photometric and spectroscopic galaxy biases, as well as magnification bias. 
The theoretical descriptions are further able to accommodate both Gaussian and non-Gaussian likelihoods and extended cosmologies with non-zero curvature, massive neutrinos, evolving dark energy, and simple forms of modified gravity. 
These theoretical descriptions that underpin CLOE will form a crucial component in revealing the true nature of the Universe with next-generation cosmological surveys such as \Euclid.
    }
%
%
%
%
   \titlerunning{CLOE. 1. Theoretical recipe}
   \authorrunning{Euclid Collaboration: V.~F.~Cardone et al.}

\maketitle


\section{\label{sc:Intro}Introduction}

The expansion of the Universe is accelerating. This is the unexpected picture that the first data of Type Ia supernova (SNIa) surveys presented to the scientific community at the end of the last millennium \citep{Riess1998, Perl1999}. Following the confirmation of the cosmic acceleration by an impressive variety of data (e.g.~\citealt{debernardis2000, wmap1, Eisenstein2005, schrabback10}), this discovery was awarded the Nobel Prize in 2011. Subsequent SNIa surveys have strengthened these first results via an increase in both the quantity and quality of the data, along with a careful analysis of possible systematic uncertainties (see e.g. \citealt{Pantheon+} and \citealt{DESY3SNeIA} for the latest samples). Cosmic microwave background (CMB) satellite data (first WMAP, then \Planck) confirmed the accelerated expansion and quantified the contribution of each ingredient 
to the energy budget \citep{WMAP2013, Planck2018}. In turn, galaxy surveys have collected large amounts of spectroscopic data to measure the clustering of galaxies, and provide through baryonic acoustic oscillations (BAO) and redshift-space distortions (RSD) yet additional evidence in favour of the pioneering results \citep{Eisenstein2005, WiggleZ2012,eBOSS2021, desi2024}. 

This remarkable variety of data pointed towards a new standard model of the cosmos. This includes the acceleration of the expansion measured by the SNIa Hubble diagram, the spatial flatness pointed out by the CMB temperature and polarisation anisotropies, and the subcritical content of matter determined by galaxy surveys. All of these observational pieces of evidence can be reconciled within a new single framework defining what is referred to as the {\it concordance} cosmological model. This is a spatially flat Universe whose energy content is currently dominated by the cosmological constant $\Lambda$, with cold dark matter (CDM) driving the formation of structures. 

This $\Lambda$CDM model has successfully passed an impressive number of tests over the years \citep{Peebles1984, Weinberg1989, Carroll2001, Frieman2008}. Yet, as data become more precise, tensions have started to emerge \citep{Verde2019, DiVa2022, Peri2022}. The first is the discrepancy between local model-independent estimates of the Hubble constant $H_0$ and those inferred by fitting cosmological data with different models. A second source of tension is the disagreement on the value of $S_8 = (\Omega_{\rm m}/0.3)^{1/2} \sigma_8$, with $\Omega_{\rm m}$ and $\sigma_8$ denoting the present-day matter density parameter and variance of the linear density perturbations on the scale of $8 \, h^{-1} \ {\rm Mpc}$. Although with less statistical significance than the $H_0$ tension, a disagreement appears between estimates based on the CMB and low-redshift tracers, such as cosmic shear \citep{KiDSAsgari2021, Amon2022}. These tensions can be evidence for the presence of undetected systematic uncertainties or signatures of deviations from the $\Lambda$CDM model. The latter possibility opens the way to a flood of alternative theories that share the same basic property of being in accordance with observational data \citep{DiVa2021, Abdalla2022}. In particular, recent results have pointed to dynamical dark energy as a favoured candidate, with the $\Lambda$CDM model disfavoured at $3.1 \sigma$ from the combination of BAO data from the Dark Energy Spectroscopic Instrument (DESI) and CMB data from the \Planck satellite, and between $2.4$--$4.2 \sigma$ when SNIa data are further added \citep{DESI2025}. Whether such an unexpected result is a hint for new physics or the outcome of residual systematics is one of the key questions of current cosmological research. 

More than two decades after the first evidence for the accelerated expansion, we still do not know what is causing it. This is not because we lack suitable models \citep{Copeland2006, YooWata2012, ReviewAmendola2018}, but because we have too many of them without a clear understanding of the fundamental physics. 
In a pessimistic scenario, there is still the possibility that we are missing a major piece of the puzzle.
Indeed, current data are consistent with distinct modified gravity models where General Relativity (GR) is considered as the small-scale and early-Universe limit of a more general theory of gravity \citep{Clifton2012, Cantata2021}. 

Discriminating among rival solutions to the same problem has become the next big challenge for the cosmological community \citep{Kazuya2016, MatteoSanti2021}. Cosmic shear and galaxy clustering have emerged as two of the most promising probes to constrain the nature of dark energy (e.g. \citealt{ReportDETF}). This has motivated the proposal, approval, preparation, and realisation of dedicated cosmological galaxy surveys. Since the first pioneering measurements of cosmic shear \citep{Bacon2000, Kaiser2000, VanWaerbeke2000, Wittman2000}, Stage-III photometric surveys such as the Kilo Degree Survey (KiDS), the Dark Energy Survey (DES), and Hyper Suprime-Cam (HSC) have demonstrated the power of self-consistent combined analyses of cosmic shear and galaxy clustering \citep{Joudaki2018, vanUitert2018, Abbott2018_DES_Y1, Heymans2021, desy3, hscy3first, hscy3second}. These surveys have highlighted the main challenges ahead, which includes the systematic uncertainties in both observation and theory, which will be described in the forthcoming sections. Beyond the significance of their results, these surveys have been of paramount importance in the preparation of the next-generation Stage-IV surveys.  

To this end, \Euclid is the most recent fully operational Stage-IV survey \citep{Laureijs11, EuclidSkyOverview}. The successful launch on July 1, 2023 from Cape Canaveral has marked the beginning of a new era of high quality and quantity of data, spectacular images, unparalleled challenges, and fascinating questions to resolve. 
As discussed in \citet{EuclidSkyOverview}, \Euclid is carrying out both an imaging and spectroscopic survey that will allow for the investigation of the dark energy with complementary information from cosmic shear and galaxy clustering. The joint analysis of these probes will allow us to both improve the cosmological constraining power and decrease the impact of the systematic uncertainties through their self-calibration.

\begin{table*}
\centering
\caption{\CLOE related publications. The first set of papers are released with the present one, while the second half will be published later.}
\begin{tabular}{
>{\raggedright\arraybackslash}m{4.0cm}
>{\raggedright\arraybackslash}m{11.0cm}}
\textbf{Paper} & \textbf{Content} \\ \hline \hline
 & \\
 Cardone et al. & theoretical modelling of photometric and spectroscopic observables \\ 
Joudaki et al. & code architecture and implementation of the theoretical recipe \\ 
Ca$\tilde{{\rm n}}$as-Herrera et al. & inference  and forecasts of cosmological and nuisance parameters \\ 
Martinelli et al. & validation through comparison against external benchmark codes \\
Goh et al. & extensions beyond standard modelling and additional systematics \\
Blot et al. & impact of systematics modelling on the inference of cosmology \\ 
 & \\
\hline 
 & \\
Crocce et al. & description of the nonlinear module and its implementation \\
Carrilho et al. & impact of uncertainties in nonlinear modelling of photometric probes \\
Moretti et al. & impact of uncertainties in nonlinear modelling of spectroscopic probes \\
Sciotti et al. & theoretical derivation and numerical computation of 3$\times$2pt covariance \\
 & \\ 
\hline
\end{tabular}
\label{tab: papers}
\end{table*}

In order to exploit the rich \Euclid data, a tool is needed that allows us to compare the model predictions against the observations. This tool is a code, which computes the theory predictions of the \Euclid observables for a given cosmological model, treatment of systematic uncertainties, and survey specifications (e.g.\ lens and source redshift distributions, number of redshift and multipoles bins, and angular scales), and then quantifies the agreement with the measurements in terms of the data covariance matrix.
Given this necessity, the Euclid Consortium (EC) created a dedicated group in 2019 to produce such a code, which
has resulted in \CLOE (Cosmology Likelihood for Observables in Euclid). 

\CLOE\footnote{In the 1972 book {\it Le citt\`a invisibili (The invisible cities)} by Italo Calvino, Cloe is the name of an imaginary city where its inhabitants can experience a new set of wonders each day. In a similar way, \CLOE aims at delivering the wonders of novel cosmological results to its users.} indeed performs the computation of the theoretical predictions of the \Euclid observables, in either harmonic or configuration space, and evaluates the likelihood 
by comparing these predictions against the \Euclid data vector given the covariance matrix. Although similar codes already exist, such as the Core Cosmology Library ({\tt CCL}; \citealt{pyccl}) for the theoretical predictions,  
\CLOE has been designed to meet the specific requirements of the EC in addition to its general-purpose capabilities. This includes computing the likelihood for both photometric and spectroscopic observables in a unified theoretical framework, taking care of the modelling choices made within the collaboration and interfacing with the data format of its internal products. In the case of external datasets, these can be included in a joint analysis together with \Euclid within the {\tt Cobaya} and {\tt CosmoSIS} frameworks as described in Paper 2.

This is the first paper in a series of five presenting \CLOE. In particular, this work details how the \Euclid observables are computed, giving all the relevant formulae and information on the adopted survey specifications. Paper 2 \citep{ISTL-P2} is dedicated to explaining the structure of the code, and how the formulae presented here are implemented. Paper 3 \citep{ISTL-P3} shows \CLOE at work to constrain cosmological parameters, while Paper 4 \citep{ISTL-P5} presents the validation of the code, i.e.\ it compares various intermediate and final quantities to the corresponding ones computed by independent benchmark codes. An extension to a class of modified gravity models, and the inclusion of some additional systematics is presented in Paper 5 \citep{ISTL-P6}. Finally, the impact of systematics and priors on the inference is investigated in Paper 6 \citep{ISTL-P4}. A second set of companion papers will provide detailed information on the nonlinear module, which includes its implementation \citep{ISTNL-P1} along with the cosmological impact of different prescriptions for the nonlinear power spectra for the photometric \citep{ISTNL-P3} and spectroscopic \citep{ISTNL-P4} probes. Finally, a dedicated paper \citep{ISTNL-P4} will describe the computation of the analytical covariance matrices. Table\,\ref{tab: papers} summarises the content of these two sets of papers. 

The outline of this paper is as follows. In Sect.\ 2, we first describe the basic cosmological quantities needed to compute the theoretical predictions of the cosmological observables. Sections 3 and 4 will then present the \CLOE recipes for the photometric and spectroscopic observables, respectively, i.e. the set of formulae used to compute the theoretical predictions in a given cosmological model. The auxiliary quantities needed to perform the computations are discussed in Sect.\ 5, which includes the modelling choices motivated by the specifications of the \Euclid survey. Section 6 then presents the visual representations of the theoretical predictions for a reference cosmological model, before Sect.\ 7 summarises the  work.  

\section{Common definitions}
\label{sec:common}

\CLOE is a software package that estimates not only the likelihood for a given set of cosmological parameters, but also all of the associated theoretical quantities. 
We dedicate this section to a brief description of the basic quantities entering the modelling of the photometric and spectroscopic observables (based on e.g. \citealt{KS84}). We partly follow \citet{Blanchard-EP7}, hereafter referred to as EP-VII.

\subsection{The energy content of the Universe} \label{sec:cosmo-bcg}

The Hubble rate as a function of the redshift $z=1/a-1$, where $a$ is the scale factor, reads
\begin{align}
\frac{H(z)}{H_0} \equiv E(z) = 
  \Bigg[&\Omega_{\gamma} (1+z)^4 + \Omega_{\rm bc} \, (1 + z)^3 + \OK \, (1 + z)^2 \; + \\ \nonumber
 & \Omega_{\nu} \frac{\rho_{\nu}(z)}{\rho_{\nu,0}} + \Omega_{\rm DE} \, \frac{\rho_{\rm DE}(z)}{\rho_{\rm DE,0}}\Bigg]^{1/2} \;.
\label{eq: hzde}
\end{align}
Here, $H_0\equiv H(z=0) = 100\,h\,\mathrm {km\,s^{-1}\,Mpc^{-1}}$ is the Hubble constant, with dimensionless factor $h$, and $(\Omega_{\gamma},\Omega_{\rm cb}, \OK, \Omega_{\nu}, \Omega_{\rm DE})$ denote the  density parameters for radiation, the sum of baryonic and cold dark matter, curvature, neutrinos, and dark energy (DE), respectively (at present, unless expressed otherwise). Here, we have defined 
$\Omega(z) \equiv \rho(z)/\rho_{\rm crit}(z)$, where $\rho$ is the energy density of a given component and $\rho_{\rm crit}$ is critical energy density, with the subscript `0' referring to the present. The matter density, $\Om$, is in turn composed of the sum of the densities of baryons, cold dark matter, and non-relativistic neutrinos. Note that $\sum_i \Omega_i(z) = \Omega_{\gamma}(z)+\Omega_{\rm cb}(z) + \OK(z) + \Omega_{\nu}(z) + \Omega_{\rm DE}(z) = 1$ at all times.

In natural units $(c = \hbar = 1)$, the total energy density of massive neutrinos is given by (e.g. \citealt{Komatsu2011})
\begin{displaymath}
\rho_{\nu}(a) = 2 \int{\frac{{\rm d}^3p}{(2 \pi)^3}
\frac{1}{{\rm e}^{p/T_{\nu}(a)} + 1}
\sum_{i}{\sqrt{p^2 + m_{\nu,i}^2}}} \ ,
\end{displaymath}
where $T_{\nu}(a)$ is the neutrino temperature and the sum is over the different neutrino species with mass $m_{\nu, i}$. Using the comoving momentum $q = p a$ and the present day temperature $T_{\nu, 0} = (4/11)^{1/3} T_{\rm CMB} = 1.945 \ {\rm K}$, the energy density can be expressed as
\begin{displaymath}
\rho_{\nu}(a) = \frac{1}{a^4} \int{\frac{q^2 {\rm d}q}{\pi^2} \frac{1}{{\rm e}^{q/T_{\nu,0}} + 1}
\sum_{i}{\sqrt{q^2 + m_{\nu,i}^2 a^2}}} \ .
\end{displaymath}
For relativistic neutrinos, one obtains
\begin{displaymath}
\rho_{\nu}(a) \ \longrightarrow \ \frac{7}{8} \left ( \frac{4}{11} \right )^3 \rho_{\gamma}(a) \simeq 0.2271 N_{\rm eff} \rho_{\gamma}(a),
\end{displaymath}
with $\rho_{\gamma}$ and and $N_{\rm eff}$ denoting the energy density of  radiation and the effective number of neutrino species, respectively. We can then rearrange the neutrino energy density to obtain
\begin{equation}
\Omega_{\nu}(a) = 0.2271 \Omega_{\gamma} N_{\rm eff} a^{-4} \  f_{\nu}(y) \, ,
\label{eq: omeganu}
\end{equation}
with
\begin{equation}
f_{\nu}(y) = \frac{120}{7 \pi^4} 
\int_{0}^{\infty}{{\rm d}x \frac{x^2 \sqrt{x^2 + y^2}}{{\rm e}^{x} + 1}}
\label{eq: fnudef}
\end{equation}
and
\begin{equation}
y = \frac{m_{\nu} a}{T_{\nu,0}} = \frac{187}{1 + z} 
\left ( \frac{\Omega_{\nu,0} ^2}{10^{-3}} \right ) \ ,
\label{eq: ynu}
\end{equation}
where we have assumed for simplicity that all neutrinos have the same mass. 
The present day neutrino density parameter is then related to the sum of the neutrino masses by
\begin{equation}
\Omega_{\nu,0} h^2 = \frac{\sum_{i}{m_{\nu,i}}}{94 \ {\rm eV}} \ .
\label{eq: omeganuzero}
\end{equation}
It can be shown that $f(y) \rightarrow 1$ for $a \rightarrow 0$, such that neutrinos are always relativistic in the early Universe. On the contrary, for the typical values of $\sum{m_{\nu,i}}$ of interest, in the late universe one can approximate
\begin{displaymath}
0.2271 \Omega_{\gamma} a^{-4} f(y) \simeq \Omega_{\nu,0} a^{-3} = \Omega_{\nu,0} (1 + z)^3 \ , 
\end{displaymath}
i.e., massive neutrinos scale with $z$ as baryons and CDM.

As the last component contributing to the expansion rate, we have considered a time-varying energy density of DE, normalised to its present-day value via
\begin{equation}
\frac{\rho_{\rm DE}(z)}{\rho_{\rm DE,0}} = 
(1 + z)^{\,3 \,(1 + w_0 + w_a)}  \exp{\left (-\frac{3 \, w_a\, z}{1 + z}\right )} \, .
\label{eq; rhodecpl}
\end{equation}
Here, we have adopted the so-called Chevallier--Polarski--Linder (CPL) parameterisation for the dark energy equation of state \citep{CP2001, Linder2003}, such that
\begin{equation}
w_{\rm DE}(z) = w_0 + w_a \,\frac{ z}{1 + z}\;,
\label{eq: wcpl}
\end{equation}
where $w_0$ denotes the present value of the equation of state and $w_a = -{\rm d}w_{\rm DE}/{\rm d}a|_{a = 1} 
$. 
The cosmological constant corresponds to $(w_0, w_a) = (-1, 0)$ and a Universe with zero curvature is obtained by setting $\OK = 0$, such that the dimensionless Hubble rate in a flat $\Lambda$CDM model with massless neutrinos can at late times be expressed as
\begin{equation}
E_{{\rm \Lambda CDM}}(z) = H_{{\rm \Lambda CDM}}(z)/H_0 = \sqrt{\Om \, (1 + z)^3 + (1 - \Om)}\;.
\label{eq: ezlcdm}
\end{equation}
In this $\Lambda$CDM scenario, the matter density reads $\Om = \Oc+\Ob$, such that it is the sum of the contributions of cold dark matter and baryons.

\subsection{Cosmological distances}
\label{sec:distances}

The radial comoving distance from an observer at $z = 0$
to an object at redshift $z$ is
\begin{equation}
r(z) = \frac{c}{H_0}\int_0^z \frac{{\rm d}z^{\prime}}{E(z^{\prime})} \;.
\label{eq: rzdef}
\end{equation}
The comoving angular diameter distance between two observers at redshifts $(z_1, z_2)$ depends on the sign of the curvature density parameter:
\begin{equation}
f_K[r(z_1), r(z_2)] = \left \{
\begin{array}{ll}
\displaystyle{
\frac{(c/H_0)\sinh{\left \{\sqrt{\OK} \left [ \tilde{r}(z_2) - \tilde{r}(z_1) \right ] \right \}}}
{\sqrt{\OK}}} & \OK > 0 \\
 & \\
r(z_2) - r(z_1) & \OK = 0 \\
 & \\
\displaystyle{
\frac{(c/H_0)\sin{\left \{ \sqrt{-\OK} \left [ \tilde{r}(z_2) - \tilde{r}(z_1) \right ] \right \}}}
{\sqrt{-\OK}}} & \OK < 0 \\
\end{array}
\right .  
\label{eq: fkz1z2}
\end{equation}
where $\tilde{r}(z) = (H_0/c) r(z)$. In the following, we will use the shortened notation $f_K[r(z)] = f_K[r(0), r(z)]$. 

The comoving volume of a region of the sky covering a solid angle $\hat{\Omega}$ and stretching between redshifts  $(z_{\rm i}, z_{\rm f})$ is given by 
\begin{equation}
V(z_{\rm i},z_{\rm f}) = {\hat{\Omega}} 
\int_{z_{\rm i}}^{z_{\rm f}} \frac{c \; {\rm d}z}{H(z)}\, f_{K}^{2}[r(z)]  \, , \label{eq:Vcomoving}
\end{equation}
where the curvature $K = -\Omega_{K} (H_0/c)^2$. This reduces to
\begin{equation}
\label{eq:survol}
V(z_{\rm i},z_{\rm f}) = \hat{\Omega}  \int_{r(z_{\rm i})}^{r(z_{\rm f})} {\rm d}r\,r^2 = \frac{\hat{\Omega}}{3} \left[ r^3(z_{\rm f}) - r^3(z_{\rm i}) \right]
\end{equation}
for a spatially flat Universe ($K = 0$).

\subsection{Linear perturbations}
\label{sec:linear_perturbation}

At any given time $t$, the (equal-time) Fourier-space power spectrum of a scalar perturbation ${\cal S}(\vec x, t)$, whose Fourier modes are $\hat {\cal S}(\vec k,t)$, is implicitly defined via
\begin{equation}
  \left\langle\hat {\cal S}(\vec k,t) \,\hat {\cal S}(\vec k^\prime,t) \right\rangle=(2\pi)^3\,\delta_{\rm D}(\vec k+\vec k^\prime)\,P_{\cal S}(k, t) \;,
  \label{eq:P_S}
\end{equation}
where $\delta_{\rm D}$ is the Dirac-delta distribution, and we have used statistical homogeneity and isotropy to restrict the scale dependence to $k=\left| \vec k \right|$. It is also useful to introduce the corresponding dimensionless power spectrum $\mathcal P_{\cal S}(k, t)=k^3/\,(2\pi^2)\,P_{\cal S}(k, t)$.

The dimensionless power spectrum of the primordial curvature perturbation $\zeta$ generated by inflation is parameterised as a power law:
\begin{equation}
 \mathcal{P}_{\rm \zeta}(k) = A_{\rm s} \left(\frac{k}{k_0}\right)^{n_{\rm s}-1} \;,
  \label{eq:P_large_scales}
\end{equation}
with amplitude $A_{\rm s}$, pivot scale $k_0$, and scalar spectral index $n_{\rm s}$. We can thereby link the late-time power spectrum of any perturbation ${\cal S}$ to the primordial curvature perturbation power spectrum via the transfer function $\mathcal T_{\cal S}$ through
\begin{equation}
 P_{\cal S}(k,z)= {2\pi^2} \,\mathcal T_{\cal S}^2(k,z)\,\mathcal{P}_{\zeta}(k)\,k\;.\label{eq:PowSpec}
\end{equation}

Let us now focus on a scalar perturbation of particular interest in the late Universe: the matter density perturbation $\delta\rho_{\rm m}$. In particular, it is useful to define the matter density contrast\footnote{
In practice, the comoving-gauge density contrast,
defined as $\Delta = \delta + 3\,(aH/ck)\,v$,
is considered in \CLOE \citep{KS84}. 
The sub-dominant second term is omitted in the expressions for simplicity.
}
\begin{equation}
    \delta_{\rm m}=\frac{\delta\rho_{\rm m}}{\bar\rho_{\rm m}} = \frac{\rho_{\rm m} - \bar{\rho}_{\rm m}}{\bar{\rho}_{\rm m}} \; .
    \label{eq:density_contrast}
\end{equation}
At linear order, in the absence of pressure perturbations and any source of anisotropic stress, it obeys the following second-order differential equation\footnote{Henceforth, as long as there is no ambiguity, we will omit the caret symbol to denote the Fourier transform.}
\be
\delta_{\rm m}^{\prime\prime}(z,\vec {k})+\left[\frac{H^\prime(z)}{H(z)}-\frac{1}{1+z}\right]\,\delta_{\rm m}^\prime(z,\vec {k})-\frac{3}{2}\,\frac{\Omz}{(1+z)^2}\,\delta_{\rm m}(z,\vec {k}) = 0 \;,
\label{eq:sec-gc-comparison-ode-delta}
\ee
where a prime denotes derivatives with respect to $z$.

At late times, we can decouple the redshift and scale-dependent components of the density contrast to good approximation, such that
\begin{equation}
 \delta_{\rm m}(z,\vec {k}) = \delta_{\rm m}(z_\star,\vec {k})\, \frac{D(z)}{D(z_\star)} \;,
 \label{eq:growth_factor}
\end{equation}
where $D(z)$ is the growth factor and $z_\star$ is an arbitrary reference redshift in the early stages of the matter-dominated era. We can moreover define the growth rate
\be\label{eq:fgrowth}
f_{\rm g}(z, k) \equiv - \frac{{\rm d}\ln\delta_{\rm m}(z,\vec {k})}{{\rm d} \ln(1+z)} \; .
\ee
Inserting Eq.~(\ref{eq:fgrowth}) into Eq.~(\ref{eq:sec-gc-comparison-ode-delta}) gives the following first-order differential equation:

\be \label{eq:ode-eff}
f_{\rm g}^{\prime}(z)-\frac{f_{\rm g}^2(z)}{1+z}-\left[\frac{2}{1+z}-\frac{H'(z)}{H(z)}\right]\,f_{\rm g}(z)+\frac{3}{2}\,\frac{\Omega_\text{m}(z)}{1+z}=0 \;,
\ee
which can be solved with initial condition $f_{\rm g}(z=z_\star) =1$. 

It should be noted that the growth factor $D(z)$ can be obtained either by first solving Eq.~(\ref{eq:sec-gc-comparison-ode-delta}) and then using Eq.~(\ref{eq:growth_factor}), or by first determining the growth rate $f(z)$ and then integrating back using the definition in Eq.~(\ref{eq:fgrowth}). In practice, we consider a different procedure in \CLOE, which relies on the fact that the linear matter power spectrum is given by 
\begin{equation}
P_{\rm lin}(k, z) = D^{2}(z) P_{\rm lin}(k, 0) \, .
\label{eq:pklinscale}
\end{equation}
We therefore compute $D(z)$ as 
\begin{equation}
D(z) = \sqrt{\frac{P_{\rm lin}(k, z)}{P_{\rm lin}(k, 0)}} \; .
\label{eq: dzpk}
\end{equation}
We note that the growth factor in the above equation has no scale dependence 
in the absence of anisotropic stress.
As will be further discussed, the linear matter power spectrum is imported from either the {\tt CAMB} or {\tt CLASS} Boltzmann codes (noting that linear matter power spectrum emulation is included in a forthcoming {\tt CLOE} version). The primary motivation for using Eq.~(\ref{eq: dzpk}) is that $D(z)$ is not directly outputted by {\tt CLASS}, and we want \CLOE to be self-consistently interfaced with both codes. 

It is worth noticing that the approximation of no 
 anisotropic stress breaks down in certain extended cosmologies, such as those with massive neutrinos or modified gravity (e.g. \citealt{Ma95, Tsujikawa07}). As a consequence, the growth factor $D(z)$ and the growth rate $f_{\rm g}(z)$ will in general be scale dependent. In practice, this is already accounted for when using Eq.~(\ref{eq: dzpk}), as the Boltzmann codes include the impact of massive neutrinos in the linear matter power spectrum. Although the dependence on $k$ is negligible for the values of interest (see e.g. Fig.\,1 in \citealt{Blanchard-EP7} for the impact on the growth rate), we stress that, unless otherwise stated, the recipe we present is the same regardless of the possible dependence of the growth factor on scale. We will suppress the dependence on $k$ in the argument of the function for sake of succinctness.

In the large-scale structure convention (e.g. \citealt{CS02}), the linear matter power spectrum is  normalised at present by requiring that the RMS variance on spheres of radius $R_8=8\,h^{-1}\,\mathrm{Mpc}$ is equal to a normalisation factor defined as
\begin{equation}
  \sigma_8 \equiv \left(\frac{1}{2 \pi^2} \int{{\rm d}k\, k^2 \, P_{\rm lin}(k, z = 0) \,\left|W_{\rm TH}(k R_8)\right|^2}\right)^{1/2} \; ,
  \label{eq: sigma8}
\end{equation}
with $W_{\rm TH}(x) = 3 \, (\sin{x} - x \cos{x})/x^3$ denoting the Fourier transform of the top hat filter in real space. In the linear regime, one can moreover define a redshift-dependent normalisation as $\sigma_8(z) = \sigma_8 D(z)$, which can be useful for some applications.

\subsection{Deviations from General Relativity}

The above equations implicitly assume that GR holds. It is possible to account for modified gravity (MG) in a phenomenological way by redefining some of the linear-order cosmological perturbations in terms of the functions $\mu(z,k)$, $\eta(z,k)$, and $\Sigma(z,k)$. These functions describe departures from the standard Poisson equation and the relation between the time-time and space-space gravitational potentials in the longitudinal gauge, $\Psi$ and $\Phi$, respectively (e.g.~\citealt{Hu2007pj,Amendola2008,Jain2007yk,Bertschinger2008zb,Daniel2008et,Zhao2008bn,Pogosian2010,cgq19,MatteoSanti2021} and references therein): 
\begin{align}
\Psi(z,\vec {k}) &= - \frac{4\pi G}{c^4} \frac{\bar\rho_{\rm m}(z)\,\delta_{\rm m}(z,\vec {k})}{k^2\,(1+z)^2}\,\mu(z,k)\;,\label{eq:mu} \\ 
 \Phi(z,\vec {k}) &= \Psi(z,\vec {k})\,\eta(z,k)\;, \label{eq:eta} \\
\Phi(z,\vec {k})+\Psi(z,\vec {k}) &= -\frac{8\pi G}{c^4} \,\frac{\bar\rho_{\rm m}(z)\,\delta_{\rm m}(z,\vec {k})}{k^2\,(1+z)^2}\,\Sigma(z,k)\;, \label{eq:sigma}
\end{align}
where we have assumed negligible anisotropic stress from matter, and $(\Psi, \Phi)$ are taken to be dimensionless. Note that only two of these functions are independent, as $\Sigma = \mu (1 + \eta)/2$ from Eqs.~(\ref{eq:mu})--(\ref{eq:sigma}). 
In practice, \CLOE imports $(\mu, \Sigma)$ or uses the Weyl potential power spectrum and modified growth rate from the modified Boltzmann code.

It is moreover worth stressing that, strictly speaking, Eqs.~(\ref{eq:mu})--(\ref{eq:sigma}) only hold in the linear regime. The details of how to deal with $(\mu, \Sigma)$ in the nonlinear regime depend on the model at hand. Recent examples can be found in \citet{Joudaki22}, which performs a comprehensive case study of Jordan--Brans--Dicke gravity, along with \cite{Casas23a} and \cite{Frusciante23}, which present forecasts for constraining modified gravity models with \Euclid data. For what concerns us here, we simply note that the recipe we present is unchanged regardless of how the nonlinearities with $(\mu, \Sigma)$ are taken into account.


An alternative purely phenomenological way to introduce deviations from GR is to change the growth rate $f_{\rm g}(z)$ by writing it as  \citep{Linder2005, LinderCahn2007}
\begin{equation}
f_{\rm MG}(z) = [\Omz]^{\gamma_{\rm g}}
\label{eq: fzgamma}
\end{equation}
with $\gamma_{\rm g} \simeq 6/11 \simeq0.545$ providing an excellent fit to the $\Lambda$CDM growth rate \citep{WS98,Gong09}. Inserting $f_{\rm MG}(z)$ into Eq.~(\ref{eq:fgrowth}) and solving with respect to $D(z)$ gives a modified growth factor $D_{\rm MG}(z)$, which can be used to rescale the matter power spectrum in the linear regime as 
\begin{equation}
P_{\rm m}^{\rm \, MG}(k, z) = \left [ 
\frac{D_{\rm MG}(z)}{D(z)} \right ]^2
P_{\rm m}^{\rm \, GR}(k, z) \; .
\label{eq: pmggr}
\end{equation}
We will also consider a case where the growth factor ratio is used to rescale the nonlinear power spectra. 
We stress that the main driver behind this simplifying assumption is to allow for a comparison with EP-VII (rather than used in the data analysis).
In the case of non-flat models, Eq.~(\ref{eq: fzgamma}) is generalised to \citep{Gong09}
\begin{equation}
f_{\rm MG}(z) = [\Omz]^{\gamma_{\rm g}} + (\gamma_{\rm g} + 4/7) \, \Omega_{K} \; .
\label{eq: fzgammanf}
\end{equation}
We will nevertheless only consider flat models when showing constraints on $\gamma_{\rm g}$. 

\subsection{Theoretical assumptions}

In the forthcoming sections, we will detail how the theoretical predictions of the main \Euclid observables are computed in \CLOE. This includes any assumptions or approximations made in obtaining the relevant expressions. 
We will comment on each as they are introduced, but also find it valuable to summarise them below.
Note that \CLOE inherits the expansion and growth histories, along with the linear matter power spectrum, from Boltzmann codes such as {\tt CAMB} and {\tt CLASS}. 
Some of the assumptions below are therefore not related to \CLOE itself, but are rather choices made when running those codes. On the other hand, some limitations are related to the version of \CLOE available at the time of writing, and will be removed in future releases of the code. With these caveats in mind, we list the main ones below.
\begin{itemize}


\item{We work in the synchronous gauge and quasi-static regime (e.g.~\citealt{Ma95, SB15}).
} \\ 

\item{The DE is modelled as a fluid with an equation of state described by the CPL formula through the two parameters $(w_0, w_a)$. In cases where the $(w_0, w_a)$ parameters 
cross the phantom divide, the parameterised post-Friedmann framework \citep{Hu2007pj, HuPPF, FangPPF} is used to compute the corresponding linear matter power spectrum.} \\



\item{Deviations from GR are phenomenologically parameterised through the two functions $(\mu, \Sigma)$. Note that it is up to the user to decide whether to impose consistency, no-ghost, or no-gradient type instability conditions \citep{Skordis2008} since \CLOE only needs as input the linear power spectra, regardless of the underlying physics involved in computing them from the modified Boltzmann code.} \\


\item{We employ the Limber approximation \citep{Limber1953, 1992ApJ...388..272K, LA2008} to compute angular power spectra. The non-Limber generalization, based on {\tt FFTLog} \citep{Fang2019}, will be included in a future {\tt CLOE} version.} \\

\item{The impact of RSDs on photometric galaxy clustering is considered only at  linear order \citep{fisher94, HT95, Tanidis-TBD}.} \\

\item{The covariance matrices of both photometric and spectroscopic probes are assumed to be cosmology-independent.} \\
\end{itemize}
It is worth stressing that this list only contains the assumptions related to the implementation of the theoretical predictions and likelihood in \CLOE. Additional assumptions employed in using \CLOE to perform Monte Carlo inferences, as detailed in for instance Paper 3, are considered as user choices that can be readily updated via the \CLOE configuration files. 

\section{The recipe for photometric observables} 
\label{sc:recipephot}

The VIS instrument on board the \Euclid satellite will perform a photometric survey imaging billions of galaxies \citep{EuclidSkyOverview, EuclidSkyVIS}. For each of these galaxies, \Euclid will deliver its position on the sky, shape, and photometric redshift (photo-$z$). These quantities are the input data for tracing both the galaxy and the shear fields, which, at the lowest non-zero order, can be characterised by studying their 2-point correlation. In particular, the availability of photo-$z$'s will make it possible to subdivide galaxies in redshift bins, hence performing a tomographic analysis. 

As tracers of the lensing and the galaxy density fields, the galaxy shapes and positions can be used to measure three 2-point correlation functions. This motivates the nomenclature of 3\texttimes2pt as reference for the collection of photometric observables. The analysis can be carried out in either harmonic or configuration space. While it is easier to measure the observables in the latter, the theoretical predictions are more straightforwardly computed in the former. 
As a result, \CLOE will first compute the theoretical predictions in harmonic space, and then transform them into configuration space. The collection of formulae used to implement the theoretical predictions is what we refer to as {\it recipe}. In the rest of this section, we detail the recipe for the 3\texttimes2pt probes.

\subsection{3\texttimes2pt observables in harmonic space}

For a tracer $A$ measured in the $i^{\rm th}$ redshift bin and a second tracer $B$ in the $j^{\rm th}$ redshift bin, the tomographic angular power spectrum is given by
\begin{equation}
\cl{\ell}[ij][AB] = \int\de z\,
\frac{c W_i^{A}(z) \, W_j^{B}(z)}
{H(z) \, f_{K}^{2}[r(z)]} \, P_{AB}\left [\frac{\ell + 1/2}{f_K[r(z)]}, \, z \right ] \; ,
\label{eq: cijabgen}
\end{equation}
where $W_i^{A}(z)$ is the radial weight function for the tracer $A$, and $P_{AB}(k, z)$ is the 3D power spectrum for the $(A, B)$ combination. It is worth stressing that Eq.\ (\ref{eq: cijabgen}) relies on the Limber approximation \citep[][further see \citealt{Kilbinger2017,Kitching2017}]{Limber1953, 1992ApJ...388..272K, LA2008} so that the power spectrum $P_{AB}(k, z)$ is evaluated at
\begin{equation}
k_{\ell}(z) = (\ell + 1/2)/\, f_K[r(z)] \; .
\label{eq: klimber}
\end{equation}
The Limber approximation 
holds for $\ell \gtrsim 100$ (depending on the tracers and redshift bins of interest). However, the need for moving beyond this approximation has already emerged in the analysis of DES-Y3 \citep{Porredon2022}, which motivates the generalization of Eq.~(\ref{eq: cijabgen}) to the beyond-Limber case 
as a main priority in the forthcoming
version of \CLOE. In the following sections, we detail the terms contributing to each observable, including the corresponding weight functions and power spectra.

\subsubsection{Cosmic shear (WL)}
\label{wlsect}

The ellipticity of galaxies in an area centred on a given position on the sky provides an unbiased yet noisy estimator of the shear field. In order to reduce the noise, averaging over a large number of galaxies is required (hence the requirement of large galaxy density that modern weak gravitational lensing surveys achieve). A second underlying assumption is that of a random orientation of the intrinsic source ellipticity. Actually, tidal effects during galaxy formation may imprint a preferred orientation. This intrinsic galaxy alignment (IA) may thus add a non-zero contribution to the measured shear. The weak lensing power spectrum will then read (e.g.\ \citealt{Kilbinger2017} and references therein)
\begin{equation}
\cl{\ell}[ij][\rm LL] = 
\cl{\ell}[ij][\rm \gamma \gamma] +
\cl{\ell}[ij][\rm \gamma I] + 
\cl{\ell}[ij][\rm II]\;,
\label{eq: cijwl}
\end{equation}
where the three terms account for the shear-shear, shear-IA, and IA-IA correlations, respectively. These are given by
\begin{equation}
\cl{\ell}[ij][\rm \gamma \gamma] = F_{\gamma}^2(\ell) 
\int{\de z \ 
\frac{c \,W_i^{\gamma}(z)\, W_j^{\gamma}(z)}
{H(z)\, f_K^{2}[r(z)]}\,
P_{\rm m}\left [k_{\ell}(z), z \right ]} \;,
\label{eq: cijgammagamma}
\end{equation}
\begin{eqnarray}
\cl{\ell}[ij][\rm \gamma I] & = &  F_{\gamma}^{2}(\ell) \nonumber \\
 & \times &  
\left \{ \int{\de z \ 
\frac{c \, W_i^{\gamma}(z) \,W_j^{{\rm IA}}(z)}
{H(z) \, f_K^{2}[r(z)]}
\, P_{\rm mI}\left [k_{\ell}(z), z \right ]} \right . \nonumber \\
 & + & \left . \int{\de z \ 
\frac{c \,W_j^{\gamma}(z) \,W_i^{{\rm IA}}(z) }
{H(z) \, f_K^{2}[r(z)]}
\, P_{\rm mI}\left [k_{\ell}(z), z \right ]} \right \} \;,
\label{eq: cijgammaI}
\end{eqnarray}
\begin{equation}
\cl{\ell}[ij][\rm II] = F_{\gamma}^{2}(\ell)   
\int{\de z \ 
\frac{c W_i^{{\rm IA}}(z) \, W_j^{{\rm IA}}(z)}
{H(z) \, f_K^{2}[r(z)]}
\, P_{\rm II}\left [k_{\ell}(z), z \right ]} \;.
\label{eq: cijII}
\end{equation}
with 
\begin{equation}
F_{\gamma}(\ell) = \left [ \frac{(\ell + 2)!}{(\ell - 2)!} \right ]^{1/2} \, \left ( \frac{2}{2 \ell + 1} \right )^2  \ .
\label{eq: fgammadef}
\end{equation}
In Eqs.~(\ref{eq: cijgammagamma})--(\ref{eq: cijII}), $P_{\rm m}(k, z)$, $P_{\rm mI}(k, z)$, $P_{\rm II}(k,z)$ are the matter-matter, matter-IA, IA-IA power spectra. We refer the reader to Sects.\,5.2 and 5.3 for their modelling. Here, we simply stress that they must be described up to very large wave numbers (i.e.\ small scales) because of the range spanned by $k_{\ell}$ in the integrals. 

The two radial weight functions $W_i^{\gamma}(z)$ and $W_i^{\rm IA}(z)$ account for how much the sources contribute at each redshift.\footnote{As a technical remark, note that the integrals in Eqs.~(\ref{eq: cijgammagamma})--(\ref{eq: cijII}) and corresponding ones in the rest of this section are carried out over the range $0 \le z \le z_{\rm max}$. In order to avoid numerical divergence due to distances vanishing at $z = 0$, we truncate the integral to $z = 0.001$, which is equivalent to assuming that $n_i(z < 0.001) = 0$. This is in practice always the case for every realistic scenario.} 
The former is given by
\begin{eqnarray}
W_i^{\gamma}(z) & = &  
\frac{3}{2} \, \left ( \frac{H_0}{c} \right )^2 \,
\Om\, (1 + z)\, f_K[r(z)] \,
\Sigma \left [ z, k_{\ell}(z) \right ]
\nonumber \\ 
&& \times
\int_{z}^{z_{\rm max}}{\de z^{\prime} \ 
\frac{f_K[r(z^{\prime}) - r(z)]}{f_K[r(z^{\prime})]}\,
\frac{n_i^{\rm S}(z^{\prime})}{\bar{n}_i^{\rm S}}} \;,
\label{eq: wigamma}
\end{eqnarray}
where $n_i^{\rm S}(z)$ is the redshift distribution of the sources in the $i$th bin, $\bar{n}_i^{\rm S}$ their mean number density in that same bin, and $z_{\rm max}$ is the maximum redshift of the survey.  A caveat is in order concerning the term $\Om (1 + z)$, which takes this form only under the assumption that matter is modelled as dust, such that its density scales with redshift as $(1 + z)^3$. Should this not be the case, one can simply replace the term $(1 + z)$ with 
\begin{equation}
\tilde{\rho}_{\rm m}(z) = \frac{\rho_{\rm m}(z)\,/\,\rho_{\rm m}(z = 0)}{(1 + z)^2} \, ,
\end{equation}
where $\rho_{\rm m}(z)$ is the redshift-dependent matter density. 

A further comment concerns the $\Sigma(k, z)$ function to quantify deviations from GR. To be consistent with the Limber approximation, we have evaluated it at $k_\ell(z)$. As a consequence, in modified gravity extensions, the weight function $W_i^{\gamma}$ depends on $(\ell, z)$ rather than $z$ alone as denoted in Eqs.\ (\ref{eq: cijgammagamma}) and (\ref{eq: cijgammaI}). 

The IA weight function reads
\begin{equation}
W_i^{\rm IA}(z) = \frac{H(z)}{c} \, \frac{n_i^{\rm S}(z)}{\bar{n}_i^{\rm S}} \;.
\label{eq: wiia}
\end{equation}
Note that there is no impact of deviations from GR on the IA weight function since we assume that IA originates from phenomena on scales where any MG theory approaches GR. To be more precise, the validity of such an assumption depends on the IA model and MG theory. In this regard, the two main physical phenomena contributing to IA are tidal alignment and tidal torquing, which both depend on the gravitational potential on galaxy scales. 
The absence of deviations from GR on such scales motivates the absence of an MG related term in the IA kernel. Although likely valid for most MG theories, we cannot exclude that there are models where deviations do manifest on such scales. As a consequence, writing the IA kernel as in Eq.~(\ref{eq: wiia}) is, strictly speaking, the consequence of a model-dependent assumption. We therefore caution the user to keep this point in mind when using \CLOE.

\subsubsection{Photometric galaxy clustering (GCph)}
\label{gcphsect}

Galaxies are biased tracers of the matter overdensity field. The number of galaxies entering an observed sample can, however, be altered by the magnification due to lensing, which increases or decreases their apparent luminosity and size, thus making a galaxy drop in or out of the sample. As a consequence, the photometric\footnote{We add the label {\it photometric} to emphasize that the redshifts of the galaxies in this sample have been estimated photometrically rather than from spectra, in contrast to the spectroscopic galaxy clustering.} galaxy power spectrum reads (e.g. \citealt{Pandey2022})
\begin{equation}
\cl{\ell}[ij][\rm GG] = \cl{\ell}[ij][\rm gg] + 
\cl{\ell}[ij][\rm g\mu] + \cl{\ell}[ij][\mu \mu] \, ,
\label{eq: cijggtot}
\end{equation}
with the three terms corresponding to the galaxy-galaxy, galaxy-magnification, and magnification-magnification correlations, respectively. These are given by
\begin{equation}
\cl{\ell}[ij][\rm gg] = 
\int{\de z \ 
\frac{c \,W_i^{\rm g}(\ell, z)\, W_j^{\rm g}(\ell, z)}
{H(z) \, f_K^{2}[r(z)]} 
\,P_{\rm gg}^{{\rm photo}}\left [ 
k_{\ell}(z), z \right ]} \;,
\label{eq: cijgg}
\end{equation}

\begin{eqnarray}
\cl{\ell}[ij][\rm g\mu] & = & 
F_{\mu}(\ell) \int{\de z \ 
\frac{c\, \left [ 
W_i^{\rm g}(\ell, z)\, W_j^{\rm \mu}(z) + 
W_i^{\rm \mu}(z) \,W_j^{\rm g}(\ell, z)
\right ]}
{H(z) \, f_K^{2}[r(z)]}} \nonumber \\ 
& & \ \ \ \ \ \ \ \ \ \ \ \ 
\,P_{\rm gm}^{{\rm photo}}\left [ 
k_{\ell}(z), z \right ] \;,
\label{eq: cijgmu}
\end{eqnarray}

\begin{equation}
\cl{\ell}[ij][\rm \mu \mu] = 
F_{\mu}^{2}(\ell)
\int{\de z \ 
\frac{c \,W_i^{\rm \mu}(z)\, W_j^{\rm \mu}(z)}
{H(z) \, f_K^{2}[r(z)]} 
\,P_{\rm m}\left [ 
k_{\ell}(z), z \right ]} \;.
\label{eq: cijmumu}
\end{equation}
having set

\begin{equation}
F_{\mu}(\ell) = \frac{\ell \, (\ell + 1)}{(\ell + 1/2)^2} \ .
\label{eq: fmudef}
\end{equation}
It is worth noting that the galaxy weight function now depends on the multipole $\ell$ as well. To understand why, let us first consider $W_i^{\rm g}(\ell, z)$. This can be split into a sum of two terms
\begin{equation}
W_i^{\rm g}(\ell, z) = W_i^{\delta}(z) + W_i^{\rm RSD}(\ell, z) \;,
\label{eq: wigsum}
\end{equation}
where the first contribution 
\begin{equation}
W_i^{\delta}(z) = \frac{H(z)}{c} \,\frac{n_i^{\rm L}(z)}{\bar{n}_i^{\rm L}}
\label{eq: widelta}
\end{equation}
accounts for density fluctuations in the number of lenses, with $n_i^{\rm L}(z)$ the lens redshift distribution for the $i$th redshift bin. The second term in Eq.\ (\ref{eq: wigsum}) accounts for the effect of RSDs at linear order. Under this approximation and the assumption of linear galaxy bias, the RSD weight function is given by \citep{Kostas2019}
\begin{equation}
W_i^{\rm RSD}(\ell, z) =  
\sum_{m = -1}^{+1}{
\frac{L_m(\ell) \, H[z_m(\ell, z)] \, f_{\rm g}[z_m(\ell, z)] \, n_i^{\rm L}[z_m(\ell, z)]}{c \, b_{\rm g}^{\rm photo}[z_m(\ell, z)]}} \, ,
\label{eq: wirsd}
\end{equation}
where the growth rate, $f_{\rm g}(z)$, and the photometric linear galaxy bias, $b_{\rm g}^{\rm photo}(z)$, are evaluated at the shifted redshift, $z_m(\ell, z)$, obtained by solving 
\begin{equation}
r[z_m(\ell, z)] = 
\frac{2 \ell + 1 + 4 m}{2 \ell + 1} \ r(z) \; .
\label{eq: zmeq}
\end{equation}
In Eq.\ (\ref{eq: wirsd}), we have also used the definitions
\begin{equation}
L_m(\ell) = \left \{
\begin{array}{ll}
\displaystyle{-\frac{\ell\,(\ell-1)}{(2\,\ell-1)\,\sqrt{(2\,\ell-3)\,(2\,\ell+1)}}} & m = -1 \\
 & \\
\displaystyle{\frac{2\,\ell^2+2\,\ell-1}{(2\,\ell-1)\,(2\,\ell+3)}} & m = 0 \\
 & \\
\displaystyle{-\frac{(\ell+1)\,(\ell+2)}{(2\,\ell+3)\,\sqrt{(2\,\ell+1)\,(2\,\ell+5)}}} & m = 1 
\end{array}
\right . \;.
\label{eq: lmdef}
\end{equation}
The importance of including the RSD contribution in a \Euclid context has been highlighted in \cite{Tanidis-TBD}, which moreover shows the limits of this approximation.

The magnification weight function is given by (e.g.~\citealt{Duncan2014, Thiele2020}) 
\begin{eqnarray}
W_i^{\mu}(z) & = &  
\frac{3}{2} \, \left ( \frac{H_0}{c} \right )^2 \,
\Om \, (1 + z) \, f_K[r(z)] \, 
\Sigma \left [ \frac{\ell + 1/2}{f_K[r(z)]}, z \right ] \, b_{\rm mag}(z)
\nonumber \\ 
&& \times
\int_{z}^{z_{\rm max}}{\de z^{\prime} \ 
\frac{f_K[r(z^{\prime}) - r(z)]}{f_K[r(z^{\prime})]} \,
\frac{n_i^{\rm L}(z^{\prime})}{\bar{n}_i^{\rm L}}} \;,
\label{eq: wimu}
\end{eqnarray}
where $b_{\rm mag}(z)$ is the magnification bias. The selection cuts used to assemble the lens sample determine the $b_{\rm mag}(z)$ profiles \citep{Schmidt2009a, Schmidt2009b}. For a selection based on the galaxies' apparent luminosity, the magnification bias would be given by 
\begin{equation}
b_{\rm mag}(z) = 5 \, s(z) - 2 \, ,
\label{eq: bmag}
\end{equation}
where $s(z)$ is the logarithmic slope of the lenses' luminosity function in the waveband used for the selection \citep[see e.g.][]{2011PhRvD..84d3516C}. Note that this expression only holds for a flux-limited sample. Should the sample be volume limited or assembled through a set of selection cuts, one needs to check how the number of selected galaxies changes because of lensing magnification. 

Not surprisingly, the magnification weight function is the same as the lensing weight function in Eq.\ (\ref{eq: wigamma}). This is because magnification is indeed a lensing effect, which also explains the presence of the phenomenological $\Sigma(k, z)$ function to account for deviations from GR. There are, however, two key differences. First, the term $b_{\rm mag}(z)$ does not only modulate the amplitude of the weight function but also its sign. If the range dominating the integral is the one where $b_{\rm mag}(z)$ is negative, it is therefore possible that magnification reduces the overall angular power spectrum $\cl{\ell}[ij][\rm GG]$ due to the negative sign of $\cl{\ell}[ij][\rm g \mu]$. A second important difference is the use of the redshift distribution of the lenses $n_i^{\rm L}(z)$ instead of the sources $n_i^{\rm S}(z)$, which enters the shear weight function. One can, indeed, use different selection criteria to assemble the lens and source samples, hence dealing with two different redshift distributions. We will come back to this point, highlighting the pros and cons of both choices.

In Eqs.~(\ref{eq: cijgg})--(\ref{eq: cijgmu}), we have introduced two additional power spectra. The first is the galaxy power spectrum $P_{\rm gg}^{\rm photo}(k, z)$, which in the linear regime can be modelled as 
\begin{equation}
P_{\rm gg}^{\rm photo}(k, z) = [b^{\rm photo}(z)]^2 P_{\rm m}(k, z) \, .
\label{eq: pggphotolin}
\end{equation}
Here, $b^{\rm photo}(z)$ is the linear galaxy bias for the photometric sample and $P_{\rm m}(k, z)$ is here meant to be the linear matter power spectrum. It is nevertheless customary to extend the use of Eq.~(\ref{eq: pggphotolin}) to the nonlinear regime by simply using the nonlinear $P_{\rm m}(k, z)$, obtaining what is at times referred to as the minimal bias method (e.g. \citealt{Sugy2022} and refs. therein). In this same limit, we can then set 
\begin{equation}
P_{\rm gm}^{\rm photo}(k, z) = b^{\rm photo}(z) P_{\rm m}(k. z)
\label{eq: pgmphotolin}
\end{equation}
for the galaxy-matter power spectrum. Additional terms are included for the full nonlinear modelling, as discussed in \cite{ISTNL-P2}.

\subsubsection{Galaxy-galaxy lensing (GGL)}
\label{gglsect}

Up to now we have correlated the lensing tracer with itself, and the galaxy tracer with itself, accounting for different tomographic bins. The cross-correlation between these tracers is referred to as galaxy-galaxy lensing, and is useful for improving the constraints on cosmological parameters and systematic uncertainties (e.g. \citealt{JK12, Isaac2020}). The corresponding angular power spectrum is the sum of four terms arising from the different contributions to the tracers. It is given by
\begin{equation}
\cl{\ell}[ij][\rm GL] = 
\cl{\ell}[ij][\rm g\gamma] +
\cl{\ell}[ij][\rm gI] + 
\cl{\ell}[ij][\rm \mu \gamma] + 
\cl{\ell}[ij][\rm \mu I] \; , 
\label{eq: cijgglsum}
\end{equation}
with
\begin{equation}
\cl{\ell}[ij][\rm g\gamma] = F_{\gamma}(\ell)
\int{\de z \ 
\frac{c \ W_i^{\rm g}(\ell, z) \, W_j^{\gamma}(z)}
{H(z) \, f_K^{2}[r(z)]} \,
P_{\rm gm}\left [ k_{\ell}(z), z \right ]
} \;, 
\label{eq: cijggamma}
\end{equation}

\begin{equation}
\cl{\ell}[ij][\rm gI] = F_{\gamma}(\ell)
\int{\de z \ 
\frac{c \ W_i^{\rm g}(\ell, z) \, W_j^{\rm IA}(z)}
{H(z) \, f_K^{2}[r(z)]} \,
P_{\rm gI}\left [ k_{\ell}(z), z \right ]
} \;, \nonumber 
\label{eq: cijgI}
\end{equation}

\begin{equation}
\cl{\ell}[ij][\rm \mu\gamma] = F_{\gamma}(\ell) F_{\mu}(\ell)
\int{\de z \ 
\frac{c \ W_i^{\rm \mu}(z)\,  W_j^{\gamma}(z)}
{H(z) \, f_K^{2}[r(z)]} \,
P_{\rm m}\left [ k_{\ell}(z), z \right ]
} \;, 
\label{eq: cijmugamma}
\end{equation}

\begin{equation}
\cl{\ell}[ij][\rm \mu I] = F_{\gamma}(\ell) F_{\mu}(\ell)
\int{\de z \ 
\frac{c \ W_i^{\rm \mu}(z) \, W_j^{\rm IA}(z)}
{H(z) \, f_K^{2}[r(z)]} \,
P_{\rm mI}\left [ k_{\ell}(z), z \right ]
} \;.
\label{eq: cijmuI}
\end{equation}
All of the relevant radial weight functions have been defined in Sects.~\ref{wlsect} and \ref{gcphsect}. We here stress that, different from the WL and GCph power spectra, the galaxy-galaxy lensing (hereafter, GGL) power spectrum\footnote{In EP-VII, we referred to this quantity with the label XC denoting cross-correlation. Here, we employ the label GGL since it is more commonly used in the current literature.} $C_{ij}^{\rm GL}(\ell)$ is not symmetric in the exchange of tomographic indices. In other words (for $i \neq j$),
\begin{equation}
\cl{\ell}[ij][\rm GL] \neq \cl{\ell}[ji][\rm GL] \; .
\end{equation}
On the contrary, 
\begin{equation}
\cl{\ell}[ij][\rm GL] = \cl{\ell}[ji][\rm LG] \;,
\end{equation}
where $C_{ij}^{LG}(\ell)$ is obtained by reverting the order of the labels in Eq.~(\ref{eq: cijgglsum}). The order of the weight functions in the different terms contributing to the full power spectrum is swapped too, thus motivating the above result.

\subsection{Impact of survey mask in harmonic space: Pseudo-$C_\ell$}

The main drawback of the harmonic space analysis is that the shear and galaxy fields are both sampled only at the positions of the corresponding sample galaxies. In a realistic data analysis, one first bins the data over a pixelised map, and then defines a mask to exclude regions where no source galaxies are present and regions that are contaminated by stars and other defects. On the sky, the field of interest is therefore multiplied by the mask, which translates into a convolution in harmonic space. As a consequence, different scales will be correlated, hence introducing a bias in the data vector and covariance that must be accounted for. A well-known method to address this issue is through pseudo-$C_\ell$ \citep{Hivon2002}, which has been routinely used to analyse CMB data.\footnote{An alternative approach, which we will not consider here, is to rely on optimal quadratic estimators (e.g.\ \citealt{2023MNRAS.520.4836M}).} Here, we briefly review the derivation of the relation between the measured pseudo-$C_\ell$ and the theory power spectra under the flat-sky approximation. To this end, we adapt  the results in \cite{BCT05}, which refer to the CMB. However, the shear field has the same properties as the polarization field, while the position field can be described analogously to the temperature field. A different approach has been considered in \cite{Asgari18} for the shear field arriving at similar results, but treating the multipole $\ell$ as a continuous variable instead of a discrete one. We prefer to rely on \cite{BCT05} since it is more correct from a mathematical point of view. 

In any realistic scenario, parts of the images are masked, and one analyses the gridded image in order to utilize Fast Fourier Transform methods. A pseudo-$C_\ell$ analysis relies on these gridded fields where pixels with no sources or affected by any problem are masked out. One can eventually apodise the mask by smoothing it to make its Fourier Transform less numerically unstable. Alternatively, one can smooth the shear field and retain the mask as it is. Since the aim of this paper is just to provide the recipe rather than its derivation, we skip all the intermediate steps and only give the final result. This is surprisingly simple being given by

\begin{equation}
\tilde{C}(\ell) = \sum_{\ell^{\prime}}{M_{\ell \ell^{\prime}} \, C(\ell^{\prime})} \;,
\label{eq: pseudocl}
\end{equation}
where we have dropped the $ij$ label to shorten the notation, and denoted with a superimposed tilde the pseudo-$C_\ell$ quantities. Under the simplifying assumptions of no $B$ modes, Eq.\ (\ref{eq: pseudocl}) can be expanded as 
\begin{equation}
\left [ 
\begin{array}{l}
\displaystyle{\tilde{C}^{\rm LL}(\ell)} \\
\\
\displaystyle{\tilde{C}^{\rm GL}(\ell)} \\
\\
\displaystyle{\tilde{C}^{\rm GG}(\ell)} \\
\end{array}
\right ]
= 
\left [ 
\begin{array}{lll}
\displaystyle{M^{\rm LL,LL}_{\ell, \ell^{\prime}}} & 0 & 0\\
\\
0 & \displaystyle{M^{\rm GL,GL}_{\ell, \ell^{\prime}}} & 0 \\
\\
0 & 0 & \displaystyle{M^{\rm GG,GG}_{\ell, \ell^{\prime}}} \\
\end{array}
\right ]
\left [ 
\begin{array}{l}
\displaystyle{C^{\rm LL}(\ell^{\prime})} \\
\\
\displaystyle{C^{\rm GL}(\ell^{\prime})} \\
\\
\displaystyle{C^{\rm GG}(\ell^{\prime})} \\
\end{array}
\right ] \;,
\label{eq: mixing}
\end{equation}
where we stress that the mixing matrix ${\bf M}$ is block diagonal because of the assumptions of no $B$ modes \citep{BCT05}. Its non-zero  elements are then given by
\begin{equation}
M^{\rm LL,LL}_{\ell, \ell^{\prime}} = \frac{2 \ell^{\prime} + 1}{8 \pi}
\sum_{\ell^{\prime \prime}}
{W_{\ell^{\prime \prime}}^{\rm LL}  
\left [ 1 + (-1)^{(\ell + \ell^{\prime} + \ell^{\prime \prime \prime})} \right ]
\left (
\begin{array}{lll}
\displaystyle{\ell} & \displaystyle{\ell^{\prime}} & \displaystyle{\ell^{\prime \prime}} \\
& & \\
\displaystyle{2} & \displaystyle{-2} & \displaystyle{0} \\
\end{array}
\right )^2} \;,
\label{eq: mllll}
\end{equation}
\begin{equation}
M^{\rm GL,GL}_{\ell, \ell^{\prime}} = \frac{2 \ell^{\prime} + 1}{4 \pi}
\sum_{\ell^{\prime \prime}} \,
{W_{\ell^{\prime \prime}}^{\rm GL}  \,
\left (
\begin{array}{lll}
\displaystyle{\ell} & \displaystyle{\ell^{\prime}} & \displaystyle{\ell^{\prime \prime}} \\
& & \\
\displaystyle{0} & \displaystyle{0} & \displaystyle{0} \\
\end{array}
\right )
\left (
\begin{array}{lll}
\displaystyle{\ell} & \displaystyle{\ell^{\prime}} & \displaystyle{\ell^{\prime \prime}} \\
& & \\
\displaystyle{2} & \displaystyle{-2} & \displaystyle{0} \\
\end{array}
\right )} \;,
\label{eq: mlglg}
\end{equation}
\begin{equation}
M^{\rm GG,GG}_{\ell, \ell^{\prime}} = \frac{2 \ell^{\prime} + 1}{4 \pi}
\sum_{\ell^{\prime \prime}}
{W_{\ell^{\prime \prime}}^{\rm GG}  
\left (
\begin{array}{lll}
\displaystyle{\ell} & \displaystyle{\ell^{\prime}} & \displaystyle{\ell^{\prime \prime}} \\
& & \\
\displaystyle{0} & \displaystyle{0} & \displaystyle{0} \\
\end{array}
\right )^2} \;,
\label{eq: mgggg}
\end{equation}
with 
\begin{equation}
\begin{array}{l}
\displaystyle{W_{\ell}^{\rm LL} = \sum_m{(\omega^{\rm L}_{\ell m}) \, (\omega^{\rm L}_{\ell m})^\ast}} \;, \\
\\
\displaystyle{W_{\ell}^{\rm GL} = \sum_m{(\omega^{\rm G}_{\ell m}) \, (\omega^{\rm L}_{\ell m})^\ast}} \;, \\
\\
\displaystyle{W_{\ell}^{\rm GG} = \sum_m{(\omega^{\rm G}_{\ell m}) \, (\omega^{\rm G}_{\ell m})^\ast}} \;. \\
\end{array}
\label{eq: wlmask}
\end{equation}
Above, we used
\begin{equation}
\omega^{\rm L/G}_{\ell m} = \int{W_{\rm L/G}({\bf \theta}) \, Y_{\ell m}^\ast({\bf \theta}) \, \diff{\bf \Omega}}
\label{eq: maskcoeff}
\end{equation}
where $W_{\rm L}({\bf \theta})$ and $W_{\rm G}({\bf \theta})$ are the masks denoting the positions on the sky of the WL and GCph sample galaxies. Note that, since different galaxies will contribute to the measurement of the WL, GCph, and GGL pseudo-$C_\ell$ for each $(i, j)$ bin combination, we will have as many masks as there are redshift bin combinations. Eq.\ (\ref{eq: mixing}) then holds for each bin combination allowing us to estimate the predicted pseudo-$C_\ell$ once the $(i, j)$ mixing matrix is given and the corresponding theoretical WL, GGL, and GCph spectra are computed.

It is worth noting that the above equation does still not account for the fact that all fields of interest are in practice pixelated. As such, the average over the angular coordinate depends on the method used and the number of pixels included. This can introduce a mismatch between the theoretical prediction and the measured pseudo-$C_\ell$, which should be accounted for. We refer the reader to \cite{Asgari18} for a possible approach, while \cite{NaMaster} presents a general algorithm to infer pseudo-$C_\ell$ from the maps, and \cite{Kohl17} for an application to KiDS-450. We refer the reader to \cite{Tessore2024} for the specific algorithm that will be used on \Euclid data to measure the pseudo-$C_\ell$.

\subsection{3\texttimes2pt observables in configuration space}

A different way to test a cosmological model through the 3\texttimes2pt probes is to rely on observables defined in the real rather than harmonic space.\footnote{Although not formally correct, we will use {\it real space} and {\it configuration space} as synonymous. We make this choice to stress the difference with respect to harmonic space.}  Although the theoretical modelling of observables is better defined in harmonic space, working in real space has the significant advantage to get rid of the window effects since these are taken into account and corrected for by the measurement algorithm. We will therefore not need to discuss the impact of the window function.

\subsubsection{2-point correlation functions}
The 2-point correlation functions (hereafter, 2PCFs) are the real-space counterparts of the harmonic-space power spectra. To define them, we first consider the two components of the shear, which can be conveniently decomposed in tangential, $\gamma_{\rm t}$, and cross, $\gamma_{\times}$, components defined as
\begin{equation}
\left \{
\begin{array}{l}
\gamma_{\rm t} = -{\rm Re}\left ( \gamma {\rm e}^{-2 i \phi} \right ) \ , \\
 \\
\gamma_{\times} = -{\rm Im}\left ( \gamma {\rm e}^{-2 i \phi} \right ) \ , \\
\end{array}
\right . \ 
\end{equation}
with $\gamma^{2} = \gamma_{1}^2 + \gamma_{2}^{2}$ and $\phi$ denoting the polar angle. The two non-null two-point correlators are then given by \citep{ME91}
\begin{equation}
\left \{
\begin{array}{l}
\xi_{+}(\theta) = 
\left \langle {\bf \gamma} {\bf \gamma}^{\star} \right \rangle(\theta) = \left \langle \gamma_t \gamma_t \right \rangle(\theta) + \left \langle \gamma_{\times} \gamma_{\times}^{\star} \right \rangle(\theta) \ , \\
 \\ 
\xi_{-}(\theta) = 
{\rm Re}\left [ \left \langle {\bf \gamma} {\bf \gamma}(\theta) {\rm e}^{-4 i \phi} \right \rangle \right ] = 
\left \langle \gamma_t \gamma_t \right \rangle(\theta) - \left \langle \gamma_{\times} \gamma_{\times} \right \rangle(\theta) \ , \\
\end{array}
\right . \ 
\end{equation}
with $\theta$ denoting the angular distance between two points on the sky.

Let us first follow Appendix A in \cite{Kitching2017} for the case of the cosmic shear. The field in a position ${\vec \vartheta}$ on the sky may be decomposed in spherical harmonics as 
\begin{equation}
\gamma_1({\vec \vartheta}) \pm {\rm i} \gamma_{2}({\vec \vartheta}) = \sum_{\ell m}{
\left ( \gamma_{\ell m}^{E} \pm {\rm i} \gamma_{\ell m}^{B} \right ) {}_{\pm 2}\, Y_{\ell m}(\vec \vartheta)} \, ,
\label{eq: gammaylm}
\end{equation}
where we have allowed the presence of both $E$ and $B$ modes to be fully general. In order to compute the correlation function, it is now convenient to consider two points that are at the same azimuthal angle, separated by an angle in the polar direction. Using now the $E$- and $B$-mode power spectra and skipping some intermediate steps, we obtain
\begin{eqnarray}
\label{corr}
\xi_{ij}^{\pm}(\theta) & = &
\sum_{\ell, m}{\left [
C_{ij}^{{EE}}(\ell)  \pm C_{ij}^{BB}(\ell) \right ] 
{}_2Y_{\ell m}({\vec n}) \, {}_{\pm 2}Y_{\ell m}^\ast({\vec n^{\prime}})} \nonumber \\
 & + & 
\sum_{\ell, m}{
{\rm i} \left [ C_{ij}^{BE}(\ell) \mp C_{ij}^{EB}(\ell) \right ] 
\, {}_2Y_{\ell m}({\vec n}) \, {}_{\pm 2}Y_{\ell m}^\ast({\vec n^{\prime}})} \nonumber \\
 & = & 
\frac{1}{2 \pi}
\sum_{\ell}{\left [
C_{ij}^{EE}(\ell)  \pm C_{ij}^{BB}(\ell) \right ] 
\, (\ell + 1/2) \, d_{\pm 2 \ 2}^{\ell}(\theta)} \nonumber \\
 & + & 
\frac{1}{2 \pi}
\sum_{\ell}{{\rm i} \, \left [ 
C_{ij}^{BE}(\ell) \mp C_{ij}^{EB}(\ell) \right ] 
\, (\ell + 1/2) \, d_{\pm 2 \ 2}^{\ell}(\theta)} \nonumber \\
 & \simeq &
\frac{1}{2 \pi} 
\sum_{\ell}{\ell \, J_{2 \mp 2}(\ell \theta) \, \left [
C_{ij}^{EE}(\ell)  \pm C_{ij}^{BB}(\ell) \right ]} \nonumber \\
 & + & 
\frac{1}{2 \pi} 
\sum_{\ell}{\ell \, J_{2 \mp 2}(\ell \theta) \, {\rm i} \, \left [ C_{ij}^{BE}(\ell) \mp C_{ij}^{EB}(\ell) \right ]} \nonumber \\
& \simeq & 
\frac{1}{2 \pi} 
\int_{\ell_{\rm low}}^{\ell_{\rm top}}{\ell \, J_{2 \mp 2}(\ell \theta) \, C_{ij}^{\rm LL}(\ell) \; \diff\ell} \;.
\label{eq: xijpm} 
\end{eqnarray}

It is worth stressing that the result up to the sixth row is correct for all $\ell$, and involves the Wigner small-$d$ matrices defined as 
\begin{eqnarray}
d_{m m^{\prime}}^{\ell}(\theta) & = & 
\left [ (\ell + m^{\prime})! \, (\ell - m^{\prime})! \, (\ell + m)! \,  (\ell - m)! \, \right ]^{1/2} \\
 && \times
\sum_{s = s_{\rm min}}^{s_{\rm max}}{
\frac{{\cal{T}}(\vartheta)} 
{(\ell + m - s)! \, s! \, (m^{\prime} - m + s)! \, (\ell - m^{\prime} - s)! \,}} \, , \nonumber 
\label{eq: smalldwigner}
\end{eqnarray}
with $s_{\rm min} = {\rm max}(0, m - m^{\prime})$ and $s_{\rm max} = {\rm min}(\ell + m, \ell - m^{\prime})$, and 
\begin{equation}
{\cal{T}(\vartheta)} = 
(-1)^{\, m^\prime - m + s} \, [\cos{(\theta/2)}]^{\, 2 \ell + m - m^{\prime} - 2 s} 
\, [\sin{(\theta/2)}]^{\, m^{\prime} - m + 2s}
\label{eq: deftfun} \;.
\end{equation}
The fifth and sixth lines are obtained by approximating the Wigner small-$d$ matrices with the Bessel functions $J_n(x)$ of order $n$ in the limit $\ell \gg 2$, so that $(\ell+1/2) \simeq \ell$. Finally, the last line is obtained assuming that there are no $B$ modes, so that the only surviving angular power spectrum is $\smash{\cl{\ell}[ij][EE]}$, which is then equal to $\smash{\cl{\ell}[ij][\rm LL]}$. We have also approximated the discrete sum with an integral, thus retrieving an expression similar to that in \cite{Joachimi2021}, with the only remarkable difference that we explicitly integrate over a limited multipole range to be consistent with the numerical implementation in \CLOE. 

It is now possible to repeat a similar derivation for the galaxy overdensity field and for the galaxy shear-position cross-correlation. Denoting with $\xi_{ij}^{\rm GG}(\theta)$ and $\xi_{ij}^{\rm GL}(\theta)$ the corresponding 2PCF, we obtain the following expressions \citep{vanUitert2018,Joachimi2021}
\begin{eqnarray}
\xi_{ij}^{\rm GG}(\theta) & = & 
\frac{1}{2 \pi}
\sum_{\ell}{
\, C_{ij}^{\rm GG}(\ell) \, (\ell + 1/2) \, d_{00}^{\ell}(\theta)} \nonumber \\
 & \simeq & 
\frac{1}{2 \pi} \int_{\ell_{\rm low}}^{\ell_{\rm top}}{\ell \, C_{ij}^{\rm GG}(\ell) \, J_{0}(\ell \theta)} \;,
\label{eq: xijgg}
\end{eqnarray}
\begin{eqnarray}
\xi_{ij}^{\rm GL}(\theta) & = & 
\frac{1}{2 \pi}
\sum_{\ell}{
\, C_{ij}^{\rm GL}(\ell) \, (\ell + 1/2) \, d_{02}^{\ell}(\theta)} \nonumber \\
 & \simeq & 
\frac{1}{2 \pi} \int_{\ell_{\rm low}}^{\ell_{\rm top}}{\ell \, C_{ij}^{\rm GL}(\ell) \, J_{2}(\ell \theta)} \;,
\label{eq: xijgl}
\end{eqnarray}
where the first line gives the exact result, and the second line is the approximation for $\ell \gg 2$, hence $\ell + 1/2 \simeq \ell$. Note that we have also explicitly cut the sum from $(2, \infty)$ to $(\ell_{\rm low}, \ell_{\rm top})$ in accordance with what is done in numerical computations.

\subsubsection{Band powers}

An alternative probe in real space is provided by band powers, introduced by \cite{Becker2013} and \cite{BR2016} in the context of weak lensing based on the analysis in \cite{Schneider2002}. Following the notation in \cite{Joachimi2021}, and assuming no $B$ modes and a negligible leakage of $E$ modes into $B$ modes, band powers are computed as an average of the shear correlation functions, hence reading
\begin{equation}
{\cal{C}}_{ij}^{\rm LL}(\ell_{\rm k}) = \frac{\pi}{\Delta_{\ell}(\ell_{\rm k})}
\int_{0}^{\infty}{\diff\theta \, \theta \, T(\theta) \left [ 
\xi_{ij}^{+}(\theta) \, g_{+}(\theta, \ell_{\rm k}) + 
\xi_{ij}^{-}(\theta) \, g_{-}(\theta, \ell_{\rm k}) \right ]} \, .
\label{eq: xibp}
\end{equation}
Here, $\Delta_{\ell}(\ell_{\rm k})$ is a normalization constant, and we have defined the kernel functions
\begin{equation}
g_{\pm}(\theta, \ell_{k}) = \int_{0}^{\infty}{\diff\ell \, \ell \, S_{k}(\ell) \, J_{0/4}(\ell \theta)} \, ,
\label{eq: gpmell}
\end{equation}
where $S_{k}(\ell)$ is the adopted response function for the multipole $\ell_{k}$. Note that, although the integral in Eq.~(\ref{eq: xibp}) formally extends from 0 to infinity, in practice the correlation functions are measured over a finite range $(\theta_{\rm min}, \theta_{\rm max})$. The function $T(\theta)$ is therefore introduced to take into account the finiteness of the range and avoid ringing effects at the borders. We adopt a Hann window setting
\begin{equation}
\left \{
\begin{array}{ll}
\displaystyle{0} & \displaystyle{x < x_{\rm min} - \Delta x/2} \\
 & \\
\displaystyle{\cos^2{\left ( \frac{\pi}{2} \, \frac{x - (x_{\rm min} + \Delta x/2)}{\Delta x} \right )}} & 
\displaystyle{x_{\rm min} - \Delta x/2 \le x < x_{\rm min} + \Delta x/2} \\
 & \\ 
\displaystyle{1} & \displaystyle{x_{\rm min} + \Delta x/2 \le x < x_{\rm max} - \Delta x /2} \\
 & \\
\displaystyle{\cos^2{\left ( \frac{\pi}{2} \, \frac{x - (x_{\rm max} - \Delta x/2)}{\Delta x} \right )}} & 
\displaystyle{x_{\rm max} - \Delta x/2 \le x < x_{\rm max} + \Delta x/2} \\
 & \\
\displaystyle{0} & 
\displaystyle{x \ge x_{\rm max} + \Delta x/2} \\
\end{array}
\right .
\label{eq: tdef}
\end{equation}
with $x = \log_{10}(\theta)$, and $\Delta x$ as the user defined log-width of the apodization. We also warn the reader that $\ell_{\rm k}$ in the above formulae should not be meant as a continuous variable, but as a discrete index denoting the band power multipole bin with limits $(\ell_{{\rm lo},k}, \ell_{{\rm up},k})$. As such, for each tomographic bin combination $(i, j)$, the angular power spectrum ${\cal{C}}_{ij}^{LL}(\ell_{\rm k})$ is a single number and not a function, and there are as many of them as the number of band powers the user defines.

Although useful for defining a measurement algorithm, it is more convenient to relate band powers directly to the harmonic-space power spectra. This can be done using \citep{vanUitert2018,Joachimi2021}
\begin{equation}
{\cal{C}}_{ij}^{AB}(\ell_k) = \frac{1}{2 \Delta_{\ell}(\ell_k)} \int_{0}^{\infty}{ 
{\cal{B}}_{AB}(\ell, \ell_k) \, C_{ij}^{AB}(\ell) \, \ell \, \diff\ell} \; .
\label{eq: bandpow}
\end{equation}
Adopting a top-hat response function, the normalisation constant in Eq.\ (\ref{eq: bandpow}) simply becomes
\begin{equation}
\Delta_\ell(\ell_k) = \ln{(\ell_{{\rm up},k}/\ell_{{\rm lo},k})} \;,
\label{eq: normband}
\end{equation}
while the auxiliary functions $\tilde{{\cal{B}}}_{AB}(\theta, \ell, \ell_k)$ can be written as
\begin{equation}
{\cal{B}}_{AB}(\ell, \ell_k) = \int_{0}^{\infty}{T(\theta)\, \tilde{{\cal{B}}}_{AB}(\theta, \ell, \ell_k) \,\theta\, \diff\theta} \;.
\label{eq: kernelband}
\end{equation}
In the case of our interest, they take the following expressions
\begin{equation}
\tilde{{\cal{B}}}_{AB}(\theta, \ell, \ell_k) = \left \{
\begin{array}{ll}
\displaystyle{J_0(\ell \theta) \, \tilde{g}_{+}(\theta, \ell_k)} & \\
 & \\ 
\displaystyle{+ J_4(\ell \theta) \, \tilde{g}_{-}(\theta, \ell_k)} & A = B = {\rm L} \\
 & \\
\displaystyle{J_{2}(\ell \theta) \, \tilde{h}(\theta, \ell_k)} & A = {\rm L} \;, \ B = {\rm G} \\
 & \\
\displaystyle{J_{0}(\ell \theta) \, \tilde{f}(\theta, \ell_k)} & A = B = {\rm G} \\ 
\end{array}
\right . \;,
\label{eq: innerkernel}
\end{equation}
where the $g_{\pm}(\theta, \ell_{\rm k})$ kernels now read
\begin{equation}
\left \{
\begin{array}{l}
\displaystyle{
\tilde{g}_{+}(\theta, \ell_k) = \frac{\theta \,\ell_{{\rm up},k} \,J_{1}(\theta \ell_{{\rm up},k}) - \theta \, \ell_{{\rm lo},k}\, J_{1}(\theta \ell_{{\rm lo},k})}{\theta^2}} = \tilde{f}(\theta, \ell_k) \\
 \\
\displaystyle{\tilde{g}_{-}(\theta, \ell_k) = \frac{{\cal{G}}_{-}(\theta \ell_{{\rm up},k}) - {\cal{G}}_{-}(\theta \ell_{{\rm lo},k})}{\theta^2}} \\
 \\
\displaystyle{
\tilde{h}(\theta, \ell_k) = 
\frac{
\theta \, \ell_{{\rm lo},k} \,J_1(\theta \ell_{{\rm up},k}) - \theta \ell_{{\rm up},k} \,J_1(\theta \ell_{{\rm up},k})}{\theta^2}} \\
 \\
 \displaystyle{\ \ \ \ \ \ \ \ \ \ \ \  \ \ + \frac{ 
2 \,J_0(\theta \ell_{{\rm lo},k}) - 2 \,J_0(\theta \ell_{{\rm up},k})}{\theta^2}} \\ 
\\
\displaystyle{
\tilde{f}(\theta, \ell_k) = 
\frac{\theta \,\ell_{{\rm up},k} \,J_1(\theta \ell_{{\rm up},k}) -  \,\theta \,\ell_{{\rm lo},k} \,J_1(\theta \ell_{{\rm lo},k})}{\theta^2}} \\
\end{array}
\right . 
\label{eq: auxfun}
\end{equation}
with ${\cal{G}}_{-}(y) = (y - 8/y) \,J_1(y) - 8 \,J_2(y)$. Note that $\ell_k$ only labels the multipole bin whose upper and lower limit in $\ell$ explicitly enter on the right-hand side. 

\subsubsection{COSEBIs}
The complete orthogonal sets of $E$/$B$ integrals (COSEBIs) have been introduced to cleanly separate $E$ and $B$ modes within a given angular range $(\theta_{\rm min}, \theta_{\rm max})$, removing any signal that cannot be clearly ascribed to either of the two cases. They form a set of discrete values that can be estimated through weighted integrals of the 2pt shear correlation function with suitable filter functions. Depending on the choice of the filter function, one can obtain different kinds of COSEBIs. We are here only concerned with their theoretical values, which can be predicted as follows
\begin{equation}
E_{n,ij} = \frac{1}{2 \pi}
\int_{0}^{\infty}{C_{ij}^{EE}(\ell) \, W_{n}(\ell) \, \ell \,\diff\ell} \;,
\label{eq: enij}
\end{equation}
\begin{equation}
B_{n,ij} = \frac{1}{2 \pi}
\int_{0}^{\infty}{C_{ij}^{BB}(\ell) \, W_{n}(\ell) \, \ell \,\diff\ell} \;,
\label{eq: bnij}
\end{equation}
where $C_{ij}^{EE}(\ell)$ and $C_{ij}^{BB}(\ell)$ are the $EE$ and $BB$ tomographic lensing power spectra $C_{ij}^{\rm LL}(\ell)$, and the weight functions are given by 
\begin{equation}
W_{n}(\ell) = \int_{\theta_{\rm min}}^{\theta_{\rm max}}{T_{+n}(\theta) \,J_0(\ell \theta) \,\theta\, \diff\theta}
= \int_{\theta_{\rm min}}^{\theta_{\rm max}}{T_{-n}(\theta)\, J_4(\ell \theta) \,\theta \,\diff\theta} \;,
\label{eq: wndef}
\end{equation}
where the $T_{\pm n}(\theta)$ functions may be built as described in \citet{SEK10}. Among the different choices, we follow \cite{KiDSAsgari2021}, and use \citep{Asgari2012}

\begin{equation}
T_{+n}(\theta) = \sqrt{\frac{2 n + 3}{2}} {\cal{P}}_{n + 1}(x)
\label{eq: tndef}
\end{equation}
with $x = 2(\theta - \bar{\theta})/(\theta_{\rm max}  -\theta_{\rm min})$, $\bar{\theta} = (\theta_{\rm min} + \theta_{\rm max})/2$, $(\theta_{\rm min}, \theta_{\rm max})$ the edge of the filter, and ${\cal{P}}_{n}(x)$ the Legendre polynomial of order $n$. The $T_{-n}(\theta)$ functions can then be computed by solving Eq-(\ref{eq: wndef}).

\subsection{The BNT transform}

Weak lensing is extremely sensitive to the nonlinear recipe adopted to model the 3D  matter power spectrum at large $k$ \citep{Pirates20}. An incorrect modelling of $P_{\rm m}(k, z)$ on small scales may dramatically bias the constraints on the cosmological parameters, such that it is mandatory to either add additional nuisance parameters to account for these uncertainties or to carry out brute force approaches running high-resolution simulations including baryons. An alternative strategy is to stay conservative, cutting those multipoles (in harmonic space) or angles (in real space) most sensitive to the small scale $P_{\rm m}(k, z)$. 

Taking the harmonic space as an example, one could naively cut all multipoles $\ell >  \ell_{\rm max}^{(ij)}$ with $\ell_{\rm max}^{(ij)} \simeq k_{\rm max} {\rm min}[r(z_i), r(z_j)]$, where $k_{\rm max}$ is the maximum wavenumber and $z_i$ is the central redshift of the $i^{\rm th}$ tomographic bin. However, the shear window function $W_i^{\gamma}(z)$ has a large width in $z$, so that even for $\ell < \ell_{\rm max}$, one is actually including contributions from $k >  k_{\rm max}$ in the $C_{ij}^{\rm LL}(\ell)$ integral. In order to avoid this, one can redefine the lensing window functions, minimising the overlap between them so that each bin only receives contributions from a well-defined range in scale. This is the basic idea motivating the Bernardeau--Nishimichi--Taruya (hereafter, BNT) nulling scheme \citep{BNT}, which has  been promoted for cosmic shear and 3\texttimes2pt analyses in both harmonic \citep{kcut1,kcut2, gu24} and configuration \citep{xcut} space. 

We refer the interested reader to the quoted papers for details, while we here only report how to implement the BNT transform in practice. Denoting with a tilde the transformed quantities, and assuming implicit summation on repeated indices, it is simply
\begin{equation}
\left \{
\begin{array}{l}
\displaystyle{\tilde{Q}_{ij}^{\rm LL}(y) = \mathfrak{M}_{ik} \,Q_{kl}^{\rm LL}(y)\, \mathfrak{M}^{\sf T}_{lj}} \\
\\
\displaystyle{\tilde{Q}_{ij}^{\rm LG}(y) = \mathfrak{M}_{ik} \,Q_{kj}^{\rm LG}(y)} \\
\\
\displaystyle{\tilde{Q}_{ij}^{\rm GG}(y) = Q_{ij}^{\rm GG}(y)} \\
\end{array}
\right .
\label{eq: bnt}
\end{equation}
where $Q_{ij}^{AB}(y) = C_{ij}^{AB}(\ell)$ in harmonic space and $Q_{ij}^{AB}(y) = \xi_{ij,\pm}^{AB}(\theta)$ in real space, while $\mathfrak{M}_{ij}$ is the BNT matrix computed as described in \cite{xcut, kcut2}.  

Some caveats are worth remembering here:
\begin{itemize}
\item[-]{Equation\ (\ref{eq: bnt}) shows that the GCph summary statistics are not affected by the BNT transform. This is a consequence of having changed only the shear window function, which does not affect the photometric galaxy clustering.} 

\item[-]{The BNT matrix $\mathfrak{M}_{ij}$ is computed neglecting the contribution from IA so that it is actually suboptimal. The effect is, however, negligible since IA accounts for at most $10\%$ of the signal and only in the lowest redshift bins.} 

\item[-]{If one wants to use the BNT transform, this same transformation must also be applied to the data in order to make a meaningful comparison; an analytical formula can be found in the appendix of \cite{Vaz21}.}

\item[-]{In an inference process, the BNT transform must be fixed from the beginning, and used to convert both the theory and data vectors, and the covariance matrix. Since the derivation of $\mathfrak{M}_{ij}$ involves the comoving distance $r(z)$, the departures from the fiducial cosmology assumed at the beginning to compute it make the transformation itself suboptimal. However, the cosmology dependence has been shown to be small, so that using the same $\mathfrak{M}$ for all models has not a significant impact on the efficiency of nonlinear scales separation \citep{BNT, xcut}.}

\end{itemize}
As a final remark, we note that the BNT transform is a recent method, which has only been applied twice to the analysis of real data  \citep{xcut,Vaz21}. As far as we know, \CLOE is the only likelihood code that has implemented it as a user option.

\section{The recipe for spectroscopic observables}
\label{sc: recipespec}

The NISP instrument onboard the \Euclid satellite \citep{EuclidSkyNISP} will perform slitless spectroscopy of H$\alpha$ emission-line galaxies over the redshift range 0.9--1.8. This will allow us to trace the clustering of galaxies with redshift measured with an accuracy better than $\sigma_{\rm err} = 0.002$, thus enabling a 3D reconstruction of the galaxy position field. As for the spectroscopic observables, the analysis can be carried out both in Fourier and configuration space, using as observable the 3D galaxy power spectrum and correlation function, respectively. The way \CLOE computes these summary statistics is referred to as the spectroscopic recipe, and is detailed in this section.  

\subsection{Spectroscopic galaxy clustering (GCsp) in harmonic space}
\label{specone}

Galaxies are biased tracers of the total matter in the Universe. In the linear regime, we can assume a linear deterministic bias model and write the relation between the galaxy and matter power spectra as
\begin{equation}
P_{\rm gg}^{\rm spectro}(\vec{k}, z) = [b^{\rm spectro}(z)]^2\,P_{\rm m}(k, z) \;,
\label{eq: pggspectrolinbias}
\end{equation} 
where $b^{\rm spectro}(z)$ is the redshift-dependent linear bias parameter for the spectroscopic galaxy sample. Note that this could (and, in general, will) be different from the bias for the photometric sample, denoted before as $b^{\rm photo}(z)$. This is because the selection criteria are different, such that the two samples are composed of galaxies with different astrophysical properties. As a consequence, we have added a superscript (`photo' or `spectro') to avoid ambiguity.

Furthermore, our observations are affected by redshift-space distortions (RSD), whose effect in the power spectrum can be described at the linear level by \citep{Kaiser1987}
\begin{equation}
P_{\rm gg}^{\rm spectro}(k, \mu_k, z) = 
\left [ b^{\rm spectro}(z) + f_{\rm g}(z) \,\mu_{k}^{2} \right ]^2 
P_{\rm m}(k, z) \, ,
\label{eq: pggkaiser}
\end{equation}
where we remind the reader that $k\equiv\ |\vec{k}|$
is the modulus of the wavevector $\vec{k}$, $\mu_k$ is the cosine of the angle between $\vec{k}$ and the line-of-sight direction, and  $f(z)$ is the growth rate. 

Equations~(\ref{eq: pggspectrolinbias})--(\ref{eq: pggkaiser}) are valid only in the very linear regime,\footnote{Note that residual leading-order effects, such as the damping of BAO features due to large-scale bulk flows \citep{EisSeoWhi2007, CroSco2008, BalMirSim2015}, are not fully captured by this simplistic model.} such that we need to consider larger $k$ values in order to extract more information on the underlying large-scale structure. To this end, we need to consider higher perturbative orders to correctly account for nonlinearities coming from the evolution of the matter density field, galaxy bias, and RSD. We will therefore not use Eqs.~(\ref{eq: pggspectrolinbias})--(\ref{eq: pggkaiser}), but rather an approach based on the effective field theory of the large-scale structure \citep[EFTofLSS, see e.g.][and references therein]{Moretti24}. What is important to stress, however, is that, different from the photometric case, $P_{\rm gg}^{\rm spectro}$ will now be a function of both $k$ and $\mu_k$.

The observable quantity we consider is, nonetheless, not the full power spectrum, but rather its projection on the Legendre multipoles. In Fourier space, they are defined as
\begin{equation}
P_{\ell}(k, z) =  \frac{2 \ell + 1}{2} 
\int_{-1}^{-1}{\de \mu_k \,L_{\ell}(\mu_k) \,P_{\rm gg}^{\rm spectro}(k, \mu_k, z)} \, ,
\label{eq: pellgg}
\end{equation}
where $L_\ell(\mu_k)$ is the Legendre polynomial of order $\ell$.

When comparing galaxy clustering theory and data, one should be aware that a fiducial cosmological model needs to be assumed to transform the observed redshifts into physical separations. A difference between the fiducial and true cosmologies leads to a rescaling of the components parallel and perpendicular to the line-of-sight direction. As a consequence, in Fourier space, the theoretical power spectrum evaluated at $(k^{\rm fid}, \mu_{k}^{\rm fid})$ actually refers to the observed power spectrum measured at $(k, \mu_k)$ with \citep{Ballinger1996}

\begin{equation}
k(\kfid, \mukfid, z) = \kfid \, q\left(\mukfid\right) \ ,
\ 
\muk(\mukfid, z) = \mukfid \, \qpara^{-1} \, q\left(\mukfid\right)^{-1} \ ,
\end{equation}
where
\begin{equation}
    q\left(\mukfid\right) = \left\{\qpara^{-2}(z)\left(\mukfid\right)^2 + \qperp^{-2}(z)\left[1-\left(\mukfid\right)^2\right]\right\}^{1/2}\; ,
\end{equation}
and $q_{\perp}(z)$ and $q_{\parallel}(z)$ are the Alcock--Paczy{\'n}ski parameters \citep{AlcPac1979}. These are meant to rescale the physical separations parallel and perpendicular to the line of sight, respectively, and are given by
\begin{equation}
 q_{\perp} = \frac{f_{K}[r(z)]}{f^{\rm fid}_{K}[r(z)]} \  \ , \ 
 q_{\parallel}  \frac{H^{\rm fid}(z)}{H(z)} \; .\label{eq:q_perp_para}
\end{equation}
with the label `fid' marking the quantities for the fiducial model. 

A rescaling of the power spectrum amplitude by a factor 
$(\qperp^2\,\qpara)^{-1}$ is an additional consequence of deviations from the fiducial cosmology. These geometric distortions must be applied to all model predictions before they are compared with real galaxy clustering measurements. The prediction for the  observed Legendre multipoles $P_{{\rm obs},\ell}(k^{\rm fid})$ including geometric distortions can then be obtained as
\begin{eqnarray}
P_{\rm obs, \ell}(\kfid, z) & = & 
\frac{1}{\qperp^2(z) \, \qpara(z)} 
\frac{2 \ell + 1}{2} \nonumber \\
 && \times
\int_{-1}^{1}{\de \mukfid \,
L_{\ell}(\mukfid) \,
P_{\rm gg}^{\rm spectro}(k, \muk, z)} \;,
\label{eq:pell_AP}
\end{eqnarray}
where, unless otherwise stated, in the rest of this section $P_{\rm gg}^{\rm spectro}(k, \muk, z)$ is meant as a function of $(\kfid, \mukfid, z)$ through the implicit relations between $(k, \muk)$ and $(\kfid, \mukfid)$.

\subsection{Impact of observational systematic uncertainties}
\label{sec:GCspectro_systematics}

Spectroscopic redshifts are measured with great accuracy, yet the effect of redshift uncertainties must be propagated to the observed galaxy power spectrum. This can be easily done by replacing the galaxy power spectrum $P_{\rm gg}^{\rm spectro}(k, \mu_k, z)$ in Eq.\ (\ref{eq:pell_AP}) with the redshift-corrected one given by
\begin{equation}
P_{\rm gg, obs}^{\rm spectro}(k, \mu_k, z) =
P_{\rm gg}^{\rm spectro}(k, \mu_k, z) \,
F_{z}(k, \mu_k, z) \, ,
\label{eq: pggobszerr}
\end{equation}
with 
\begin{equation}
F_z(k, \mu_k, z) = {\rm e}^{-k^2 \mu_{k}^{2} \sigma_{r}^{2}(z)}
\label{eq; fzdef}
\end{equation}
defined as the (unbiased) scatter in the measured galaxy redshifts. This term causes the smearing of the observed galaxy density field (and therefore a suppression of the measured galaxy power spectrum and correlation function) along the line-of-sight direction $k_\parallel = k \,\muk$ due to possible redshift errors. This error propagates into a comoving-distance error as 
\begin{equation}
\sigma_r(z) = \frac{{\rm d}r}{{\rm d}z} \sigma_z(z) 
= \frac{c \,\sigma_{\rm err}(z)}{H(z)} \;,
\label{eq:error-z-gc}
\end{equation}
where $\sigma_{\rm err}(z)$ is the redshift error. We consider a constant error by setting $\sigma_{\rm err} = 0.002$, since this is the expected performance of the NISP spectrograph. Note that the redshift error term $F_z$ must be calculated in the fiducial cosmology to not artificially include a cosmology dependence (i.e.\ the signal) into the noise (i.e.\ the error-correction term). 

An additional correction comes from the presence of random outliers in the sample used to estimate the clustering properties. Let us denote with $n_{\rm true}(z)$ the true redshift distribution of included objects\footnote{Note that the number density we refer to for spectroscopic galaxy clustering is the comoving (volumetric) one, not to be confused with the angular one used in photometric galaxy clustering.}, and let $f_{\rm inc} = 1 - f_{\rm comp}$ be the fraction of objects correctly included in the clustering sample, with $f_{\rm comp}$ representing the redshift sample efficacy given the instrumental observation limit. The number of clustering objects reads as
\begin{equation}
n_{\rm C}(z) = n_{\rm true}(z) \, (1 - f_{\rm inc}) \;,
\label{eq: ncvsz}
\end{equation}
while the number of non-clustering ones is 
\begin{equation}
n_{\rm NC}(z) = n_{\rm C}(z) \,\frac{f_{\rm out}}{1 - f_{\rm out}} = n_{\rm true}(z) \, \frac{f_{\rm out}\,(1 - f_{\rm inc})}{1 - f_{\rm out}} \;,
\label{eq: nncvsz}
\end{equation}
where $f_{\rm out} = 1 - f_{\rm pur}$ is the fraction of outliers and $f_{\rm pur}$ is the purity of the redshift sample. This can deviate from unity due to the occurrence of spectral line misidentification (line interlopers) and catastrophic failure in the redshift measurement (noise interlopers). In summary, the total number density of observed objects is then given as
\begin{equation}
n_{\rm tot} = n_{\rm C}+n_{\rm NC} = \frac{n_{\rm C}}{(1-f_{\rm out})}=n_{\rm true}\,\frac{1-f_{\rm inc}}{1-f_{\rm out}} \;.
\label{eq:ntot}
\end{equation}
The presence of outliers causes further damping of the clustering signal by a factor $1 - f_{\rm amp}$ with
\begin{equation}
f_{\rm amp} = \frac{n_{\rm NC}}{n_{\rm C} + n_{\rm NC}} = f_{\rm out} = 1 - f_{\rm pur}  \;. 
\label{eq: famp}
\end{equation}

The final power spectrum corrected for redshift errors and random outliers is then given by
\begin{eqnarray}
P_{\rm gg, obs}^{\rm spectro}(k, \mu_k, z) & = & 
(1 - f_{\rm out})^2 \,P_{\rm gg}^{\rm spectro}(k, \mu_k, z) 
\nonumber \\ 
 && \times\exp{\left \{ - \left [ \frac{c \,k \,\mu_k \,\sigma_{\rm err}(z)}{H(z)} \right ]^2\right \} } \;.
\label{eq: pggobsend}
\end{eqnarray}
Two remarks are in order about Eq.\ (\ref{eq: pggobsend}). First, we have implicitly assumed that $f_{\rm out}$ is redshift independent. In practice, it is sufficient that the purity $f_{\rm pur}$ (which is the parameter varied in \CLOE) is constant within the redshift bin considered, and allowed to be different from one bin to another. Second, and much more importantly, the above derivation is only valid in the case of completely random outliers. In fact, there could be interlopers with their own clustering properties, which enter the clustering sample, hence biasing the measurement of the power spectrum. A dedicated recipe to account for them will be presented in forthcoming work and then included in \CLOE.  

As described in Sect.~\ref{specone}, the foundation of the spectroscopic model implemented within \CLOE is the projection on the Legendre-multipole basis given in Eq.\ (\ref{eq:pell_AP}), which, accounting for the observational errors described in this paragraph, now becomes
\begin{eqnarray}
P_{\rm obs, \ell}^{\rm spectro}(k^{\rm fid}, z) & = & 
\frac{1}{q_{\perp}^2(z) q_{\parallel}(z)} \,
\frac{2 \ell + 1}{2} \nonumber \\
 && \times
\int_{-1}^{1}{\de \mu_{k}^{\rm fid} \,
L_{\ell}(\mu_{k}^{\rm fid}) \,
P_{\rm gg, obs}^{\rm spectro}(k, \mu_k, z)} \;.
\label{eq:pell_APObs}
\end{eqnarray}
This is the summary statistics in Fourier space, but one can trivially pass it to configuration space. Here, one works with the observed $2$-point correlation function multipoles that can be computed from the observed power spectrum multipoles as 
\begin{equation}
\xi_{{\rm obs},\ell}(s^{\rm fid};z) = \frac{{\rm i}^{\ell}}{2\pi^2}\int^{\infty}_{0}{\rm d}k^{\rm fid}\,\left(k^{\rm fid}\right)^2 \,P_{{\rm obs},\ell}^{\rm spectro}(k^{\rm fid};z) \,
j_{\ell}(k^{\rm fid}s^{\rm fid}) \; ,
\label{eq:pl2xil}
\end{equation}
where $j_{\ell}(x)$ is the spherical Bessel function of order $\ell$. Note that, depending on the Fourier convention, the factor $1/(2\pi^2)$ can be substituted by a factor $4\pi$. In practice, we use {\tt CAMB} or {\tt CLASS} to generate the linear matter power spectrum, for which Eq.~(\ref{eq:pl2xil}) has the correct normalisation.

It is worth noting that the presence of systematic outliers in the observed galaxy sample must also be taken into account when deriving the nominal covariance matrix associated with each data vector. Although the covariance matrix is not computed by \CLOE, but imported as an external input, we provide a brief discussion below for completeness. Under the Gaussian approximation, the generic per-mode covariance for each combination of multipoles $\ell_1$ and $\ell_2$ can be written as \citep{GriSanSal2016}
\begin{equation}
    \begin{split}
        \sigma^2_{\ell_1\ell_2}& (k,z) = \frac{(2\ell_1+1)(2\ell_2+1)}{V_{\rm sur}} \,\\
        & \times  \int_{-1}^{+1}\left[P_{\rm gg}^{\rm spectro}(k,\muk,z)+\frac{1}{n_{\rm true}(z)}\right]^{\,2} L_{\ell_1}(\muk) L_{\ell_2}(\muk) \, \de\muk\; ,
    \end{split}
\end{equation}
where $V_{\rm sur}$ represents the volume of the comoving shell of the redshift bin under consideration. To get to the bin-averaged covariance, the previous formula must be averaged over the specific $k$ bins assumed for the data vectors
\begin{equation}
    C_{\ell_1\ell_2}(k_i,k_j)=\frac{2(2\pi)^4}{V_{k_i}^2}\delta_{{\rm K}_{ij}}\int_{k_i-\Delta k/2}^{k_i+\Delta k/2}\sigma^2_{\ell_1\ell_2}(k)\,k^2\,\de k\; ,
    \label{eq:gcspec_cov}
\end{equation}
where $\Delta k$ corresponds to the size of the $k$ bins, $V_{k_i}\equiv4\pi[(k_i+\Delta k/2)^3-(k_i-\Delta k/2)^3]/3$ is the volume of the spherical $k$ shell centred at $k_i$, and $\delta_{{\rm K}_{ij}}$ is the Kronecker delta highlighting the independence of the density field at different $k$ modes in the Gaussian approximation. For readability, we have omitted from Eq.~(\ref{eq:gcspec_cov}) the dependence on the redshift. To take into account the impact of systematic effects, the galaxy power spectrum $P_{\rm gg}^{\rm spectro}$ and number density $n_{\rm true}$ should be substituted with the corresponding quantities from Eqs.~(\ref{eq:ntot}) and (\ref{eq: pggobsend}).
Moreover, the GCsp covariance matrix must account for systematic uncertainties that cannot accurately be modelled analytically. A numerical covariance will therefore be used in the fiducial analysis of \Euclid data as described in {\citet{PiGiCov}.}

\subsection{Window function}
\label{sec:WF}

We have so far assumed that we are able to perform measurements of the relevant observables over the full sky. In fact, this is far from reality because of missing regions (e.g. \citealt{ross25}). One has therefore to account for the effect of the survey mask on the observables just as we did for the 3$\times$2pt probes. To this end, we implement in \CLOE the method adopted in for instance \cite{DAmico_etal2020}, convolving the Legendre multipoles $P_{\rm obs, \ell}^{\rm spectro}(k^{\rm fid}, z)$ with the mixing matrix $\mathcal{W}_{\ell \ell'} (k, k')$. We note that this mixing matrix is not computed by \CLOE, but provided by the user. Dropping some labels to avoid cluttering the notation, the observed Legendre multipoles then read
\begin{equation}
\tilde{P}_\ell(k) = \frac{{\rm i}^{2 \ell' - \ell}}{2 \pi^2} \int \diff k' ~ k'^2 \mathcal{W}_{\ell \ell'}(k, k')\, P_{\ell '} (k')\;,
\label{eq: pellwf}
\end{equation}
where there is the implicit sum over the repeated index $\ell^{\prime}$, and  the mixing matrix is given by
\begin{equation}
\mathcal{W}_{\ell \ell'} (k, k') = 4 \pi (-{\rm i})^{\ell'} \int \diff s ~ s^2\, j_{\ell'}(k's)\, \,j_\ell(ks) \,Q_{\ell,\ell'}(s) \;,
\label{eq: gcspmixmatrix} 
\end{equation}
with $Q_{\ell \ell'}(s)$ denoting the configuration-space window functions. As noted, the function ${\cal{W}}_{\ell \ell^{\prime}}(k, k^{\prime})$ is not computed by \CLOE, but rather imported as an external input generated by a dedicated processing function. However, for completeness, we report below the main steps to obtain it. First, a random catalogue matching the survey mask is generated and used to compute the window function multipoles as 
\begin{equation}
Q_\ell(s) = (2\ell+1) \int \frac{\diff^2 \hat{s}}{4\pi} \int\diff^3 s_1 \,W(\vec{s}_1) \,
W(\vec{s}+\vec{s}_1) \,
\mathcal{L}_\ell(\vec{s} \cdot \vec{n}) \;,
\end{equation}
with $W({\vec s})$ denoting the survey window function, which depends on the angular mask and redshift depth of the survey. Assuming summation for repeated indexes, one can then set
\begin{equation}
Q_{\ell,\ell'}(s) \equiv C_{\ell,\ell',\ell''}\, Q_{\ell''}(s) \;,
\end{equation}
with
\begin{equation}
{\cal{C}}_{\ell,\ell',\ell''} \equiv (2\ell+1) \,\sum_{\ell''} 
\begin{pmatrix}
\ell & \ell' & \ell'' \\
0 & 0 & 0
\end{pmatrix}^2\;.
\end{equation}
For the multipoles of our interest with $\ell = (0, 2, 4)$, this becomes 
\begin{eqnarray}
{\cal{C}}_{0,\ell',\ell''} & = &  
\begin{pmatrix}
1 & 0 & 0 \\
0 & \tfrac{1}{5} & 0 \\
0 & 0 & \tfrac{1}{9}
\end{pmatrix}_{\ell',\ell''} \;, \\ 
{\cal{C}}_{2,\ell',\ell''} & = &  
\begin{pmatrix}
0 & 1 & 0 \\
1 & \tfrac{2}{7} & \tfrac{2}{7} \\
0 & \tfrac{2}{7} & \tfrac{100}{693}
\end{pmatrix}_{\ell',\ell''} \;,  \\ 
{\cal{C}}_{4,\ell',\ell''} & =  & 
\begin{pmatrix}
0 & 0 & 1 \\
0 & \tfrac{18}{35} & \tfrac{20}{77} \\
1 & \tfrac{20}{77} & \tfrac{162}{1001}
\end{pmatrix}_{\ell',\ell''} \;.
\end{eqnarray}
In practice, however, the $Q_\ell(s)$ are rather computed as the Hankel transform of the window power spectrum multipoles $P_{W,\ell}$, hence reading 
\begin{equation}
{Q}_\ell(s) = \frac{{\rm i}^\ell (2 \ell + 1)}{2 \pi^2} 
\int{\diff k \,k^2 \,P_{W,\ell}(k) \,j_\ell(ks)} \;.
\end{equation}
To this end, an {\tt FFTlog} algorithm is used to speed up the numerical computation.

\section{Recipe ingredients}
The above theoretical formulae allow us to compute the photometric and spectroscopic observables provided one specifies a set of input quantities and modelling choices. We collectively refer to them as {\it recipe ingredients}. In this section, we will sketch the choices made for generating the figures shown in the rest of the paper. More details on specific topics can be found in companion papers, mainly Paper 5 for the impact on parameter constraints from the assumptions on systematic uncertainties, \cite{ISTNL-P2} for the choice of the photometric galaxy power spectrum, and \cite{ISTNL-P3} for the choice of the spectroscopic galaxy power spectrum.

\subsection{Redshift distributions for the photometric probes}\label{sec:redshift_distribution}

With the aim to use a realistic galaxy distribution for our samples, we have based our analysis on the \Euclid Flagship simulation v2.1 \citep{EuclidSkyFlagship}. In more detail, we have first selected a patch with right ascension between 190 and 210 degrees and declination between 40 and 60 degrees. We have explicitly verified that the galaxy distributions obtained from this patch are perfectly compatible with the distributions obtained from another patch in a different region of the sky and with half the area; therefore, we are confident that the $n(z)$ obtained can be safely extrapolated to the entire \Euclid footprint. In addition to the spatial selection, we have applied a magnitude cut of $\IE\leq 24.5$. Once the selection of the sample has been made, we have considered the first mode of the probability density function for the photometric redshift of each object, which we denote as photo-$z$ here for simplicity. These photo-$z$ have been obtained using the internal \Euclid pipeline that will be applied to the real measurements. 

\begin{figure}
    \centering
    \includegraphics[scale=0.4]{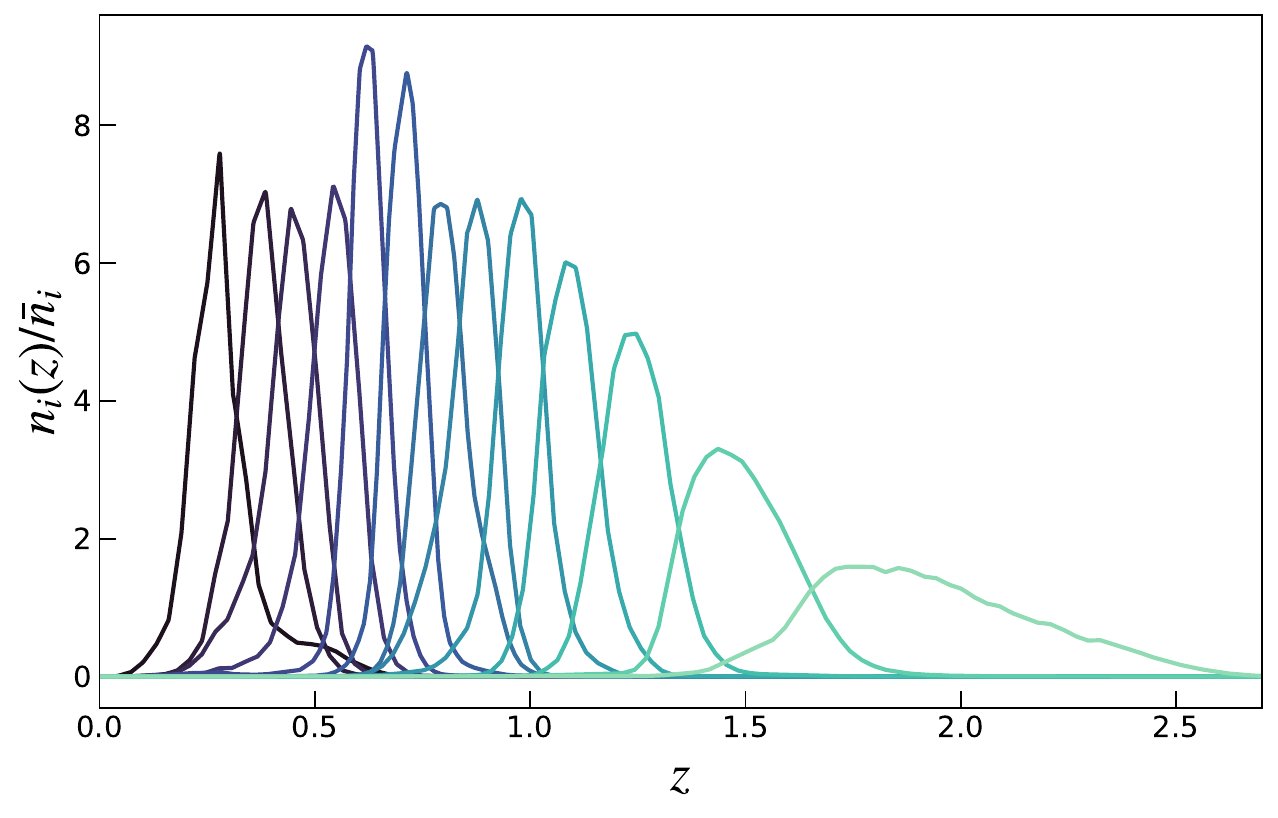}
    \caption{Normalised galaxy distribution for each of the 13 baseline tomographic bins extracted from the \Euclid Flagship simulation (Sect.~\ref{sec:redshift_distribution}). Each bin has a number density of $1.8674$\,galaxies/arcmin$^2$.}
    \label{fig:nofz}
\end{figure}

We further limit our sample to objects for which the photo-$z$ can be properly determined and consider only objects with photo-$z$ above $z=0.2$ and below $z=2.5$ \citep{Desprez-EP10}. The lower limit is given by the poor photo-$z$ performance and the absence of a significant lensing signal, while the upper limit is added both to avoid going too close to the edge of the simulation and the low number of galaxies with spectroscopic redshift usable as calibrators. We then classify all the remaining objects of the sample into tomographic bins according to their photo-$z$. After exploring different configurations and numbers of tomographic bins, the baseline choice for the future final data release of \Euclid is 13 bins with the same number of galaxies per bin \citep{EuclidSkyOverview}. Once the galaxies have been classified, we directly compute the $n(z)$ in each of the bins by building the histogram of the observed true redshifts, which are shown in Fig. \ref{fig:nofz}. Given that we consider the same number of galaxies in each bin, we have a number density of $1.8674$\,galaxies/arcmin$^2$ per bin, leading to a total of $24.3$\,galaxies/arcmin$^2$.

There are significant ongoing efforts within the collaboration to provide the best possible estimates for the photo-$z$ \citep{Desprez-EP10, EP-Paltani, dAssignies25}, but given the complexity of the issue, it is possible that some residual discrepancies compared to the true redshifts remain. In order to avoid biasing our cosmological results because of these residual systematic effects, one typically parametrises the $n(z)$ used in the prediction with some nuisance parameters and allows them to vary to absorb the residual discrepancies. Here, we introduce a free parameter that allows for a shift of the mean of the $n(z)$, such that we perform the replacement (e.g. \citealt{Joudaki17, Troxel2018_DES_Y1})
\begin{equation}
n_i(z) \longrightarrow n_i(z - \Delta z_i) \, , 
\end{equation}
hence introducing an additional nuisance parameter $\Delta z_i$ for each bin. Note that, in Paper 3, motivated by the requirements in the Euclid Red Book \citep{Laureijs11}, we use a Gaussian prior of width 
\begin{displaymath}
\sigma(\Delta z_i) = 2 \times 10^{-3} \,(1 + \bar{z}_i) \, ,
\end{displaymath}
with $\bar{z}_i$ denoting the effective redshift of the $i$-th bin. 

In general, one could choose a different tomographic binning and have different redshift distributions for the lens and  source galaxies in the 3\texttimes2pt photometric analysis (e.g. \citealt{desy3}). 
However, given that we add a nuisance parameter to the mean of each redshift distribution (and will in a forthcoming \CLOE version allow for changes in the widths), using two different samples would significantly increase the number of nuisance parameters. After careful consideration of different binning schemes for the two samples, the baseline choice remains to use the same sample for both lenses and sources, i.e.\ 13 tomographic bins with the same number of galaxies in each bin. In addition to reducing the number of free parameters, a major benefit is the increased ability to self-calibrate the photometric redshift distributions (see e.g. \citealt{Johnston24}).

\subsection{Nonlinear matter power spectrum}

The large width of the weak lensing kernel makes the cosmic shear harmonic power spectra extremely sensitive to the nonlinearities in the matter power spectrum. Although this is an opportunity, since it allows us to probe the growth of structures in the highly informative nonlinear regime, it comes with a significant cost. Indeed, the modelling of $P_{\rm m}(k, z)$ in the large-$k$ region to the $\sim 1\%$ accuracy necessary to not bias the estimate of $C_{ij}^{AB}(\ell)$ is still an open challenge. In order to account for the uncertainties in this critical ingredient, \CLOE has a nonlinear module, which allows the user to select among different options. These options span from semi-empirical models such as the revised {\tt HaloFit} version \citep{HaloFit} and two versions of {\tt HMCode} \citep{HMCode2016, HMCode2020}, to emulators such as \texttt{EE2} \citep{Knabenhans-EP9} and \texttt{BACCO} \citep{BACCO}. There is also the possibility to account for baryons either with the baryon-corrected {\tt HMCode} version, or emulators of the baryonification method such as {\tt BCemu} \citep{BCemu}. We refer the reader to \citet{ISTNL-P2} for a detailed description of the available options, a comparison of the harmonic power spectra, and an investigation of the bias on the estimate of cosmological parameters due to mismatches in the adopted nonlinear prescription. 

As default in the rest of this paper (and in Papers 3 and 4 on the forecasting and code validation, respectively), we will adopt the latest version of {\tt HMCode} where the effect of baryons is modulated by the single parameter $\logten{(T_{\rm AGN}/{\rm K})}$. Here, $T_{\rm AGN}$ controls the temperature to which gas is heated in the subgrid prescription for thermal AGN feedback from \cite{BS2009}. This was used in the BAHAMAS simulations \citep{McCarthy2017} to which {\tt HMCode} is fit. We refer the reader to \cite{HMCode2020} for further details.

\subsection{Intrinsic galaxy alignments}

IAs are one of the main astrophysical contaminants of the weak lensing signal. Modelling this systematic uncertainty correctly is therefore of primary importance to avoid biasing the inference of the cosmological parameters. There is a vast literature on estimating IAs from observations to inform the theoretical modeling \citep[see e.g. the reviews][and references therein]{IArev1, IArev2, IArev3, TI2015, Lamman2024}.

In order to account for this systematic uncertainty, \CLOE allows the user to choose between two different models. The first is referred to as the extended nonlinear linear alignment model (eNLA; \citealt{HS2004, BK2007, Joachimi2011}), which allows for a power-law redshift and luminosity dependence of the IA amplitude. As a second case, we consider the TATT (tidal alignment and tidal torquing) model \citep{Blazek2019, Samuroff2023} where one goes beyond the linear galaxy alignments to include effects related to the tidal torquing, which should dominate in spiral galaxies.\footnote{Turning off the tidal torquing terms, the TATT model reduces to the eNLA model, so that the latter is actually nested in the former. We will nevertheless consider them as separate models since the way the IA amplitude is parameterised is different.} Both eNLA and TATT models are implemented in \CLOE to allow the user to investigate how constraints on the cosmological parameters depend on the adopted IA model. We will address this point in Paper 6.

The results from Stage-III cosmic shear surveys and the comparison with simulations have still not been able to clearly determine whether the tidal torquing terms present in TATT are  needed \citep{DESY3Secco2022, Hoffmann2022}. On the contrary, there is a non-negligible possibility that the increased number of parameters leads to overfitting \citep{Joachimi2021}. We refer the interested reader to \cite{ISTNL-P2} for details on the models, and their pros and cons. In the following discussion, we will consider the zNLA model, which is obtained from the eNLA model by neglecting the dependence of the IA amplitude on the mean luminosity. The IA 3D power spectra can then be written as
\begin{equation}
\left \{
\begin{array}{l}
\displaystyle{P_{\rm mI}(k, z) = -A_{\rm IA}(z) \,P_{\rm m}(k, z)} \, , \\
 \\
\displaystyle{P_{\rm II}(k, z) = A_{\rm IA}^{2}(z) \,P_{\rm m}(k, z)} \, , \\
 \\
\displaystyle{P_{\rm gI}(k, z) = -A_{\rm IA}(z) \,b^{\rm photo}(z) \,P_{\rm m}(k, z)} \, , \\
\end{array}
\; \right .
\label{eq: pkznla}
\end{equation}
where the IA amplitude is modelled as
\begin{equation}
A_{\rm IA}(z) = \frac{{\cal{A}}_{\rm IA}{\cal{C}}_{\rm IA} \Om}{D(z)}\, \left ( \frac{1 + z}{1 + z_0} \right )^{\eta_{\rm IA}} \, ,
\label{eq: aiaznla}
\end{equation}
with ${\cal{C}}_{\rm IA} = 0.0139$  referring to a dimensionless amplitude, $z_0$ to a pivot redshift (here arbitrarily set to zero), 
and $({\cal{A}}_{\rm IA}, \eta_{\rm IA})$ to the model parameters. 

\begin{figure*}[htbp!]
\centering
\includegraphics[width=0.9\textwidth]{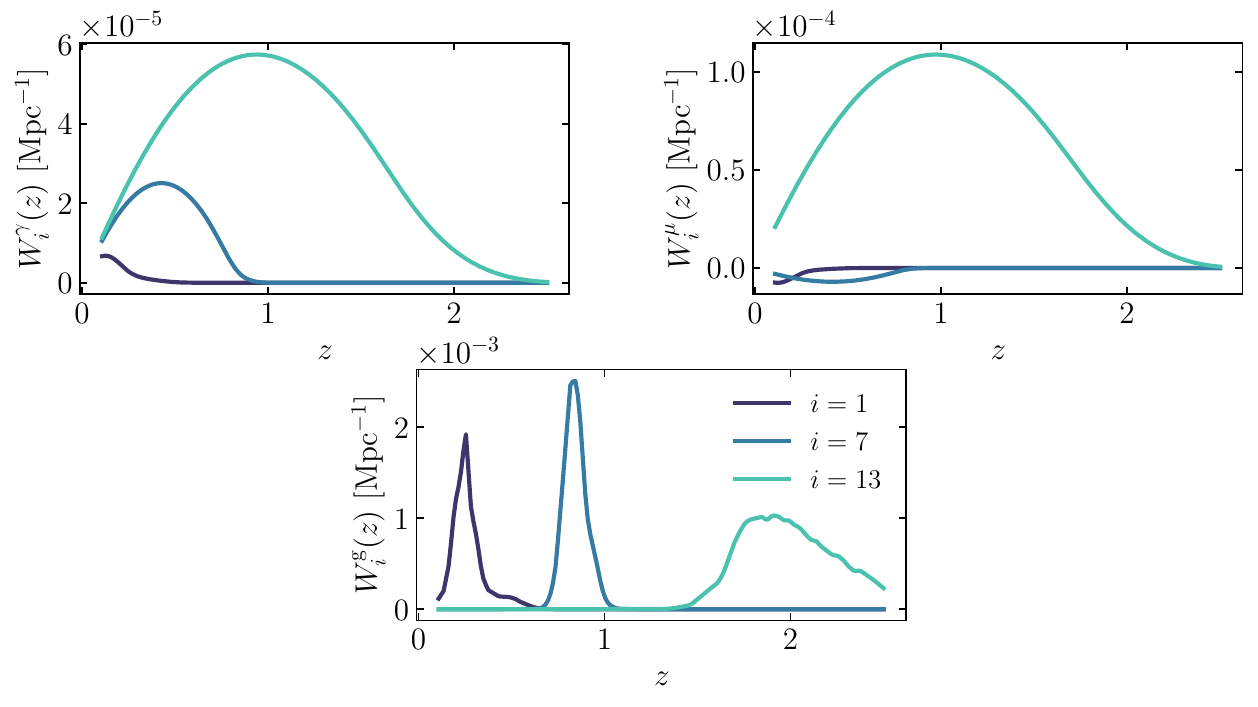}
\caption{Shear (\emph{top left}), magnification (\emph{top right}), and galaxy (\emph{bottom}) kernels for the 3\texttimes2pt harmonic power spectra for the case with 13 equipopulated redshift bins. For increased clarity, we show the kernels for three representative bins only. We do not separately show the IA kernel since it equivalent to the galaxy kernel given our use of the same galaxy sample for both lenses and sources and the absence of photometric RSD.}
\label{fig: photkernels}
\end{figure*}

\subsection{Galaxy and magnification bias}

Two other important ingredients of the recipe are the galaxy bias and the magnification bias for the lenses. We again extract both of them from the \Euclid Flagship simulation. Following Sect.\ \ref{sec:redshift_distribution}, once we have classified galaxies into different tomographic bins according to their photo-$z$, we can obtain the synthetic galaxy catalogue for each of the bins. Starting with the magnification bias, in each one of the bins we have measured the slope, $s$, of the cumulative number density of galaxies as a function of magnitude, evaluated at the magnitude cut $\IE=24.5$. Given the slope, we compute the magnification bias in the corresponding bin as $b_{\rm mag}=5s-2$. We note that we assume this magnification bias to be constant within a bin. However, given our choice of 13 tomographic bins, this introduces 13 additional nuisance parameters. Given the smooth evolution of the measured magnification biases as a function of redshift (we assign the bias of each bin at the median redshift of the corresponding distribution), we have also considered a polynomial fit given by
\begin{equation}
    b_{\rm mag}(z) = \sum_{i=0}^3 b_{{\rm mag},i}\,z^i\;,
\end{equation}
with $(b_{\rm mag,0},b_{\rm mag,1},b_{\rm mag,2},b_{\rm mag,3})=(-1.51, 1.35, 0.08, 0.04)$. These values are only used as fiducial when generating the synthetic data for the Monte Carlo forecasts in Papers 3 and 5. In the fitting procedure, they are let free to vary and then marginalised over.

For what concerns the (linear) galaxy bias, we have started from the same synthetic galaxy catalogues for the different bins. For each one of them, we have computed a theory prediction for the angular two-point correlation function and compared it to the measurements obtained with \texttt{treecorr}\,\citep{treecorr} using the Landy--Szalay estimator. We have not performed a proper $\chi^2$ fit but just a minimisation of the distance between predictions and measurements. The main reason behind this choice is speed, which has allowed us to test on the order of a hundred different tomographic-binning selections. Despite the approximations done when estimating the galaxy bias, we have verified that the results are compatible with a proper $\chi^2$ minimisation using the entire Flagship simulation. Moreover, given that we always marginalise over the galaxy bias, these estimates are just used as fiducial values and any small variation has a negligible impact on the constraints on the cosmological parameters. As was the case for the magnification bias, we need 13 additional nuisance parameters for the galaxy bias, which we either assume to be constant within each tomographic bin or allow for a smooth polynomial evolution with redshift
\begin{equation}
    b^{\rm photo}(z) = \sum_{i=0}^3 b_{{\rm gph}, i}\,z^{\, i}\;,
\end{equation}
with $(b_{\rm gph,0},b_{\rm gph,1},b_{\rm gph,2},b_{\rm gph,3})=(0.83, 1.19, -0.93, 0.42)$. As for the magnification bias, the galaxy bias coefficients will be allowed to vary when fitting the synthetic data. 

\subsection{Shear multiplicative bias}

As accurate as modern shape measurement codes can be, they are still affected by systematic uncertainties that propagate to the final shear estimate. A calibration is therefore needed to convert the measured shear into the true shear. Motivated by the typical performances of modern codes, it is customary to write the calibration relation as
\begin{equation}
\hat{\vec{\gamma}} = (1 + m_0) \vec{\gamma} + m_4 \vec{\gamma}^{\star} + \vec{c} \, ,
\label{eq: shearbiasfull}
\end{equation}
where $\hat{\vec{\gamma}}$ and $\vec{\gamma}$ are the measured and the true spin-2 complex shear fields, $\vec{\gamma}^{\star}$ is the complex coniugate, $m_0$ and $m_4$ are the complex spin-0 and spin-4 multiplicative biases, and $\vec{c}$ is the additive bias. Once $(m_0, m_4, \vec{c})$ have been determined (see e.g. \citealt{Cragg2023}), the corrected shear is used to compute the summary statistics (such as the harmonic power spectra and correlation functions). However, there could be a residual shear multiplicative bias that must be accounted for. In the most general case, this could be spatially dependent \citep{Tom2019} or coupled with the mask \citep{Tom2020}. However, under realistic assumptions, we can safely assume that the residual bias is so small that both these effects can be neglected, and rely on a first-order Taylor expansion\footnote{\cite{TomAnu2022} have shown how to propagate the shear bias if a quadratic term is needed.}
\begin{equation}\label{eq:shear_bias_def}
\hat{\vec{\gamma}} = (1+m) \, \vec{\gamma} \, ,
\end{equation}
where $m$ is a scalar quantity.
We consider therefore the $m_i$ parameters as constant (one for each tomographic bin), with a fiducial value of zero in all bins. To include these nuisance parameters, one has to update the different angular power spectra as 
\begin{equation}
\left \{
\begin{array}{l}
\displaystyle{C_{ij}^{\rm LL}(\ell) \rightarrow (1 + m_i) \,(1 + m_j)\, C_{ij}^{\rm LL}(\ell)} \, , \\
 \\ 
\displaystyle{C_{ij}^{\rm GL}(\ell) \rightarrow (1 + m_j) \,C_{ij}^{\rm GL}(\ell)} \, , \\
 \\
\displaystyle{C_{ij}^{\rm GG}(\ell) \rightarrow C_{ij}^{\rm GG}(\ell)} \, , \\
\end{array}
\right .
\label{eq: cijbias}
\end{equation}
where 
the GCph spectrum is unchanged since it does not include a shear term. 

\begin{table*}
\centering
\caption{Model assumptions and survey specifics for the estimate of quantities shown in the figures.}
\begin{tabular}{
>{\raggedright\arraybackslash}m{5.0cm}
>{\raggedright\arraybackslash}m{12.0cm}}
\hline \hline
 & \\
cosmological parameters & $(\Om, \Ob, w_0, w_a, h, n_{\rm s}, \sigma_8) = (0.32, 0.05, -1.0, 0.0, 0.6737, 0.966, 0.8155)$ \\ 
nonlinear recipe & {\tt HMCode2020} with baryon feedback for $\log_{10}{(T_{\rm AGN}/{\rm K})} = 7.75$ \\
IA model & zNLA with $({\cal{A}}_{\rm IA}, \eta_{\rm IA}) = (0.16, 1.66)$ \\
 & \\ 
\hline 
 & \\
number of redshift bins for 3$\times$2pt & 13 equipopulated bins over the range $0.2 \le z \le 2.5$ \\
number of redshift bins for GCsp & 4 equispaced bins over the range $0.9 \le z \le 1.8$ \\
number of samples & same sample for both lenses and sources \\ 
 & \\
\hline
\end{tabular}
\label{tab: modelsummary}
\end{table*}

\section{Theoretical predictions}
The above recipe has been implemented in \CLOE to make it possible to compute the observables of interest for any given set of cosmological and nuisance parameters. These theoretical predictions constitute one of the three building blocks that the likelihood computation relies on (the other two being the measurements and the covariance). In the following subsections, we will use \CLOE to evaluate relevant quantities that allow for an improved understanding of the need for the different terms that are included in the theoretical recipe. 

We will fix the model parameters to the same fiducial values used in Paper 3 to generate the synthetic data, noting that the following discussion qualitatively holds independently of the model parameter values. A short summary of the assumptions on cosmology and survey specifications is provided in Table\,\ref{tab: modelsummary}, and we refer the reader to Paper 3 for the full list of nuisance parameters. 

\begin{figure}[htbp!]
\centering
\includegraphics[width=0.5\textwidth]{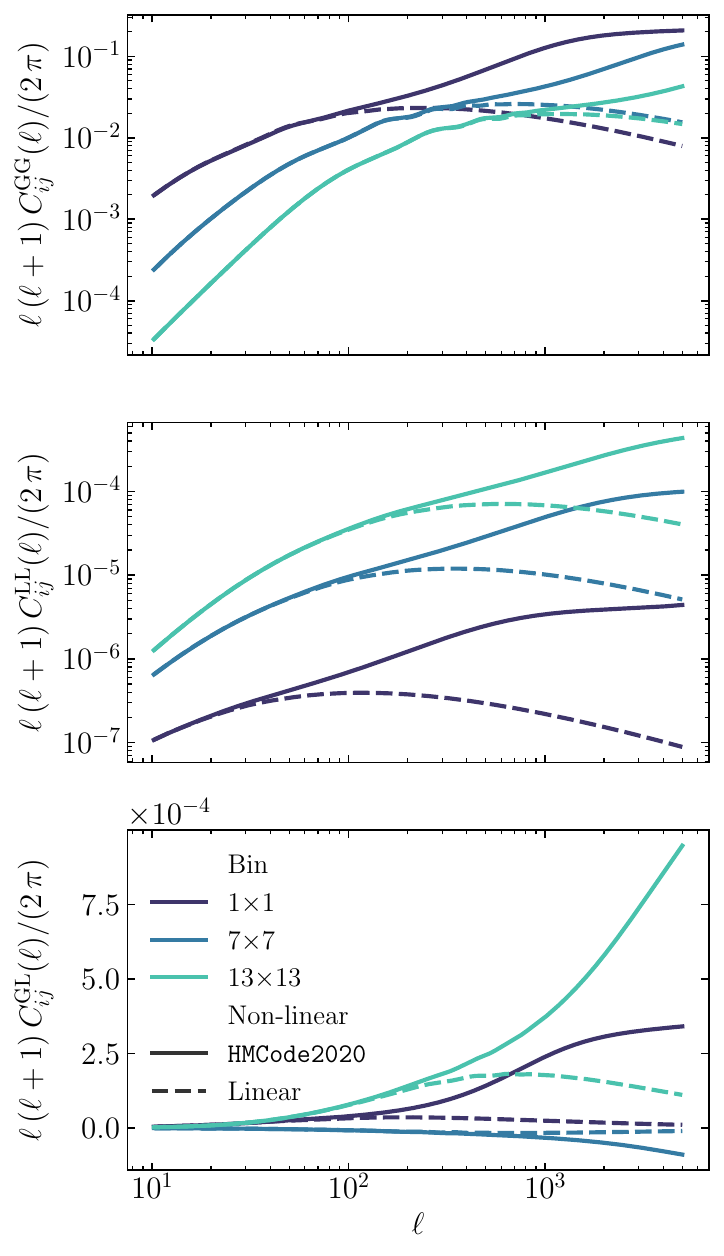}
\caption{Position-position (\emph{top}), shear-shear (\emph{centre}), and shear-position (\emph{bottom}) harmonic power spectra for representative autocorrelation bins. Dashed and solid lines refer to the use of linear and nonlinear ({\tt HMCode2020}) matter power spectra, respectively, as input.}
\label{fig: photspectra}
\end{figure}

\subsection{3\texttimes2pt observables}
We begin with the photometric observables in harmonic space. To this end, it is instructive to inspect Fig. \ref{fig: photkernels}, which illustrates the shear, magnification, galaxy, and IA kernel functions entering the integrals of the harmonic power spectra. Here, the IA and galaxy kernels 
are equivalent given our use of the same galaxy sample for both lenses and sources and the absence of photometric RSD. Most of the qualitative behaviour of the $C_{ij}^{AB}(\ell)$ functions can be inferred from the associated $W_i^{A}(z)$ kernels.

\begin{figure*}[htbp!]
\centering
\includegraphics[width=0.9\textwidth]{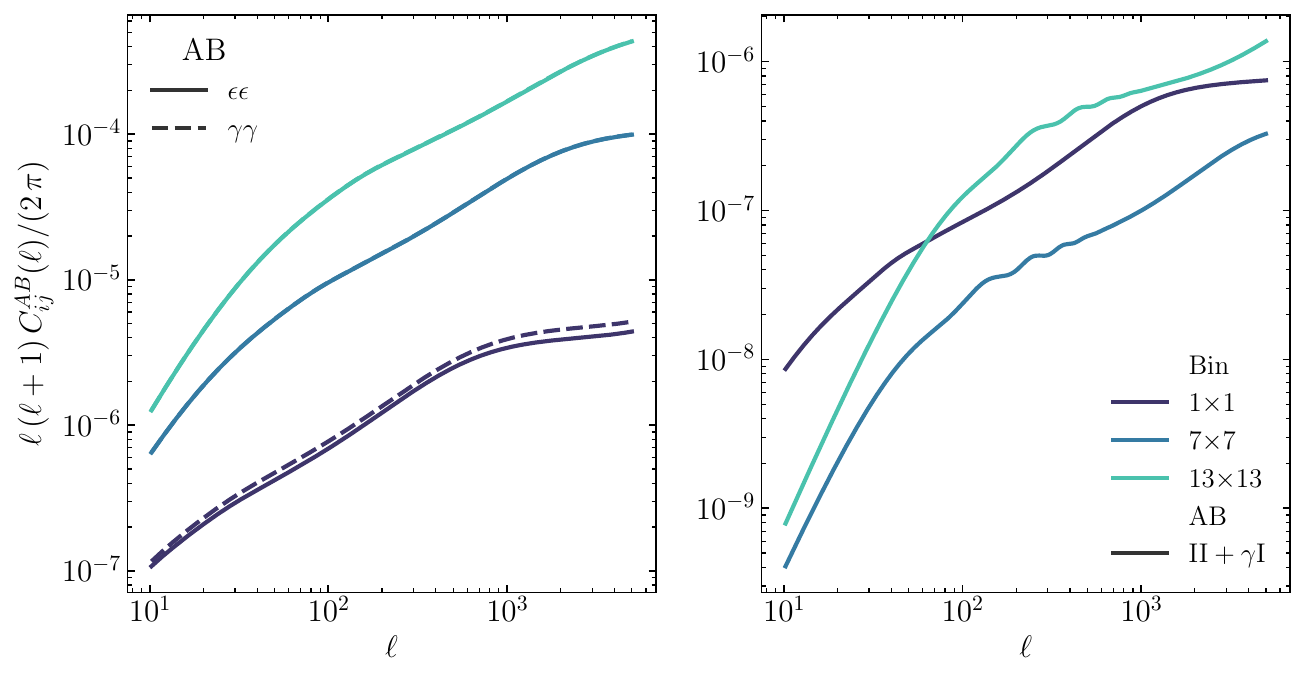}
\caption{{\it Left.} WL harmonic power spectra for three representative autocorrelations with (solid) and without (dashed) IA contributions. {\it Right.} Total IA harmonic power spectra for the same bin combinations as in the left panel.}
\label{fig: photIA}
\end{figure*}

\begin{figure*}[htbp!]
\centering
\includegraphics[width=0.9\textwidth]{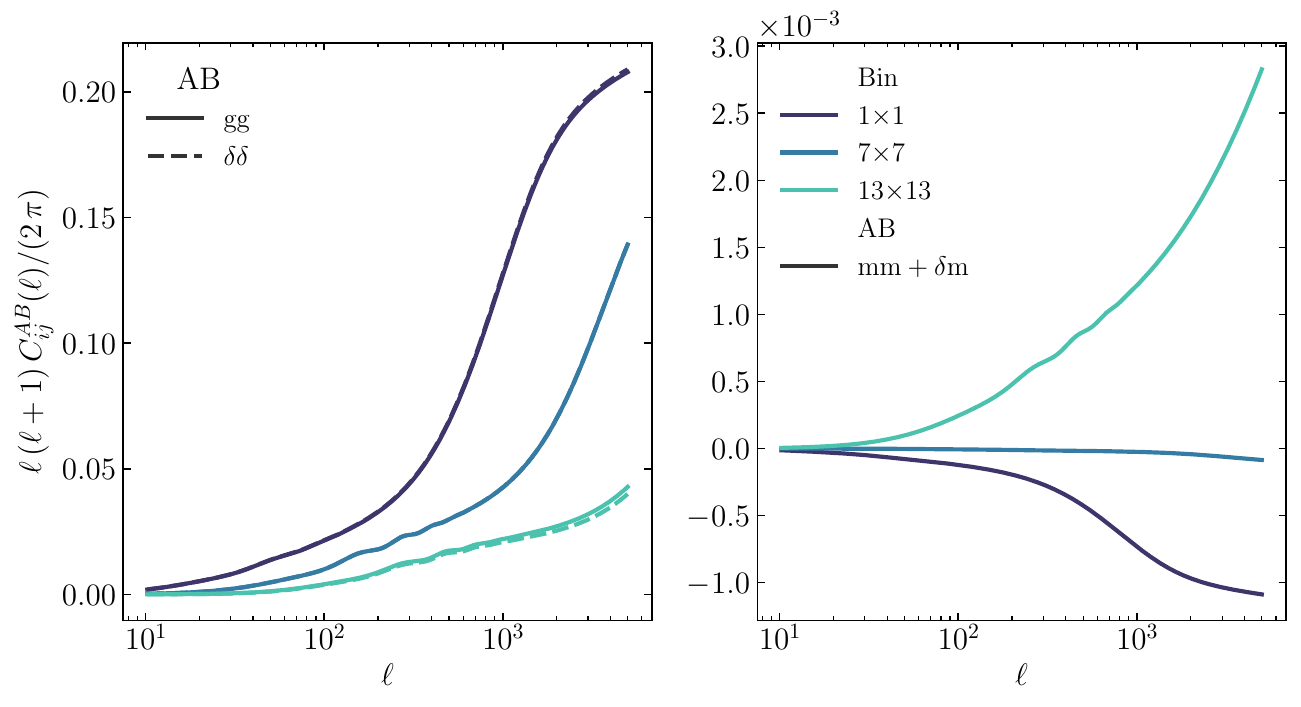}
\caption{{\it Left.} GCph harmonic power spectra for three representative autocorrelations with (solid) and without (dashed) the magnification contributions. {\it Right.} Total magnification harmonic power spectra for the same bin combinations as in the left panel.}
\label{fig: photmag}
\end{figure*}

If we consider the width of the shear $W_i^{\gamma}(z)$ and galaxy $W_i^{\rm g}(z)$ kernels, the former 
has a broader distribution for the higher redshift bins in particular,
as 
lensing probes the matter distribution along the full line of sight to the source galaxies. This difference in width has two immediate consequences that can be seen in the top and centre panels of Fig.\ \ref{fig: photspectra}. First, we notice that the fractional change in the amplitude of the signal from one tomographic bin to another is larger for $C_{ij}^{\rm LL}(\ell)$ than for $C_{ij}^{\rm GG}(\ell)$, as a consequence of the increase in the width of the shear kernel with the bin number. Note that this also motivates the inversion in the ordering of the curves along the vertical axis, with $C_{ij}^{\rm LL}(\ell)$ being larger for the higher-redshift bins, while it is the reverse for $C_{ij}^{\rm GG}(\ell)$. 
Again, this is related to the different physics underlying the cosmic shear and photometric galaxy clustering, as the former is indeed a probe of the matter distribution along the line of sight, while the latter is determined by the clustering among close galaxies. The GGL spectra inherit the combined effects of the shear and galaxy kernels, hence exhibiting a range of behaviours depending on the bin combination. This is illustrated in the bottom panel of Fig.\ \ref{fig: photspectra}, where  $C_{ij}^{\rm GL}(\ell)$ can become negative as a consequence of the 
contributions from IA and magnification. 

The different widths of the kernels also explain why WL is more sensitive to the nonlinear recipe than GCph. This can be  seen in Fig.\ \ref{fig: photspectra}, where the dashed lines (obtained by using the linear power spectra) depart from the solid ones (corresponding to the use of {\tt HMCode2020}) at smaller $\ell$ for $C_{ij}^{\rm LL}(\ell)$ than for $C_{ij}^{\rm GG}(\ell)$. This can be qualitatively understood by considering that, due to the Limber approximation, the power spectra entering the integrals defining $C_{ij}^{AB}(\ell)$ are evaluated at $k_{\ell}(z)$ from Eq.~(\ref{eq: klimber}) with $z$ spanning an effective range $(z_{i,{\mathcal L}}, z_{i,{\mathcal U}})$ depending on the redshift bin (where the subscripts ${\mathcal L}$ and ${\mathcal U}$ refer to the lower and upper boundaries). Since $W_i^{\gamma}(z)$ is non-zero also at small $z$, $k_{\ell}(z)$ can reach large values even at small $\ell$. In contrast, this is less signififant for GCph, given the narrower width of $W_i^{\rm g}(z)$. This argument applies even if one is not using the Limber approximation, although it is less immediate to grasp.

\begin{figure}[htbp!]
\centering
\includegraphics[width=0.45\textwidth]{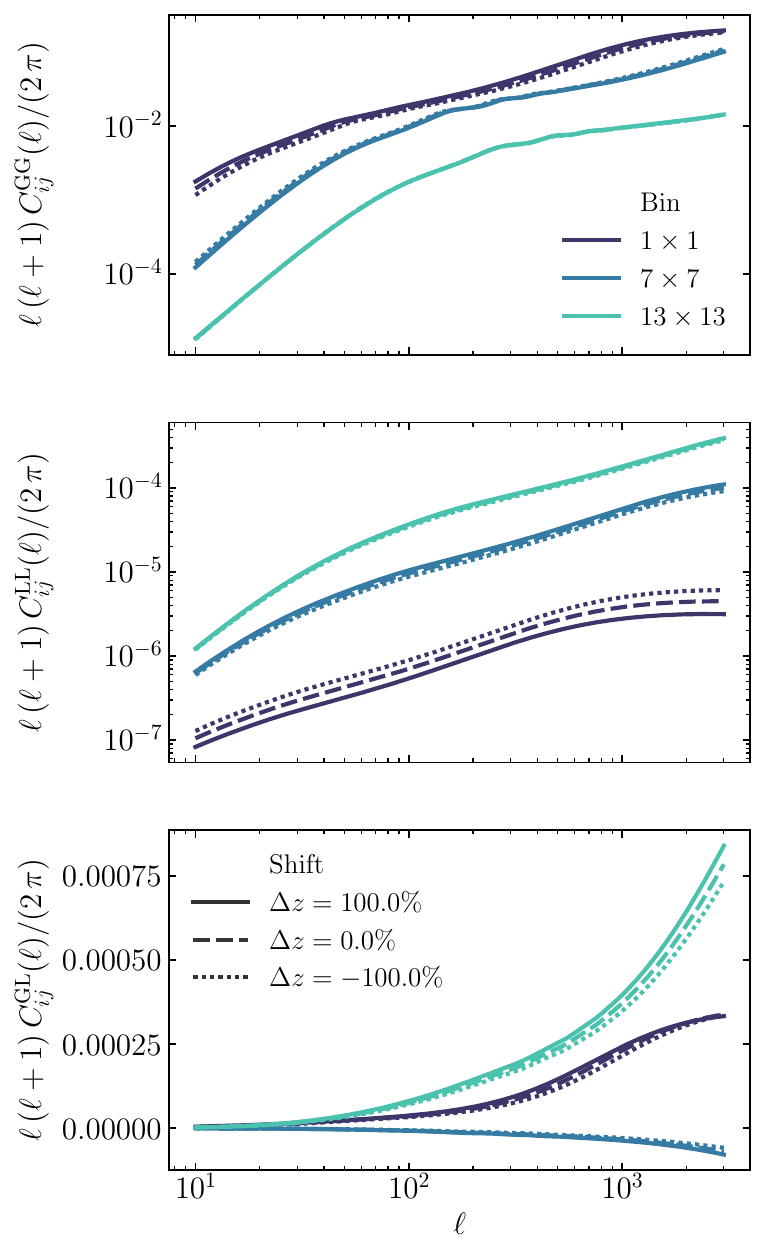}
\caption{Impact of varying the shift parameters $\Delta z_i$ on the GCph, WL, and GGL harmonic power spectra. Here, we either consider the fiducial $\Delta z_i$ shifts used in Papers 3 and 5, or change them by $\pm100\%$ (i.e, doubling or halving their value with respect to the fiducial one).}
\label{fig: photshift}
\end{figure}

Figure \ref{fig: photspectra} shows the total harmonic power spectra, but it is interesting to inspect the contribution of the contaminating signals, that is\ those terms that are of astrophysical rather than cosmological origin. In the case of WL, one of the main contaminants is the IA signal so that, in the left panel of Fig.\ \ref{fig: photIA}, we show the WL power spectra with (solid) and without (dashed) the IA contribution. The solid and dashed lines are so well superimposed when going to high redshift bins that there is no noticeable difference. This is due to the smaller amplitude of the IA signal, as can be seen from the total IA spectra in the right panel. Although the figures are restricted to the auto-correlations, the results are qualitatively the same for the cross-correlation spectra. The IA signal can have a larger contribution due to the GI term at low redshift, but it remains largely subdominant. It is worth highlighting that the overall amplitude of the IA signal is a complicated interplay between the IA kernel and the redshift dependency of $A_{\rm IA}(z)$ in Eq.\ (\ref{eq: aiaznla}). Changing the model parameters can give rise to a plethora of different scalings with bin combination, even changing the sign of the total IA signal depending on which of the II and $\gamma$I terms dominates. However, it is always true that IA only meaningfully contributes at low redshift. In this regard, proper modelling is of utmost importance in order to disentangle the impact of the IA from that of the cosmological parameters.
We refer the reader to Paper 6 and \cite{ISTNL-P2} for further details. 

\begin{figure}[htbp!]
\centering
\includegraphics[width=0.45\textwidth]{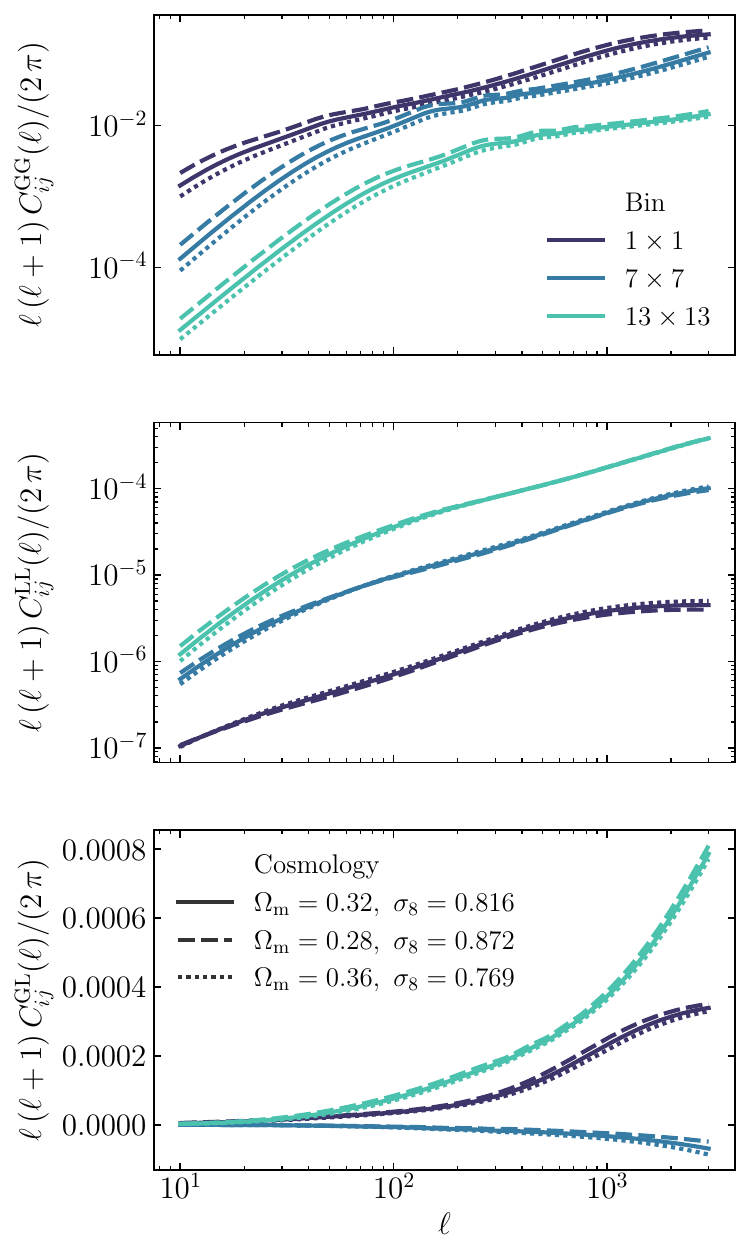}
\caption{Impact of varying $(\Omega_{\rm M}, \sigma_8)$ on the GCph, WL, and GGL harmonic power spectra.}
\label{fig: photsigma8}
\end{figure}

A similar analysis can be carried out for the contribution of lensing magnification to the GCph signal. The left panel of Fig.~\ref{fig: photmag} shows $C_{ij}^{\rm GG}(\ell)$ with (solid) and without (dashed) magnification contribution (setting the RSD term to null), while the right panel of the figure shows the total magnification power spectra. Analogously to IAs, magnification is always subdominant, which can be explained by noting that the $W_i^{\mu}(z)$ kernel is at least an order of magnitude smaller than the galaxy kernel $W_i^{\rm g}(z)$. This is both a consequence of magnification being related to lensing 
and its impact being modulated by the $b_{\rm mag}(z)$ functions. 
Note also that the magnification contribution can become negative due to the change of sign of $b_{\rm mag}(z)$. However, we warn the reader that the shape of the $b_{\rm mag}(z)$ function (and hence the sign of the magnification contribution) depends on how the lens sample is selected and how luminosity and size magnification induced by lensing change the number of galaxies entering the sample. The harmonic power spectra in Fig.\ \ref{fig: photmag} should therefore be considered as a mere example of a wider phenomenology, the only stable result being that $C_{ij}^{gg} \gg C_{ij}^{g \mu}(\ell) + C_{ij}^{\mu g}(\ell) + C_{ij}^{\mu \mu}(\ell)$.

Furthermore, there are two main observational systematic uncertainties affecting the 3$\times$2pt observables. First, it is the shear multiplicative bias, which is straightforward to understand. As Eq.~(\ref{eq: cijbias}) shows, the effect is a naive rescaling of the WL and GGL harmonic power spectrum amplitudes. As a consequence, $m_i$ are degenerate with both $\sigma_8$ and the galaxy bias. It is indeed the combination of the three probes that partially breaks the degeneracy. A more subtle effect is played by the shift $\Delta z_i$ of the mean redshift of each tomographic bin. Figure~\ref{fig: photshift} shows the impact of varying the shift $\Delta z_i$ from its fiducial values. Although the overall impact is so small that we need to use an unrealistically large shift to make it visible in the figure, it is worth noticing that this is similar to an overall rescaling of the signal amplitude. We therefore find a further degeneracy with parameters such as $\sigma_8$, the shear multiplicative bias $m_i$, and the galaxy bias. In practice, however, priors on $\Delta z_i$ are usually available, thus helping to alleviate this degeneracy.

We do not discuss here the dependence of the harmonic power spectra on the cosmological parameters, which has been demonstrated in past publications (e.g. \citealt{JK12}).
As an example, however, we show in Fig.~\ref{fig: photsigma8} the harmonic power spectra for three representative values of the $(\Omega_{\rm m}, \sigma_8)$ parameters, all corresponding to the same value of $S_8 = \sigma_8 \,  (\Omega_{\rm m}/0.3)^{1/2}$. This choice highlights the usefulness of combining WL with GCph and GGL. Indeed, apart from low $\ell$ (where the measurement error is larger), the WL harmonic power spectra for the three models are almost superimposed as a consequence of the fact that WL is mainly sensitive to $S_8$, rather than the individual  $(\Omega_{\rm m}, \sigma_8)$ values. This is not the case for GCph, and in part for GGL. Should the galaxy bias be known, one could then break the $\Omega_{\rm m}$--$\sigma_8$ degeneracy thanks to the different scalings of the GCph and WL signals with these parameters. In practice, galaxy bias must be added to the list of quantities to be constrained, but the different ways in which it enters GCph and GGL allows for an alleviation of the degeneracy with the cosmological parameters.

\subsection{Spectroscopic galaxy clustering}
\begin{figure}
    \centering
    \includegraphics[width=0.5\textwidth]{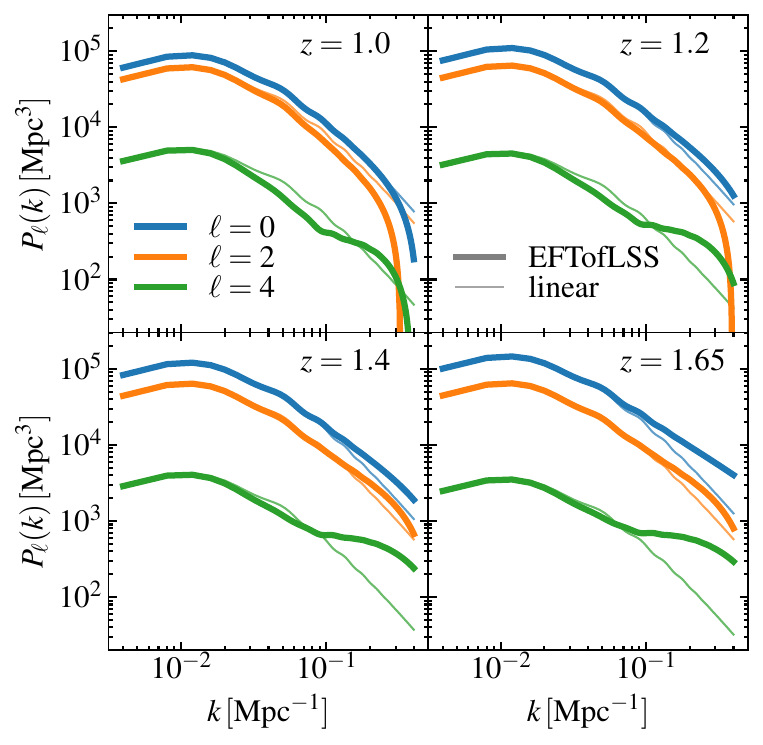}
    \caption{Spectroscopic power spectrum multipoles at the four redshifts adopted in the forecasted analyses of Papers 3 and 5, specified in the upper right corner of each panel. Different colours correspond to different Legendre multipoles, with blue, orange, and green corresponding to the monopole, quadrupole, and hexadecapole, respectively. Thicker lines show predictions for the EFTofLSS model, while thinner lines mark linear theory predictions.}
    \label{fig:GCspectro_pk}
\end{figure}

\begin{figure}
    \centering
    \includegraphics[width=0.5\textwidth]{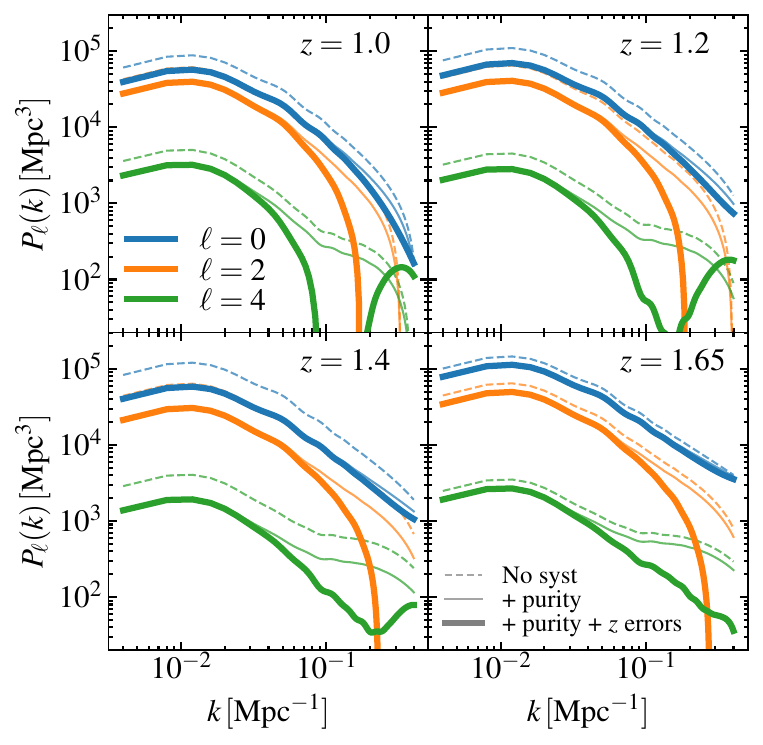}
    \caption{Impact of the different observational systematic uncertainties on the amplitude of the power spectrum Legendre multipoles (here assuming EFTofLSS predictions). The dashed, solid, and thicker solid lines respectively correspond to the cases without observational systematics, with the purity correction (assuming the fiducial value for the redshift bin centred at the individual redshift bin), and with the small-scale damping induced by the spectroscopic redshift errors additionally included (assuming $\sigma_{0,z}=0.002$). As in Fig.\ \ref{fig:GCspectro_pk}, the different colours mark different Legendre multipoles.}
    \label{fig:GCspectro_syst}
\end{figure}

\begin{figure}
    \centering
    \includegraphics[width=0.5\textwidth]{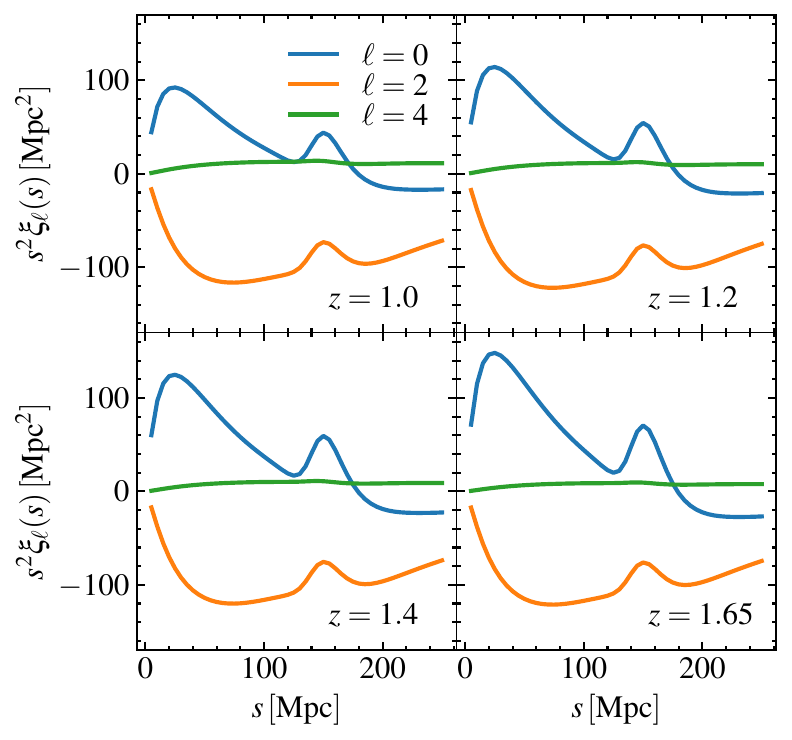}
    \caption{Same as in Fig.\ \ref{fig:GCspectro_pk}, but showing the linear-theory spectroscopic 2-point correlation function multipoles.}
    \label{fig:GCspectro_xi}
\end{figure}

For the spectroscopic probes, we show in Fig.\ \ref{fig:GCspectro_pk} the Legendre multipoles of order $\ell=0,2,4$ evaluated at the mean redshift of the four spectroscopic bins adopted in \citet{EuclidSkyOverview} and the forecasted analyses of Papers 3 and 5. Predictions from linear theory and the EFTofLSS formalism introduced in Sect.\ \ref{sc: recipespec} are represented by the thin and thick lines, respectively. The parameters used to generate the EFTofLSS predictions are also consistent with the fiducial ones employed in Papers 3 and 5. The values have been calibrated from comoving snapshots of H$\alpha$ galaxies \citep{EP-Pezzotta} obtained from a previous version of the Flagship Simulation \citep{EuclidSkyFlagship}. The reference scale cuts corresponding to these fits are redshift-dependent and become progressively more conservative (from $\sim0.3\,h\,{\rm Mpc}^{-1}$ to $\sim0.2\,h\,{\rm Mpc}^{-1}$) at lower redshifts. This is justified by the fact that galaxy clustering becomes increasingly nonlinear as redshift decreases, leading to a more rapid deterioration of the EFTofLSS model accuracy compared to higher redshifts.

In Fig.\ \ref{fig:GCspectro_pk}, the Poisson noise limit, defined as the inverse of the mean number density within the corresponding redshift bin (in this case, we do not include the impact of systematic uncertainties and thus consider $n_{\rm true}$), has been subtracted from the anisotropic galaxy power spectrum $P_{\rm gg}(k,\mu)$. Since this enters only as an isotropic correction, the only multipole affected is the monopole $P^{\rm spectro}_0(k)$, which corresponds to the spherical average of $P_{\rm gg}(k,\mu)$.

The impact of the observational systematic uncertainties described in Sect.\ \ref{sec:GCspectro_systematics} is shown in Fig.\ \ref{fig:GCspectro_syst}. In this figure, we progressively plot the Legendre multipoles under the assumption of having no systematics, including the effect of the sample purity, and finally including the damping induced by the spectroscopic-redshift uncertainty. For the last two cases, we adopt values for the purity and redshift errors calibrated with dedicated \Euclid-like simulations.
Finally, in Fig.\ \ref{fig:GCspectro_xi}, we show the linear predictions for the 2-point correlation function multipoles $\xi_{\ell}(s)$, obtained by carrying out the Hankel transformation in Eq.~(\ref{eq:pl2xil}).

\section{Conclusions}

\CLOE is 
a software package for the computation of the theoretical predictions and likelihood of large-scale structure observables, and the resulting parameter constraints for a given cosmological model and treatment of systematic uncertainties.
\CLOE is currently the only code of its kind that is designed for both photometric and spectroscopic probes in both harmonic and configuration space. This unique feature is necessary to deal with the dataset of \Euclid, which is the first Stage-IV experiment to employ both photometric and spectroscopic observables.

This first paper in a series of five has presented what we refer to as the {\it recipe}. We have described the different formulae needed to compute the theoretical predictions of the cosmological observables. We have explained the building blocks that \CLOE is based on so that every user can replicate the results. Their implementation is detailed in Paper 2, while the benchmarking against independent codes is presented in Paper 4.   

Moving forward, there are additional features planned for \CLOE. This includes higher-order effects in the computation of the lensing power spectrum $C_{ij}^{\rm LL}(\ell)$ to prevent biasing the constraints on the cosmological parameters \citep{Deshpande-EP28}.
To this end, source magnification bias, source-lens clustering \citep{Laila2024}, source obscuration, and local Universe terms are the major candidates to be implemented.
However, note that these effects are not as pertinent for the analysis of \Euclid's first data release (DR1) given the smaller survey area, conservative scale cuts, and use of the joint 3$\times$2pt and spectroscopic galaxy clustering observables.
As an additional example, the recipe to account for sample purity needs to be improved. This is because we have assumed that the contaminants do not cluster, which may not hold and will be updated in a future version of \CLOE.

Hence, as \Euclid proceeds to collect an increasing amount of data and decrease the statistical and systematic measurement uncertainties, it is of paramount importance to ensure that the theoretical recipe underpinning \CLOE continues to be refined. 
This will allow \CLOE to perform the precision cosmological inferences needed to unveil the mystery behind the accelerated expansion of the Universe.

%
%

\begin{acknowledgements}
  
\AckEC  

VFC, MM, DS acknowledge funding by the Agenzia Spaziale Italiana (\textsc{asi}) under agreement no. 2018-23-HH.0. VFC, MM, DS, SC, SD, SDD are also financially supported by INFN/Euclid.
SJ acknowledges the Ram\'{o}n y Cajal Fellowship (RYC2022-036431-I) from the Spanish Ministry of Science and the Dennis Sciama Fellowship at the University of Portsmouth.
SC acknowledges support from the Italian Ministry of University and Research (\textsc{mur}), PRIN 2022 `EXSKALIBUR – Euclid-Cross-SKA: Likelihood Inference Building for Universe's Research', Grant No.\ 20222BBYB9, CUP D53D2300252 0006, from the Italian Ministry of Foreign Affairs and International
Cooperation (\textsc{maeci}), Grant No.\ ZA23GR03, and from the European Union -- Next Generation EU. During part of this work, AMCLB was supported by a Paris Observatory-PSL University Fellowship, hosted at the Paris Observatory. The authors acknowledge the contribution of the Lorentz Center (Leiden), and of the European Space Agency (ESA), where the workshop "Making \CLOE shine" and the "\CLOE workshop 2023" were held.
\end{acknowledgements}

%

\bibliography{biblio, Euclid}

\begin{thebibliography}{172}
\expandafter\ifx\csname natexlab\endcsname\relax\def\natexlab#1{#1}\fi

\bibitem[{{Abbott} {et~al.}(2018){Abbott}, {Abdalla}, {Alarcon}, {Aleksi{\'c}}, {Allam}, {Allen}, {Amara}, {Annis}, {Asorey}, {Avila}, {Bacon}, {Balbinot}, {Banerji}, {Banik}, {Barkhouse}, {Baumer}, {Baxter}, {Bechtol}, {Becker}, {Benoit-L{\'e}vy}, {Benson}, {Bernstein}, {Bertin}, {Blazek}, {Bridle}, {Brooks}, {Brout}, {Buckley-Geer}, {Burke}, {Busha}, {Campos}, {Capozzi}, {Carnero Rosell}, {Carrasco Kind}, {Carretero}, {Castander}, {Cawthon}, {Chang}, {Chen}, {Childress}, {Choi}, {Conselice}, {Crittenden}, {Crocce}, {Cunha}, {D'Andrea}, {da Costa}, {Das}, {Davis}, {Davis}, {De Vicente}, {DePoy}, {DeRose}, {Desai}, {Diehl}, {Dietrich}, {Dodelson}, {Doel}, {Drlica-Wagner}, {Eifler}, {Elliott}, {Elsner}, {Elvin-Poole}, {Estrada}, {Evrard}, {Fang}, {Fernandez}, {Fert{\'e}}, {Finley}, {Flaugher}, {Fosalba}, {Friedrich}, {Frieman}, {Garc{\'\i}a-Bellido}, {Garcia-Fernandez}, {Gatti}, {Gaztanaga}, {Gerdes}, {Giannantonio}, {Gill}, {Glazebrook}, {Goldstein}, {Gruen}, {Gruendl}, {Gschwend}, {Gutierrez}, {Hamilton},
  {Hartley}, {Hinton}, {Honscheid}, {Hoyle}, {Huterer}, {Jain}, {James}, {Jarvis}, {Jeltema}, {Johnson}, {Johnson}, {Kacprzak}, {Kent}, {Kim}, {King}, {Kirk}, {Kokron}, {Kovacs}, {Krause}, {Krawiec}, {Kremin}, {Kuehn}, {Kuhlmann}, {Kuropatkin}, {Lacasa}, {Lahav}, {Li}, {Liddle}, {Lidman}, {Lima}, {Lin}, {MacCrann}, {Maia}, {Makler}, {Manera}, {March}, {Marshall}, {Martini}, {McMahon}, {Melchior}, {Menanteau}, {Miquel}, {Miranda}, {Mudd}, {Muir}, {M{\"o}ller}, {Neilsen}, {Nichol}, {Nord}, {Nugent}, {Ogando}, {Palmese}, {Peacock}, {Peiris}, {Peoples}, {Percival}, {Petravick}, {Plazas}, {Porredon}, {Prat}, {Pujol}, {Rau}, {Refregier}, {Ricker}, {Roe}, {Rollins}, {Romer}, {Roodman}, {Rosenfeld}, {Ross}, {Rozo}, {Rykoff}, {Sako}, {Salvador}, {Samuroff}, {S{\'a}nchez}, {Sanchez}, {Santiago}, {Scarpine}, {Schindler}, {Scolnic}, {Secco}, {Serrano}, {Sevilla-Noarbe}, {Sheldon}, {Smith}, {Smith}, {Smith}, {Soares-Santos}, {Sobreira}, {Suchyta}, {Tarle}, {Thomas}, {Troxel}, {Tucker}, {Tucker}, {Uddin}, {Varga},
  {Vielzeuf}, {Vikram}, {Vivas}, {Walker}, {Wang}, {Wechsler}, {Weller}, {Wester}, {Wolf}, {Yanny}, {Yuan}, {Zenteno}, {Zhang}, {Zhang}, {Zuntz}, \& {Dark Energy Survey Collaboration}}]{Abbott2018_DES_Y1}
{Abbott}, T.~M.~C., {Abdalla}, F.~B., {Alarcon}, A., {et~al.} 2018, \prd, 98, 043526

\bibitem[{{Abbott} {et~al.}(2022){Abbott}, {Aguena}, {Alarcon}, {Allam}, {Alves}, {Amon}, {Andrade-Oliveira}, {Annis}, {Avila}, {Bacon}, {Baxter}, {Bechtol}, {Becker}, {Bernstein}, {Bhargava}, {Birrer}, {Blazek}, {Brandao-Souza}, {Bridle}, {Brooks}, {Buckley-Geer}, {Burke}, {Camacho}, {Campos}, {Carnero Rosell}, {Carrasco Kind}, {Carretero}, {Castander}, {Cawthon}, {Chang}, {Chen}, {Chen}, {Choi}, {Conselice}, {Cordero}, {Costanzi}, {Crocce}, {da Costa}, {da Silva Pereira}, {Davis}, {Davis}, {De Vicente}, {DeRose}, {Desai}, {Di Valentino}, {Diehl}, {Dietrich}, {Dodelson}, {Doel}, {Doux}, {Drlica-Wagner}, {Eckert}, {Eifler}, {Elsner}, {Elvin-Poole}, {Everett}, {Evrard}, {Fang}, {Farahi}, {Fernandez}, {Ferrero}, {Fert{\'e}}, {Fosalba}, {Friedrich}, {Frieman}, {Garc{\'\i}a-Bellido}, {Gatti}, {Gaztanaga}, {Gerdes}, {Giannantonio}, {Giannini}, {Gruen}, {Gruendl}, {Gschwend}, {Gutierrez}, {Harrison}, {Hartley}, {Herner}, {Hinton}, {Hollowood}, {Honscheid}, {Hoyle}, {Huff}, {Huterer}, {Jain}, {James}, {Jarvis},
  {Jeffrey}, {Jeltema}, {Kovacs}, {Krause}, {Kron}, {Kuehn}, {Kuropatkin}, {Lahav}, {Leget}, {Lemos}, {Liddle}, {Lidman}, {Lima}, {Lin}, {MacCrann}, {Maia}, {Marshall}, {Martini}, {McCullough}, {Melchior}, {Mena-Fern{\'a}ndez}, {Menanteau}, {Miquel}, {Mohr}, {Morgan}, {Muir}, {Myles}, {Nadathur}, {Navarro-Alsina}, {Nichol}, {Ogando}, {Omori}, {Palmese}, {Pandey}, {Park}, {Paz-Chinch{\'o}n}, {Petravick}, {Pieres}, {Plazas Malag{\'o}n}, {Porredon}, {Prat}, {Raveri}, {Rodriguez-Monroy}, {Rollins}, {Romer}, {Roodman}, {Rosenfeld}, {Ross}, {Rykoff}, {Samuroff}, {S{\'a}nchez}, {Sanchez}, {Sanchez}, {Sanchez Cid}, {Scarpine}, {Schubnell}, {Scolnic}, {Secco}, {Serrano}, {Sevilla-Noarbe}, {Sheldon}, {Shin}, {Smith}, {Soares-Santos}, {Suchyta}, {Swanson}, {Tabbutt}, {Tarle}, {Thomas}, {To}, {Troja}, {Troxel}, {Tucker}, {Tutusaus}, {Varga}, {Walker}, {Weaverdyck}, {Wechsler}, {Weller}, {Yanny}, {Yin}, {Zhang}, {Zuntz}, \& {DES Collaboration}}]{desy3}
{Abbott}, T.~M.~C., {Aguena}, M., {Alarcon}, A., {et~al.} 2022, \prd, 105, 023520

\bibitem[{{Abdalla} {et~al.}(2022){Abdalla}, {Abell{\'a}n}, {Aboubrahim}, {Agnello}, {Akarsu}, {Akrami}, {Alestas}, {Aloni}, {Amendola}, {Anchordoqui}, {Anderson}, {Arendse}, {Asgari}, {Ballardini}, {Barger}, {Basilakos}, {Batista}, {Battistelli}, {Battye}, {Benetti}, {Benisty}, {Berlin}, {de Bernardis}, {Berti}, {Bidenko}, {Birrer}, {Blakeslee}, {Boddy}, {Bom}, {Bonilla}, {Borghi}, {Bouchet}, {Braglia}, {Buchert}, {Buckley-Geer}, {Calabrese}, {Caldwell}, {Camarena}, {Capozziello}, {Casertano}, {Chen}, {Chluba}, {Chen}, {Chen}, {Chudaykin}, {Cicoli}, {Copi}, {Courbin}, {Cyr-Racine}, {Czerny}, {Dainotti}, {D'Amico}, {Davis}, {de Cruz P{\'e}rez}, {de Haro}, {Delabrouille}, {Denton}, {Dhawan}, {Dienes}, {Di Valentino}, {Du}, {Eckert}, {Escamilla-Rivera}, {Fert{\'e}}, {Finelli}, {Fosalba}, {Freedman}, {Frusciante}, {Gazta{\~n}aga}, {Giar{\`e}}, {Giusarma}, {G{\'o}mez-Valent}, {Handley}, {Harrison}, {Hart}, {Hazra}, {Heavens}, {Heinesen}, {Hildebrandt}, {Hill}, {Hogg}, {Holz}, {Hooper}, {Hosseininejad}, {Huterer},
  {Ishak}, {Ivanov}, {Jaffe}, {Jang}, {Jedamzik}, {Jimenez}, {Joseph}, {Joudaki}, {Kamionkowski}, {Karwal}, {Kazantzidis}, {Keeley}, {Klasen}, {Komatsu}, {Koopmans}, {Kumar}, {Lamagna}, {Lazkoz}, {Lee}, {Lesgourgues}, {Levi Said}, {Lewis}, {L'Huillier}, {Lucca}, {Maartens}, {Macri}, {Marfatia}, {Marra}, {Martins}, {Masi}, {Matarrese}, {Mazumdar}, {Melchiorri}, {Mena}, {Mersini-Houghton}, {Mertens}, {Milakovi{\'c}}, {Minami}, {Miranda}, {Moreno-Pulido}, {Moresco}, {Mota}, {Mottola}, {Mozzon}, {Muir}, {Mukherjee}, {Mukherjee}, {Naselsky}, {Nath}, {Nesseris}, {Niedermann}, {Notari}, {Nunes}, {{\'O} Colg{\'a}in}, {Owens}, {{\"O}z{\"u}lker}, {Pace}, {Paliathanasis}, {Palmese}, {Pan}, {Paoletti}, {Perez Bergliaffa}, {Perivolaropoulos}, {Pesce}, {Pettorino}, {Philcox}, {Pogosian}, {Poulin}, {Poulot}, {Raveri}, {Reid}, {Renzi}, {Riess}, {Sabla}, {Salucci}, {Salzano}, {Saridakis}, {Sathyaprakash}, {Schmaltz}, {Sch{\"o}neberg}, {Scolnic}, {Sen}, {Sehgal}, {Shafieloo}, {Sheikh-Jabbari}, {Silk}, {Silvestri}, {Skara},
  {Sloth}, {Soares-Santos}, {Sol{\`a} Peracaula}, {Songsheng}, {Soriano}, {Staicova}, {Starkman}, {Szapudi}, {Teixeira}, {Thomas}, {Treu}, {Trott}, {van de Bruck}, {Vazquez}, {Verde}, {Visinelli}, {Wang}, {Wang}, {Wang}, {Watkins}, {Watson}, {Webb}, {Weiner}, {Weltman}, {Witte}, {Wojtak}, {Yadav}, {Yang}, {Zhao}, \& {Zumalac{\'a}rregui}}]{Abdalla2022}
{Abdalla}, E., {Abell{\'a}n}, G.~F., {Aboubrahim}, A., {et~al.} 2022, Journal of High Energy Astrophysics, 34, 49

\bibitem[{{Alam} {et~al.}(2021){Alam}, {Aubert}, {Avila}, {Balland}, {Bautista}, {Bershady}, {Bizyaev}, {Blanton}, {Bolton}, {Bovy}, {Brinkmann}, {Brownstein}, {Burtin}, {Chabanier}, {Chapman}, {Choi}, {Chuang}, {Comparat}, {Cousinou}, {Cuceu}, {Dawson}, {de la Torre}, {de Mattia}, {Agathe}, {des Bourboux}, {Escoffier}, {Etourneau}, {Farr}, {Font-Ribera}, {Frinchaboy}, {Fromenteau}, {Gil-Mar{\'\i}n}, {Le Goff}, {Gonzalez-Morales}, {Gonzalez-Perez}, {Grabowski}, {Guy}, {Hawken}, {Hou}, {Kong}, {Parker}, {Klaene}, {Kneib}, {Lin}, {Long}, {Lyke}, {de la Macorra}, {Martini}, {Masters}, {Mohammad}, {Moon}, {Mueller}, {Mu{\~n}oz-Guti{\'e}rrez}, {Myers}, {Nadathur}, {Neveux}, {Newman}, {Noterdaeme}, {Oravetz}, {Oravetz}, {Palanque-Delabrouille}, {Pan}, {Paviot}, {Percival}, {P{\'e}rez-R{\`a}fols}, {Petitjean}, {Pieri}, {Prakash}, {Raichoor}, {Ravoux}, {Rezaie}, {Rich}, {Ross}, {Rossi}, {Ruggeri}, {Ruhlmann-Kleider}, {S{\'a}nchez}, {S{\'a}nchez}, {S{\'a}nchez-Gallego}, {Sayres}, {Schneider}, {Seo}, {Shafieloo},
  {Slosar}, {Smith}, {Stermer}, {Tamone}, {Tinker}, {Tojeiro}, {Vargas-Maga{\~n}a}, {Variu}, {Wang}, {Weaver}, {Weijmans}, {Y{\`e}che}, {Zarrouk}, {Zhao}, {Zhao}, \& {Zheng}}]{eBOSS2021}
{Alam}, S., {Aubert}, M., {Avila}, S., {et~al.} 2021, \prd, 103, 083533

\bibitem[{{Albrecht} {et~al.}(2006){Albrecht}, {Bernstein}, {Cahn}, {Freedman}, {Hewitt}, {Hu}, {Huth}, {Kamionkowski}, {Kolb}, {Knox}, {Mather}, {Staggs}, \& {Suntzeff}}]{ReportDETF}
{Albrecht}, A., {Bernstein}, G., {Cahn}, R., {et~al.} 2006, astro-ph/0609591

\bibitem[{Alcock \& Paczyński(1979)}]{AlcPac1979}
Alcock, C. \& Paczyński, B. 1979, Nature, 281, 358

\bibitem[{{Alonso} {et~al.}(2019){Alonso}, {Sanchez}, {Slosar}, \& {LSST Dark Energy Science Collaboration}}]{NaMaster}
{Alonso}, D., {Sanchez}, J., {Slosar}, A., \& {LSST Dark Energy Science Collaboration}. 2019, \mnras, 484, 4127

\bibitem[{{Amendola} {et~al.}(2018){Amendola}, {Appleby}, {Avgoustidis}, {Bacon}, {Baker}, {Baldi}, {Bartolo}, {Blanchard}, {Bonvin}, {Borgani}, {Branchini}, {Burrage}, {Camera}, {Carbone}, {Casarini}, {Cropper}, {de Rham}, {Dietrich}, {Di Porto}, {Durrer}, {Ealet}, {Ferreira}, {Finelli}, {Garc{\'\i}a-Bellido}, {Giannantonio}, {Guzzo}, {Heavens}, {Heisenberg}, {Heymans}, {Hoekstra}, {Hollenstein}, {Holmes}, {Hwang}, {Jahnke}, {Kitching}, {Koivisto}, {Kunz}, {La Vacca}, {Linder}, {March}, {Marra}, {Martins}, {Majerotto}, {Markovic}, {Marsh}, {Marulli}, {Massey}, {Mellier}, {Montanari}, {Mota}, {Nunes}, {Percival}, {Pettorino}, {Porciani}, {Quercellini}, {Read}, {Rinaldi}, {Sapone}, {Sawicki}, {Scaramella}, {Skordis}, {Simpson}, {Taylor}, {Thomas}, {Trotta}, {Verde}, {Vernizzi}, {Vollmer}, {Wang}, {Weller}, \& {Zlosnik}}]{ReviewAmendola2018}
{Amendola}, L., {Appleby}, S., {Avgoustidis}, A., {et~al.} 2018, Living Reviews in Relativity, 21, 2

\bibitem[{{Amendola} {et~al.}(2008){Amendola}, {Kunz}, \& {Sapone}}]{Amendola2008}
{Amendola}, L., {Kunz}, M., \& {Sapone}, D. 2008, JCAP, 04, 013

\bibitem[{{Amon} {et~al.}(2023){Amon}, {Robertson}, {Miyatake}, {Heymans}, {White}, {DeRose}, {Yuan}, {Wechsler}, {Varga}, {Bocquet}, {Dvornik}, {More}, {Ross}, {Hoekstra}, {Alarcon}, {Asgari}, {Blazek}, {Campos}, {Chen}, {Choi}, {Crocce}, {Diehl}, {Doux}, {Eckert}, {Elvin-Poole}, {Everett}, {Fert{\'e}}, {Gatti}, {Giannini}, {Gruen}, {Gruendl}, {Hartley}, {Herner}, {Hildebrandt}, {Huang}, {Huff}, {Joachimi}, {Lee}, {MacCrann}, {Myles}, {Navarro-Alsina}, {Nishimichi}, {Prat}, {Secco}, {Sevilla-Noarbe}, {Sheldon}, {Shin}, {Tr{\"o}ster}, {Troxel}, {Tutusaus}, {Wright}, {Yin}, {Aguena}, {Allam}, {Annis}, {Bacon}, {Bilicki}, {Brooks}, {Burke}, {Carnero Rosell}, {Carretero}, {Castander}, {Cawthon}, {Costanzi}, {da Costa}, {Pereira}, {de Jong}, {De Vicente}, {Desai}, {Dietrich}, {Doel}, {Ferrero}, {Frieman}, {Garc{\'\i}a-Bellido}, {Gerdes}, {Gschwend}, {Gutierrez}, {Hinton}, {Hollowood}, {Honscheid}, {Huterer}, {Kannawadi}, {Kuehn}, {Kuropatkin}, {Lahav}, {Lima}, {Maia}, {Marshall}, {Menanteau}, {Miquel}, {Mohr},
  {Morgan}, {Muir}, {Paz-Chinch{\'o}n}, {Pieres}, {Plazas Malag{\'o}n}, {Porredon}, {Rodriguez-Monroy}, {Roodman}, {Sanchez}, {Serrano}, {Shan}, {Suchyta}, {Swanson}, {Tarle}, {Thomas}, {To}, \& {Zhang}}]{Amon2022}
{Amon}, A., {Robertson}, N.~C., {Miyatake}, H., {et~al.} 2023, \mnras, 518, 477

\bibitem[{{Angulo} {et~al.}(2021){Angulo}, {Zennaro}, {Contreras}, {Aric{\`o}}, {Pellejero-Iba{\~n}ez}, \& {St{\"u}cker}}]{BACCO}
{Angulo}, R.~E., {Zennaro}, M., {Contreras}, S., {et~al.} 2021, \mnras, 507, 5869

\bibitem[{{Aric{\`o}} {et~al.}(2021){Aric{\`o}}, {Angulo}, {Contreras}, {Ondaro-Mallea}, {Pellejero-Iba{\~n}ez}, \& {Zennaro}}]{BCemu}
{Aric{\`o}}, G., {Angulo}, R.~E., {Contreras}, S., {et~al.} 2021, \mnras, 506, 4070

\bibitem[{{Asgari} {et~al.}(2021){Asgari}, {Lin}, {Joachimi}, {Giblin}, {Heymans}, {Hildebrandt}, {Kannawadi}, {St{\"o}lzner}, {Tr{\"o}ster}, {van den Busch}, {Wright}, {Bilicki}, {Blake}, {de Jong}, {Dvornik}, {Erben}, {Getman}, {Hoekstra}, {K{\"o}hlinger}, {Kuijken}, {Miller}, {Radovich}, {Schneider}, {Shan}, \& {Valentijn}}]{KiDSAsgari2021}
{Asgari}, M., {Lin}, C.-A., {Joachimi}, B., {et~al.} 2021, \aap, 645, A104

\bibitem[{{Asgari} {et~al.}(2012){Asgari}, {Schneider}, \& {Simon}}]{Asgari2012}
{Asgari}, M., {Schneider}, P., \& {Simon}, P. 2012, \aap, 542, A122

\bibitem[{{Asgari} {et~al.}(2018){Asgari}, {Taylor}, {Joachimi}, \& {Kitching}}]{Asgari18}
{Asgari}, M., {Taylor}, A., {Joachimi}, B., \& {Kitching}, T.~D. 2018, \mnras, 479, 454

\bibitem[{{Bacon} {et~al.}(2000){Bacon}, {Refregier}, \& {Ellis}}]{Bacon2000}
{Bacon}, D.~J., {Refregier}, A.~R., \& {Ellis}, R.~S. 2000, \mnras, 318, 625

\bibitem[{Baldauf {et~al.}(2015)Baldauf, Mirbabayi, Simonovi\ifmmode~\acute{c}\else \'{c}\fi{}, \& Zaldarriaga}]{BalMirSim2015}
Baldauf, T., Mirbabayi, M., Simonovi\ifmmode~\acute{c}\else \'{c}\fi{}, M., \& Zaldarriaga, M. 2015, Phys. Rev. D, 92, 043514

\bibitem[{{Ballinger} {et~al.}(1996){Ballinger}, {Peacock}, \& {Heavens}}]{Ballinger1996}
{Ballinger}, W.~E., {Peacock}, J.~A., \& {Heavens}, A.~F. 1996, \mnras, 282, 877

\bibitem[{{Becker}(2013)}]{Becker2013}
{Becker}, M.~R. 2013, \mnras, 435, 1547

\bibitem[{{Becker} \& {Rozo}(2016)}]{BR2016}
{Becker}, M.~R. \& {Rozo}, E. 2016, \mnras, 457, 304

\bibitem[{{Bennett} {et~al.}(2013){Bennett}, {Larson}, {Weiland}, {Jarosik}, {Hinshaw}, {Odegard}, {Smith}, {Hill}, {Gold}, {Halpern}, {Komatsu}, {Nolta}, {Page}, {Spergel}, {Wollack}, {Dunkley}, {Kogut}, {Limon}, {Meyer}, {Tucker}, \& {Wright}}]{WMAP2013}
{Bennett}, C.~L., {Larson}, D., {Weiland}, J.~L., {et~al.} 2013, \apjs, 208, 20

\bibitem[{{Bernardeau} {et~al.}(2014){Bernardeau}, {Nishimichi}, \& {Taruya}}]{BNT}
{Bernardeau}, F., {Nishimichi}, T., \& {Taruya}, A. 2014, \mnras, 445, 1526

\bibitem[{{Bertschinger} \& {Zukin}(2008)}]{Bertschinger2008zb}
{Bertschinger}, E. \& {Zukin}, P. 2008, \prd, 78, 024015

\bibitem[{{Blazek} {et~al.}(2019){Blazek}, {MacCrann}, {Troxel}, \& {Fang}}]{Blazek2019}
{Blazek}, J.~A., {MacCrann}, N., {Troxel}, M.~A., \& {Fang}, X. 2019, \prd, 100, 103506

\bibitem[{{Booth} \& {Schaye}(2009)}]{BS2009}
{Booth}, C.~M. \& {Schaye}, J. 2009, \mnras, 398, 53

\bibitem[{{Bridle} \& {King}(2007)}]{BK2007}
{Bridle}, S. \& {King}, L. 2007, New Journal of Physics, 9, 444

\bibitem[{{Brown} {et~al.}(2005){Brown}, {Castro}, \& {Taylor}}]{BCT05}
{Brown}, M.~L., {Castro}, P.~G., \& {Taylor}, A.~N. 2005, \mnras, 360, 1262

\bibitem[{{Carroll}(2001)}]{Carroll2001}
{Carroll}, S.~M. 2001, Living Reviews in Relativity, 4, 1

\bibitem[{{Casas} {et~al.}(2023){Casas}, {Cardone}, {Sapone}, {et~al.}}]{Casas23a}
{Casas}, S., {Cardone}, V.~F., {Sapone}, D., {et~al.} 2023, \aap, submitted, arXiv:2306.11053

\bibitem[{{Challinor} \& {Lewis}(2011)}]{2011PhRvD..84d3516C}
{Challinor}, A. \& {Lewis}, A. 2011, \prd, 84, 043516

\bibitem[{{Chevallier} \& {Polarski}(2001)}]{CP2001}
{Chevallier}, M. \& {Polarski}, D. 2001, International Journal of Modern Physics D, 10, 213

\bibitem[{{Chisari} {et~al.}(2019){Chisari}, {Alonso}, {Krause}, {Leonard}, {Bull}, {Neveu}, {Villarreal}, {Singh}, {McClintock}, {Ellison}, {Du}, {Zuntz}, {Mead}, {Joudaki}, {Lorenz}, {Tr{\"o}ster}, {Sanchez}, {Lanusse}, {Ishak}, {Hlozek}, {Blazek}, {Campagne}, {Almoubayyed}, {Eifler}, {Kirby}, {Kirkby}, {Plaszczynski}, {Slosar}, {Vrastil}, {Wagoner}, \& {LSST Dark Energy Science Collaboration}}]{pyccl}
{Chisari}, N.~E., {Alonso}, D., {Krause}, E., {et~al.} 2019, \apjs, 242, 2

\bibitem[{{Clifton} {et~al.}(2012){Clifton}, {Ferreira}, {Padilla}, \& {Skordis}}]{Clifton2012}
{Clifton}, T., {Ferreira}, P.~G., {Padilla}, A., \& {Skordis}, C. 2012, \physrep, 513, 1

\bibitem[{{Cooray} \& {Sheth}(2002)}]{CS02}
{Cooray}, A. \& {Sheth}, R. 2002, \physrep, 372, 1

\bibitem[{Copeland {et~al.}(2006)Copeland, Sami, \& Tsujikawa}]{Copeland2006}
Copeland, E.~J., Sami, M., \& Tsujikawa, S. 2006, Int. J. Mod. Phys. D, 15, 1753

\bibitem[{{Cragg} {et~al.}(2023){Cragg}, {Duncan}, {Miller}, \& {Alonso}}]{Cragg2023}
{Cragg}, C., {Duncan}, C. A.~J., {Miller}, L., \& {Alonso}, D. 2023, \mnras, 518, 4909

\bibitem[{Crocce \& Scoccimarro(2008)}]{CroSco2008}
Crocce, M. \& Scoccimarro, R. 2008, Phys. Rev. D, 77, 023533

\bibitem[{{d'Amico} {et~al.}(2020){d'Amico}, {Gleyzes}, {Kokron}, {Markovic}, {Senatore}, {Zhang}, {Beutler}, \& {Gil-Mar{\'\i}n}}]{DAmico_etal2020}
{d'Amico}, G., {Gleyzes}, J., {Kokron}, N., {et~al.} 2020, JCAP, 05, 005

\bibitem[{{Daniel} {et~al.}(2008){Daniel}, {Caldwell}, {Cooray}, \& {Melchiorri}}]{Daniel2008et}
{Daniel}, S.~F., {Caldwell}, R.~R., {Cooray}, A., \& {Melchiorri}, A. 2008, \prd, 77, 103513

\bibitem[{{d'Assignies Doumerg} {et~al.}(2025){d'Assignies Doumerg}, {Manera}, {Padilla}, {Ilbert}, {Hildebrandt}, {Reynolds}, {Chaves-Montero}, {Wright}, {Tallada-Cresp{\'\i}}, {Eriksen}, {Carretero}, {Roster}, {Kang}, {Naidoo}, {Miquel}, {Altieri}, {Amara}, {Andreon}, {Auricchio}, {Baccigalupi}, {Bagot}, {Baldi}, {Balestra}, {Bardelli}, {Battaglia}, {Biviano}, {Branchini}, {Brescia}, {Camera}, {Capobianco}, {Carbone}, {Cardone}, {Casas}, {Castander}, {Castellano}, {Castignani}, {Cavuoti}, {Chambers}, {Cimatti}, {Colodro-Conde}, {Congedo}, {Conselice}, {Conversi}, {Copin}, {Courbin}, {Courtois}, {Crocce}, {Da Silva}, {Degaudenzi}, {de la Torre}, {De Lucia}, {Douspis}, {Dupac}, {Ealet}, {Escoffier}, {Farina}, {Faustini}, {Ferriol}, {Finelli}, {Fosalba}, {Fotopoulou}, {Frailis}, {Franceschi}, {Fumana}, {Galeotta}, {George}, {Gillis}, {Giocoli}, {G{\'o}mez-Alvarez}, {Gracia-Carpio}, {Grazian}, {Grupp}, {Holmes}, {Hook}, {Hornstrup}, {Jahnke}, {Jhabvala}, {Joachimi}, {Keih{\"a}nen}, {Kermiche}, {Kiessling},
  {Kubik}, {K{\"u}mmel}, {Kunz}, {Kurki-Suonio}, {Lahav}, {Le Brun}, {Ligori}, {Lilje}, {Lindholm}, {Lloro}, {Mainetti}, {Maino}, {Maiorano}, {Mansutti}, {Marcin}, {Marggraf}, {Markovic}, {Martinelli}, {Martinet}, {Marulli}, {Massey}, {Masters}, {Medinaceli}, {Mei}, {Melchior}, {Mellier}, {Meneghetti}, {Merlin}, {Meylan}, {Mora}, {Moresco}, {Moscardini}, {Neissner}, {Niemi}, {Paltani}, {Pasian}, {Pedersen}, {Pettorino}, {Pires}, {Polenta}, {Poncet}, {Popa}, {Pozzetti}, {Raison}, {Rebolo}, {Renzi}, {Rhodes}, {Riccio}, {Romelli}, {Roncarelli}, {Rossetti}, {Saglia}, {Sakr}, {Sapone}, {Sartoris}, {Schewtschenko}, {Schneider}, {Schrabback}, {Secroun}, {Sefusatti}, {Seidel}, {Seiffert}, {Serrano}, {Simon}, {Sirignano}, {Sirri}, {Spurio Mancini}, {Stanco}, {Steinwagner}, {Tavagnacco}, {Taylor}, {Teplitz}, {Tereno}, {Tessore}, {Toft}, {Toledo-Moreo}, {Torradeflot}, {Tsyganov}, {Tutusaus}, {Valenziano}, {Valiviita}, {Vassallo}, {Verdoes Kleijn}, {Wang}, {Weller}, {Zamorani}, {Zucca}, {Bolzonella}, {Burigana},
  {Gabarra}, {Mart{\'\i}n-Fleitas}, {Risso}, {Scottez}, \& {Viel}}]{dAssignies25}
{d'Assignies Doumerg}, W., {Manera}, M., {Padilla}, C., {et~al.} 2025, arXiv e-prints, arXiv:2505.10416

\bibitem[{{de Bernardis} {et~al.}(2000){de Bernardis}, {Ade}, {Bock}, {Bond}, {Borrill}, {Boscaleri}, {Coble}, {Crill}, {De Gasperis}, {Farese}, {Ferreira}, {Ganga}, {Giacometti}, {Hivon}, {Hristov}, {Iacoangeli}, {Jaffe}, {Lange}, {Martinis}, {Masi}, {Mason}, {Mauskopf}, {Melchiorri}, {Miglio}, {Montroy}, {Netterfield}, {Pascale}, {Piacentini}, {Pogosyan}, {Prunet}, {Rao}, {Romeo}, {Ruhl}, {Scaramuzzi}, {Sforna}, \& {Vittorio}}]{debernardis2000}
{de Bernardis}, P., {Ade}, P.~A.~R., {Bock}, J.~J., {et~al.} 2000, \nat, 404, 955

\bibitem[{{DESI Collaboration} {et~al.}(2025){DESI Collaboration}, {Abdul-Karim}, {Aguilar}, {Ahlen}, {Alam}, {Allen}, {Allende Prieto}, {Alves}, {Anand}, {Andrade}, {Armengaud}, {Aviles}, {Bailey}, {Baltay}, {Bansal}, {Bault}, {Behera}, {BenZvi}, {Bianchi}, {Blake}, {Brieden}, {Brodzeller}, {Brooks}, {Buckley-Geer}, {Burtin}, {Calderon}, {Canning}, {Carnero Rosell}, {Carrilho}, {Casas}, {Castander}, {Cereskaite}, {Charles}, {Chaussidon}, {Chaves-Montero}, {Chebat}, {Chen}, {Claybaugh}, {Cole}, {Cooper}, {Cuceu}, {Dawson}, {de la Macorra}, {de Mattia}, {Deiosso}, {Della Costa}, {Demina}, {Dey}, {Dey}, {Ding}, {Doel}, {Edelstein}, {Eisenstein}, {Elbers}, {Fagrelius}, {Fanning}, {Fernandez-Garcia}, {Ferraro}, {Font-Ribera}, {Forero-Romero}, {Frenk}, {Garcia-Quintero}, {Garrison}, {Gazta$\tilde{n}$aga}, {Gil-Marin}, {Gontcho}, {Gonzalez}, {Gonzalez-Morales}, {Gordon}, {Green}, {Gutierrez}, {Guy}, {Hadzhiyska}, {Hahn}, {He}, {Herbold}, {Herrera-Alcantar}, {Ho}, {Honscheid}, {Howlett}, {Huterer}, {Ishak},
  {Juneau}, {Kamble}, {Karacayli}, {Kehoe}, {Kent}, {Kim}, {Kirkby}, {Kisner}, {Koposov}, {Kremin}, {Krolewski}, {Lahav}, {Lamman}, {Landriau}, {Lang}, {Lasker}, {Le Goff}, {Le Guillou}, {Leauthaud}, {Levi}, {Li}, {Li}, {Lodha}, {Lokken}, {Lozano-Rodriguez}, {Magneville}, {Manera}, {Martini}, {Matthewson}, {Meisner}, {Mena-Fernandez}, {Menegas}, {Mergulhao}, {Miquel}, {Moustakas}, {Munoz-Gutierrez}, {Munoz-Santos}, {Myers}, {Nadathur}, {Naidoo}, {Napolitano}, {Newman}, {Niz}, {Noriega}, {Paillas}, {Palanque-Delabrouille}, {Pan}, {Peacock}, {Pellejero Ibanez}, {Percival}, {Perez-Fernandez}, {Perez-Rafols}, {Pieri}, {Poppett}, {Prada}, {Rabinowitz}, {Raichoor}, {Ramirez-Perez}, {Rashkovetskyi}, {Ravoux}, {Rich}, {Rocher}, {Rockosi}, {Rohlf}, {Roman-Herrera}, {Ross}, {Rossi}, {Ruggeri}, {Ruhlmann-Kleider}, {Samushia}, {Sanchez}, {Sanders}, {Schlegel}, {Schubnell}, {Seo}, {Shafieloo}, {Sharples}, {Silber}, {Sinigaglia}, {Sprayberry}, {Tan}, {Tarle}, {Taylor}, {Turner}, {Ure$\tilde{n}$a-Lopez}, {Vaisakh},
  {Valdes}, {Valogiannis}, {Vargas-Maga$\tilde{n}$a}, {Verde}, {Walther}, {Weaver}, {Weinberg}, {White}, {Wolfson}, {Y'eche}, {Yu}, {Zaborowski}, {Zarrouk}, {Zhai}, {Zhang}, {Zhao}, {Zhao}, {Zhou}, \& {Zou}}]{DESI2025}
{DESI Collaboration}, {Abdul-Karim}, M., {Aguilar}, J., {et~al.} 2025, arXiv e-prints, arXiv:2503.14738

\bibitem[{{DESI Collaboration} {et~al.}(2024){DESI Collaboration}, {Adame}, {Aguilar}, {Ahlen}, {Alam}, {Alexander}, {Alvarez}, {Alves}, {Anand}, {Andrade}, {Armengaud}, {Avila}, {Aviles}, {Awan}, {Bahr-Kalus}, {Bailey}, {Baltay}, {Bault}, {Behera}, {BenZvi}, {Bera}, {Beutler}, {Bianchi}, {Blake}, {Blum}, {Brieden}, {Brodzeller}, {Brooks}, {Buckley-Geer}, {Burtin}, {Calderon}, {Canning}, {Carnero Rosell}, {Cereskaite}, {Cervantes-Cota}, {Chabanier}, {Chaussidon}, {Chaves-Montero}, {Chen}, {Chen}, {Claybaugh}, {Cole}, {Cuceu}, {Davis}, {Dawson}, {de la Macorra}, {de Mattia}, {Deiosso}, {Dey}, {Dey}, {Ding}, {Doel}, {Edelstein}, {Eftekharzadeh}, {Eisenstein}, {Elliott}, {Fagrelius}, {Fanning}, {Ferraro}, {Ereza}, {Findlay}, {Flaugher}, {Font-Ribera}, {Forero-S{\'a}nchez}, {Forero-Romero}, {Frenk}, {Garcia-Quintero}, {Gazta{\~n}aga}, {Gil-Mar{\'\i}n}, {Gontcho}, {Gonzalez-Morales}, {Gonzalez-Perez}, {Gordon}, {Green}, {Gruen}, {Gsponer}, {Gutierrez}, {Guy}, {Hadzhiyska}, {Hahn}, {Hanif}, {Herrera-Alcantar},
  {Honscheid}, {Howlett}, {Huterer}, {Ir{\v{s}}i{\v{c}}}, {Ishak}, {Juneau}, {Kara{\c{c}}ayl{\i}}, {Kehoe}, {Kent}, {Kirkby}, {Kremin}, {Krolewski}, {Lai}, {Lan}, {Landriau}, {Lang}, {Lasker}, {Le Goff}, {Le Guillou}, {Leauthaud}, {Levi}, {Li}, {Linder}, {Lodha}, {Magneville}, {Manera}, {Margala}, {Martini}, {Maus}, {McDonald}, {Medina-Varela}, {Meisner}, {Mena-Fern{\'a}ndez}, {Miquel}, {Moon}, {Moore}, {Moustakas}, {Mudur}, {Mueller}, {Mu{\~n}oz-Guti{\'e}rrez}, {Myers}, {Nadathur}, {Napolitano}, {Neveux}, {Newman}, {Nguyen}, {Nie}, {Niz}, {Noriega}, {Padmanabhan}, {Paillas}, {Palanque-Delabrouille}, {Pan}, {Penmetsa}, {Percival}, {Pieri}, {Pinon}, {Poppett}, {Porredon}, {Prada}, {P{\'e}rez-Fern{\'a}ndez}, {P{\'e}rez-R{\`a}fols}, {Rabinowitz}, {Raichoor}, {Ram{\'\i}rez-P{\'e}rez}, {Ramirez-Solano}, {Ravoux}, {Rashkovetskyi}, {Rezaie}, {Rich}, {Rocher}, {Rockosi}, {Roe}, {Rosado-Marin}, {Ross}, {Rossi}, {Ruggeri}, {Ruhlmann-Kleider}, {Samushia}, {Sanchez}, {Saulder}, {Schlafly}, {Schlegel}, {Schubnell}, {Seo},
  {Shafieloo}, {Sharples}, {Silber}, {Slosar}, {Smith}, {Sprayberry}, {Tan}, {Tarl{\'e}}, {Taylor}, {Trusov}, {Ure{\~n}a-L{\'o}pez}, {Vaisakh}, {Valcin}, {Valdes}, {Vargas-Maga{\~n}a}, {Verde}, {Walther}, {Wang}, {Wang}, {Weaver}, {Weaverdyck}, {Wechsler}, {Weinberg}, {White}, {Yu}, {Yu}, {Yuan}, {Y{\`e}che}, {Zaborowski}, {Zarrouk}, {Zhang}, {Zhao}, \& {Zhao}}]{desi2024}
{DESI Collaboration}, {Adame}, A.~G., {Aguilar}, J., {et~al.} 2024, arXiv e-prints, arXiv:2404.03002

\bibitem[{{Di Valentino} {et~al.}(2021){Di Valentino}, {Mena}, {Pan}, {Visinelli}, {Yang}, {Melchiorri}, {Mota}, {Riess}, \& {Silk}}]{DiVa2021}
{Di Valentino}, E., {Mena}, O., {Pan}, S., {et~al.} 2021, Classical and Quantum Gravity, 38, 153001

\bibitem[{{Di Valentino}(2022)}]{DiVa2022}
{Di Valentino}, E.~D. 2022, Universe, 8, 399

\bibitem[{{Duncan} {et~al.}(2014){Duncan}, {Joachimi}, {Heavens}, {Heymans}, \& {Hildebrandt}}]{Duncan2014}
{Duncan}, C. A.~J., {Joachimi}, B., {Heavens}, A.~F., {Heymans}, C., \& {Hildebrandt}, H. 2014, \mnras, 437, 2471

\bibitem[{Eisenstein {et~al.}(2007)Eisenstein, Seo, \& White}]{EisSeoWhi2007}
Eisenstein, D.~J., Seo, H.-J., \& White, M. 2007, \apj, 664, 660

\bibitem[{{Eisenstein} {et~al.}(2005){Eisenstein}, {Zehavi}, {Hogg}, {Scoccimarro}, {Blanton}, {Nichol}, {Scranton}, {Seo}, {Tegmark}, {Zheng}, {Anderson}, {Annis}, {Bahcall}, {Brinkmann}, {Burles}, {Castander}, {Connolly}, {Csabai}, {Doi}, {Fukugita}, {Frieman}, {Glazebrook}, {Gunn}, {Hendry}, {Hennessy}, {Ivezi{\'c}}, {Kent}, {Knapp}, {Lin}, {Loh}, {Lupton}, {Margon}, {McKay}, {Meiksin}, {Munn}, {Pope}, {Richmond}, {Schlegel}, {Schneider}, {Shimasaku}, {Stoughton}, {Strauss}, {SubbaRao}, {Szalay}, {Szapudi}, {Tucker}, {Yanny}, \& {York}}]{Eisenstein2005}
{Eisenstein}, D.~J., {Zehavi}, I., {Hogg}, D.~W., {et~al.} 2005, \apj, 633, 560

\bibitem[{{Euclid Collaboration: Blanchard} {et~al.}(2020){Euclid Collaboration: Blanchard}, {Camera}, {Carbone}, {et~al.}}]{Blanchard-EP7}
{Euclid Collaboration: Blanchard}, A., {Camera}, S., {Carbone}, C., {et~al.} 2020, \aap, 642, A191

\bibitem[{{Euclid Collaboration: Blot} {et~al.}(2025){Euclid Collaboration: Blot}, {Tanidis}, {Ca$\tilde{\rm n}$as-Herrera}, {Carrilho}, {Bonici}, {Camera}, {Cardone}, {Casas}, {Davini}, {Di Domizio}, {Farrens}, {Goh}, {Gouyou Beauchamps}, {Ilic}, {Joudaki}, {Keil}, {Le Brun}, {Martinelli}, {Moretti}, {Pettorino}, {Pezzotta}, {Sakr}, {Sanchez}, {Sciotti}, {Tutusaus}, {Ajani}, {Crocce}, {Deshpande}, {Fumagalli}, {Giocoli}, {Legrand}, {Lembo}, {Lesci}, {Navarro-Giron\'{e}s}, {Nouri-Zonoz}, {Pamuk}, {Pourtsidou}, {Tsedrik}, {Bel}, {Carbone}, {D'Amico}, {Duncan}, {Kilbinger}, {Kitching}, {Sapone}, {Sellentin}, {Taylor}, {et~al.}}]{ISTL-P4}
{Euclid Collaboration: Blot}, L., {Tanidis}, K., {Ca$\tilde{\rm n}$as-Herrera}, G., {et~al.} 2025, A\&A submitted

\bibitem[{{Euclid Collaboration: Carrilho} {et~al.}(2025)}]{ISTNL-P2}
{Euclid Collaboration: Carrilho}, P. {et~al.} 2025, in preparation

\bibitem[{{Euclid Collaboration: Castander} {et~al.}(2025){Euclid Collaboration: Castander}, {Fosalba}, {Stadel}, {et~al.}}]{EuclidSkyFlagship}
{Euclid Collaboration: Castander}, F., {Fosalba}, P., {Stadel}, J., {et~al.} 2025, A\&A, 697, A5

\bibitem[{{Euclid Collaboration: Ca$\tilde{{\rm n}}$as-Herrera} {et~al.}(2025){Euclid Collaboration: Ca$\tilde{{\rm n}}$as-Herrera}, {Goh}, {Blot}, {Bonici}, {Camera}, {Cardone}, {Carrilho}, {Casas}, {Davini}, {Di Domizio}, {Farrens}, {Gouyou Beauchamps}, {Ilic}, {Joudaki}, {Keil}, {Le Brun}, {Martinelli}, {Moretti}, {Pettorino}, {Pezzotta}, {Sakr}, {Sanchez}, {Sciotti}, {Tanidis}, {Tutusaus}, {Ajani}, {Alvi}, {Crocce}, {Deshpande}, {Fumagalli}, {Giocoli}, {Ferrari}, {Kou}, {Legrand}, {Lembo}, {Lesci}, {Navarro-Giron\'{e}s}, {Nouri-Zonoz}, {Pagano}, {Pamuk}, {Pourtsidou}, {Tsedrik}, {Arcari}, {Bel}, {Carbone}, {Claramunt Gonzalez}, {D'Amico}, {Duncan}, {Kilbinger}, {Kitching}, {Lacasa}, {Porredon}, {Sapone}, {Sellentin}, {Taylor}, {Tessore}, {et~al.}}]{ISTL-P3}
{Euclid Collaboration: Ca$\tilde{{\rm n}}$as-Herrera}, G., {Goh}, L.~W.~K., {Blot}, L., {et~al.} 2025, A\&A submitted

\bibitem[{{Euclid Collaboration: Crocce} {et~al.}(2025)}]{ISTNL-P1}
{Euclid Collaboration: Crocce}, M. {et~al.} 2025, in preparation

\bibitem[{{Euclid Collaboration: Cropper} {et~al.}(2025){Euclid Collaboration: Cropper}, {Al-Bahlawan}, {Amiaux}, {et~al.}}]{EuclidSkyVIS}
{Euclid Collaboration: Cropper}, M., {Al-Bahlawan}, A., {Amiaux}, J., {et~al.} 2025, A\&A, 697, A2

\bibitem[{{Euclid Collaboration: Deshpande} {et~al.}(2024){Euclid Collaboration: Deshpande}, {Kitching}, {Hall}, {et~al.}}]{Deshpande-EP28}
{Euclid Collaboration: Deshpande}, A.~C., {Kitching}, T., {Hall}, A., {et~al.} 2024, \aap, 684, A138

\bibitem[{{Euclid Collaboration: Desprez} {et~al.}(2020){Euclid Collaboration: Desprez}, {Paltani}, {Coupon}, {et~al.}}]{Desprez-EP10}
{Euclid Collaboration: Desprez}, G., {Paltani}, S., {Coupon}, J., {et~al.} 2020, \aap, 644, A31

\bibitem[{{Euclid Collaboration: Goh} {et~al.}(2025){Euclid Collaboration: Goh}, {Nouri-Zonoz}, {Pamuk}, {Ballardini}, {Bose}, {Ca$\tilde{n}$as-Herrera}, {Casas}, {Franco Abell\'an}, {Il\'ic}, {Keil}, {Kunz}, {Le Brun}, {Lepori}, {Martinelli}, {Sakr}, {Sorrenti}, {Teixeira}, {Tutusaus}, {Blot}, {Bonici}, {Bonvin}, {Camera}, {Cardone}, {Carrilho}, {Di Domizio}, {Durrer}, {Farrens}, {Gouyou Beauchamps}, {Joudaki}, {Moretti}, {Pezzotta}, {Sanchez}, {Sciotti}, {Tanidis}, {et~al.}}]{ISTL-P6}
{Euclid Collaboration: Goh}, L.~W.~K., {Nouri-Zonoz}, A., {Pamuk}, S., {et~al.} 2025, A\&A submitted

\bibitem[{{Euclid Collaboration: Jahnke} {et~al.}(2025){Euclid Collaboration: Jahnke}, {Gillard}, {Schirmer}, {et~al.}}]{EuclidSkyNISP}
{Euclid Collaboration: Jahnke}, K., {Gillard}, W., {Schirmer}, M., {et~al.} 2025, A\&A, 697, A3

\bibitem[{{Euclid Collaboration: Joudaki} {et~al.}(2025){Euclid Collaboration: Joudaki}, {Pettorino}, {Blot}, {Bonici}, {Camera}, {Ca\~nas-Herrera}, {Cardone}, {Carrilho}, {Casas}, {Davini}, {Di~Domizio}, {Farrens}, {Goh}, {Gouyou~Beauchamps}, {Ili\'c}, {Keil}, {Le~Brun}, {Martinelli}, {Moretti}, {Pezzotta}, {Sakr}, {S\'anchez}, {Sciotti}, {Tanidis}, {Tutusaus}, {Ajani}, {Alvi}, {Crocce}, {Fumagalli}, {Giocoli}, {Ferrari}, {Kou}, {Legrand}, {Lembo}, {Lesci}, {Navarro~Girones}, {Nouri-Zonoz}, {Pamuk}, {Pagano}, {Tsedrik}, {Arcari}, {Artis}, {Ballardini}, {Bel}, {Carbone}, {Costanzi}, {De~Caro}, {Duncan}, {Fabbian}, {Kilbinger}, {Kitching}, {Lacasa}, {Lattanzi}, {Olivares-Miranda}, {Salvati}, {Sapone}, {Sartoris}, {Sellentin}, {Taylor}, {et~al.}}]{ISTL-P2}
{Euclid Collaboration: Joudaki}, S., {Pettorino}, V., {Blot}, L., {et~al.} 2025, A\&A submitted

\bibitem[{{Euclid Collaboration: Knabenhans} {et~al.}(2021){Euclid Collaboration: Knabenhans}, {Stadel}, {Potter}, {et~al.}}]{Knabenhans-EP9}
{Euclid Collaboration: Knabenhans}, M., {Stadel}, J., {Potter}, D., {et~al.} 2021, \mnras, 505, 2840

\bibitem[{{Euclid Collaboration: Martinelli} {et~al.}(2025){Euclid Collaboration: Martinelli}, {Pezzotta}, {Sciotti}, {Bonici}, {Camera}, {Ca\~nas-Herrera}, {Cardone}, {Carrilho}, {Casa}, {Davini}, {Di Domizio}, {Farrens}, {Goh}, {Gouyou~Beauchamps}, {Joudaki}, {Keil}, {Le~Brun}, {Moretti}, {Pettorino}, {Sakr}, {Tanidis}, {Tutusaus}, {Ajani}, {Giocoli}, {Legrand}, {Lembo}, {Lesci}, {Navarro~Girones}, {Nouri-Zonoz}, {Pamuk}, {Tsedrik}, {Bel}, {Carbone}, {Duncan}, {Kilbinger}, {Lacasa}, {Lattanzi}, {Sapone}, {Sellentin}, {et~al.}}]{ISTL-P5}
{Euclid Collaboration: Martinelli}, M., {Pezzotta}, A., {Sciotti}, D., {et~al.} 2025, A\&A submitted

\bibitem[{{Euclid Collaboration: Mellier} {et~al.}(2025){Euclid Collaboration: Mellier}, {Abdurro'uf}, {Acevedo~Barroso}, {et~al.}}]{EuclidSkyOverview}
{Euclid Collaboration: Mellier}, Y., {Abdurro'uf}, {Acevedo~Barroso}, J., {et~al.} 2025, A\&A, 697, A1

\bibitem[{{Euclid Collaboration: Monaco} {et~al.}(2025)}]{PiGiCov}
{Euclid Collaboration: Monaco}, P. {et~al.} 2025, in preparation

\bibitem[{{Euclid Collaboration: Moretti} {et~al.}(2025)}]{ISTNL-P3}
{Euclid Collaboration: Moretti}, C. {et~al.} 2025, in preparation

\bibitem[{{Euclid Collaboration: Paltani} {et~al.}(2024){Euclid Collaboration: Paltani}, {Coupon}, {Hartley}, {et~al.}}]{EP-Paltani}
{Euclid Collaboration: Paltani}, S., {Coupon}, J., {Hartley}, W.~G., {et~al.} 2024, \aap, 681, A66

\bibitem[{{Euclid Collaboration: Pezzotta} {et~al.}(2024){Euclid Collaboration: Pezzotta}, {Moretti}, {Zennaro}, {et~al.}}]{EP-Pezzotta}
{Euclid Collaboration: Pezzotta}, A., {Moretti}, C., {Zennaro}, M., {et~al.} 2024, \aap, 687, A216

\bibitem[{{Euclid Collaboration: Sciotti} {et~al.}(2025)}]{ISTNL-P4}
{Euclid Collaboration: Sciotti}, D. {et~al.} 2025, in preparation

\bibitem[{{Euclid Collaboration: Tanidis} {et~al.}(2024){Euclid Collaboration: Tanidis}, {Cardone}, {Martinelli}, {et~al.}}]{Tanidis-TBD}
{Euclid Collaboration: Tanidis}, K., {Cardone}, V.~F., {Martinelli}, M., {et~al.} 2024, \aap, 683, A17

\bibitem[{{Euclid Collaboration: Tessore} {et~al.}(2024){Euclid Collaboration: Tessore}, {Joachimi}, {Loureiro}, {Hall}, {Ca{\~n}as-Herrera}, {Tutusaus}, {Jeffrey}, {Naidoo}, {McEwen}, {Amara}, {Andreon}, {Auricchio}, {Baccigalupi}, {Baldi}, {Bardelli}, {Bernardeau}, {Bonino}, {Branchini}, {Brescia}, {Brinchmann}, {Caillat}, {Camera}, {Capobianco}, {Carbone}, {Cardone}, {Carretero}, {Casas}, {Castellano}, {Castignani}, {Cavuoti}, {Cimatti}, {Colodro-Conde}, {Congedo}, {Conselice}, {Conversi}, {Copin}, {Courbin}, {Courtois}, {Cropper}, {Da Silva}, {Degaudenzi}, {De Lucia}, {Dinis}, {Dubath}, {Duncan}, {Dupac}, {Dusini}, {Farina}, {Farrens}, {Faustini}, {Ferriol}, {Frailis}, {Franceschi}, {Fumana}, {Galeotta}, {Gillard}, {Gillis}, {Giocoli}, {G{\'o}mez-Alvarez}, {Grazian}, {Grupp}, {Guzzo}, {Haugan}, {Hoekstra}, {Holmes}, {Hormuth}, {Hornstrup}, {Hudelot}, {Jahnke}, {Jhabvala}, {Keih{\"a}nen}, {Kermiche}, {Kiessling}, {Kubik}, {K{\"u}mmel}, {Kunz}, {Kurki-Suonio}, {Ligori}, {Lilje}, {Lindholm}, {Lloro},
  {Mainetti}, {Maiorano}, {Mansutti}, {Marggraf}, {Martinelli}, {Martinet}, {Marulli}, {Massey}, {Medinaceli}, {Mei}, {Melchior}, {Mellier}, {Meneghetti}, {Merlin}, {Meylan}, {Mohr}, {Moresco}, {Morin}, {Moscardini}, {Munari}, {Nakajima}, {Niemi}, {Padilla}, {Paltani}, {Pasian}, {Pedersen}, {Percival}, {Pettorino}, {Pires}, {Polenta}, {Poncet}, {Popa}, {Raison}, {Renzi}, {Rhodes}, {Riccio}, {Romelli}, {Roncarelli}, {Rossetti}, {Saglia}, {Sakr}, {S{\'a}nchez}, {Sapone}, {Sartoris}, {Schirmer}, {Schneider}, {Schrabback}, {Secroun}, {Seidel}, {Seiffert}, {Serrano}, {Sirignano}, {Sirri}, {Stanco}, {Steinwagner}, {Tallada-Cresp{\'\i}}, {Taylor}, {Tereno}, {Toledo-Moreo}, {Torradeflot}, {Valenziano}, {Vassallo}, {Wang}, {Weller}, {Zamorani}, {Zucca}, {Biviano}, {Bolzonella}, {Boucaud}, {Bozzo}, {Burigana}, {Calabrese}, {Di Ferdinando}, {Escartin Vigo}, {Finelli}, {Gracia-Carpio}, {Matthew}, {Mauri}, {Pezzotta}, {P{\"o}ntinen}, {Scottez}, {Spurio Mancini}, {Tenti}, {Viel}, {Wiesmann}, {Akrami}, {Anselmi},
  {Archidiacono}, {Atrio-Barandela}, {Balaguera-Antolinez}, {Ballardini}, {Benielli}, {Blanchard}, {Blot}, {B{\"o}hringer}, {Borgani}, {Bruton}, {Cabanac}, {Calabro}, {Camacho Quevedo}, {Cappi}, {Caro}, {Carvalho}, {Castro}, {Chambers}, {Cooray}, {de la Torre}, {Desprez}, {D{\'\i}az-S{\'a}nchez}, {Di Domizio}, {Dole}, {Escoffier}, {Ferrari}, {Ferreira}, {Ferrero}, {Finoguenov}, {Fontana}, {Fornari}, {Gabarra}, {Ganga}, {Garc{\'\i}a-Bellido}, {Gasparetto}, {Gaztanaga}, {Giacomini}, {Gianotti}, {Gozaliasl}, {Gutierrez}, {Hartley}, {Hildebrandt}, {Hjorth}, {Jimenez Mu{\~n}oz}, {Joudaki}, {Kajava}, {Kansal}, {Karagiannis}, {Kirkpatrick}, {Kruk}, {Lacasa}, {Lattanzi}, {Le Brun}, {Le Graet}, {Legrand}, {Lesgourgues}, {Liaudat}, {Macias-Perez}, {Magliocchetti}, {Mannucci}, {Maoli}, {Mart{\'\i}n-Fleitas}, {Martins}, {Maurin}, {Metcalf}, {Miluzio}, {Monaco}, {Montoro}, {Moretti}, {Morgante}, {Murray}, {Nadathur}, {Walton}, {Patrizii}, {Popa}, {Potter}, {Reimberg}, {Risso}, {Rocci}, {Rollins}, {Sahl{\'e}n}, {Sarpa},
  {Schneider}, {Sereno}, {Simon}, {Tanidis}, {Tao}, {Testera}, {Teyssier}, {Toft}, {Tosi}, {Troja}, {Tucci}, {Valieri}, {Valiviita}, {Vergani}, {Verza}, {Vielzeuf}, {Brown}, \& {Sellentin}}]{Tessore2024}
{Euclid Collaboration: Tessore}, N., {Joachimi}, B., {Loureiro}, A., {et~al.} 2024, arXiv e-prints, arXiv:2408.16903

\bibitem[{{Fang} {et~al.}(2008){Fang}, {Hu}, \& {Lewis}}]{FangPPF}
{Fang}, W., {Hu}, W., \& {Lewis}, A. 2008, \prd, 78, 087303

\bibitem[{Fang {et~al.}(2020)Fang, Krause, Eifler, \& MacCrann}]{Fang2019}
Fang, X., Krause, E., Eifler, T., \& MacCrann, N. 2020, JCAP, 05, 010

\bibitem[{{Fisher} {et~al.}(1994){Fisher}, {Scharf}, \& {Lahav}}]{fisher94}
{Fisher}, K.~B., {Scharf}, C.~A., \& {Lahav}, O. 1994, \mnras, 266, 219

\bibitem[{Frieman {et~al.}(2008)Frieman, Turner, \& Huterer}]{Frieman2008}
Frieman, J., Turner, M., \& Huterer, D. 2008, ARA\&A, 46, 385

\bibitem[{{Frusciante} {et~al.}(2024){Frusciante}, {Pace}, {Cardone}, {et~al.}}]{Frusciante23}
{Frusciante}, N., {Pace}, F., {Cardone}, V.~F., {et~al.} 2024, \aap, 690, A133

\bibitem[{{Garcia-Quintero} {et~al.}(2019){Garcia-Quintero}, {Ishak}, {Fox}, \& {Dossett}}]{cgq19}
{Garcia-Quintero}, C., {Ishak}, M., {Fox}, L., \& {Dossett}, J. 2019, \prd, 100, 103530

\bibitem[{{Gong} {et~al.}(2009){Gong}, {Ishak}, \& {Wang}}]{Gong09}
{Gong}, Y., {Ishak}, M., \& {Wang}, A. 2009, \prd, 80, 023002

\bibitem[{Grieb {et~al.}(2016)Grieb, Sánchez, Salazar-Albornoz, \& Dalla Vecchia}]{GriSanSal2016}
Grieb, J.~N., Sánchez, A.~G., Salazar-Albornoz, S., \& Dalla Vecchia, C. 2016, \mnras, 457, 1577

\bibitem[{{Gu} {et~al.}(2024){Gu}, {van Waerbeke}, {Bernardeau}, \& {Dalal}}]{gu24}
{Gu}, S., {van Waerbeke}, L., {Bernardeau}, F., \& {Dalal}, R. 2024, arXiv e-prints, arXiv:2412.14704

\bibitem[{{Heavens} \& {Taylor}(1995)}]{HT95}
{Heavens}, A.~F. \& {Taylor}, A.~N. 1995, \mnras, 275, 483

\bibitem[{{Heymans} {et~al.}(2021){Heymans}, {Tr{\"o}ster}, {Asgari}, {Blake}, {Hildebrandt}, {Joachimi}, {Kuijken}, {Lin}, {S{\'a}nchez}, {van den Busch}, {Wright}, {Amon}, {Bilicki}, {de Jong}, {Crocce}, {Dvornik}, {Erben}, {Fortuna}, {Getman}, {Giblin}, {Glazebrook}, {Hoekstra}, {Joudaki}, {Kannawadi}, {K{\"o}hlinger}, {Lidman}, {Miller}, {Napolitano}, {Parkinson}, {Schneider}, {Shan}, {Valentijn}, {Verdoes Kleijn}, \& {Wolf}}]{Heymans2021}
{Heymans}, C., {Tr{\"o}ster}, T., {Asgari}, M., {et~al.} 2021, \aap, 646, A140

\bibitem[{{Hirata} \& {Seljak}(2004)}]{HS2004}
{Hirata}, C.~M. \& {Seljak}, U. 2004, \prd, 70, 063526

\bibitem[{{Hivon} {et~al.}(2002){Hivon}, {G{\'o}rski}, {Netterfield}, {Crill}, {Prunet}, \& {Hansen}}]{Hivon2002}
{Hivon}, E., {G{\'o}rski}, K.~M., {Netterfield}, C.~B., {et~al.} 2002, \apj, 567, 2

\bibitem[{{Hoffmann} {et~al.}(2022){Hoffmann}, {Secco}, {Blazek}, {Crocce}, {Tallada-Cresp{\'\i}}, {Samuroff}, {Prat}, {Carretero}, {Fosalba}, {Gazta{\~n}aga}, {Castander}, \& {DES Collaboration}}]{Hoffmann2022}
{Hoffmann}, K., {Secco}, L.~F., {Blazek}, J., {et~al.} 2022, \prd, 106, 123510

\bibitem[{{Hu}(2008)}]{HuPPF}
{Hu}, W. 2008, \prd, 77, 103524

\bibitem[{{Hu} \& {Sawicki}(2007)}]{Hu2007pj}
{Hu}, W. \& {Sawicki}, I. 2007, \prd, 76, 104043

\bibitem[{{Jain} \& {Zhang}(2008)}]{Jain2007yk}
{Jain}, B. \& {Zhang}, P. 2008, \prd, 78, 063503

\bibitem[{{Jarvis} {et~al.}(2004){Jarvis}, {Bernstein}, \& {Jain}}]{treecorr}
{Jarvis}, M., {Bernstein}, G., \& {Jain}, B. 2004, \mnras, 352, 338

\bibitem[{{Joachimi} {et~al.}(2015){Joachimi}, {Cacciato}, {Kitching}, {Leonard}, {Mandelbaum}, {Sch{\"a}fer}, {Sif{\'o}n}, {Hoekstra}, {Kiessling}, {Kirk}, \& {Rassat}}]{IArev1}
{Joachimi}, B., {Cacciato}, M., {Kitching}, T.~D., {et~al.} 2015, \ssr, 193, 1

\bibitem[{{Joachimi} {et~al.}(2021){Joachimi}, {Lin}, {Asgari}, {Tr{\"o}ster}, {Heymans}, {Hildebrandt}, {K{\"o}hlinger}, {S{\'a}nchez}, {Wright}, {Bilicki}, {Blake}, {van den Busch}, {Crocce}, {Dvornik}, {Erben}, {Getman}, {Giblin}, {Hoekstra}, {Kannawadi}, {Kuijken}, {Napolitano}, {Schneider}, {Scoccimarro}, {Sellentin}, {Shan}, {von Wietersheim-Kramsta}, \& {Zuntz}}]{Joachimi2021}
{Joachimi}, B., {Lin}, C.~A., {Asgari}, M., {et~al.} 2021, \aap, 646, A129

\bibitem[{{Joachimi} {et~al.}(2011){Joachimi}, {Mandelbaum}, {Abdalla}, \& {Bridle}}]{Joachimi2011}
{Joachimi}, B., {Mandelbaum}, R., {Abdalla}, F.~B., \& {Bridle}, S.~L. 2011, \aap, 527, A26

\bibitem[{{Johnston} {et~al.}(2024){Johnston}, {Chisari}, {Joudaki}, {Reischke}, {St{\"o}lzner}, {Loureiro}, {Mahony}, {Unruh}, {Wright}, {Asgari}, {Bilicki}, {Burger}, {Dvornik}, {Georgiou}, {Giblin}, {Heymans}, {Hildebrandt}, {Joachimi}, {Kuijken}, {Li}, {Linke}, {Porth}, {Shan}, {Tr{\"o}ster}, {van den Busch}, {von Wietersheim-Kramsta}, {Yan}, \& {Zhang}}]{Johnston24}
{Johnston}, H., {Chisari}, N.~E., {Joudaki}, S., {et~al.} 2024, arXiv e-prints, arXiv:2409.17377

\bibitem[{{Joudaki} {et~al.}(2017){Joudaki}, {Blake}, {Heymans}, {Choi}, {Harnois-Deraps}, {Hildebrandt}, {Joachimi}, {Johnson}, {Mead}, {Parkinson}, {Viola}, \& {van Waerbeke}}]{Joudaki17}
{Joudaki}, S., {Blake}, C., {Heymans}, C., {et~al.} 2017, \mnras, 465, 2033

\bibitem[{{Joudaki} {et~al.}(2018){Joudaki}, {Blake}, {Johnson}, {Amon}, {Asgari}, {Choi}, {Erben}, {Glazebrook}, {Harnois-D{\'e}raps}, {Heymans}, {Hildebrandt}, {Hoekstra}, {Klaes}, {Kuijken}, {Lidman}, {Mead}, {Miller}, {Parkinson}, {Poole}, {Schneider}, {Viola}, \& {Wolf}}]{Joudaki2018}
{Joudaki}, S., {Blake}, C., {Johnson}, A., {et~al.} 2018, \mnras, 474, 4894

\bibitem[{{Joudaki} {et~al.}(2022){Joudaki}, {Ferreira}, {Lima}, \& {Winther}}]{Joudaki22}
{Joudaki}, S., {Ferreira}, P.~G., {Lima}, N.~A., \& {Winther}, H.~A. 2022, \prd, 105, 043522

\bibitem[{{Joudaki} \& {Kaplinghat}(2012)}]{JK12}
{Joudaki}, S. \& {Kaplinghat}, M. 2012, \prd, 86, 023526

\bibitem[{{Kaiser}(1987)}]{Kaiser1987}
{Kaiser}, N. 1987, \mnras, 227, 1

\bibitem[{{Kaiser}(1992)}]{1992ApJ...388..272K}
{Kaiser}, N. 1992, \apj, 388, 272

\bibitem[{Kaiser {et~al.}(2000)Kaiser, Wilson, \& Luppino}]{Kaiser2000}
Kaiser, N., Wilson, G., \& Luppino, G.~A. 2000, astro-ph/0003338

\bibitem[{{Kiessling} {et~al.}(2015){Kiessling}, {Cacciato}, {Joachimi}, {Kirk}, {Kitching}, {Leonard}, {Mandelbaum}, {Sch{\"a}fer}, {Sif{\'o}n}, {Brown}, \& {Rassat}}]{IArev2}
{Kiessling}, A., {Cacciato}, M., {Joachimi}, B., {et~al.} 2015, \ssr, 193, 67

\bibitem[{{Kilbinger} {et~al.}(2017){Kilbinger}, {Heymans}, {Asgari}, {Joudaki}, {Schneider}, {Simon}, {Van Waerbeke}, {Harnois-D{\'e}raps}, {Hildebrandt}, {K{\"o}hlinger}, {Kuijken}, \& {Viola}}]{Kilbinger2017}
{Kilbinger}, M., {Heymans}, C., {Asgari}, M., {et~al.} 2017, \mnras, 472, 2126

\bibitem[{{Kirk} {et~al.}(2015){Kirk}, {Brown}, {Hoekstra}, {Joachimi}, {Kitching}, {Mandelbaum}, {Sif{\'o}n}, {Cacciato}, {Choi}, {Kiessling}, {Leonard}, {Rassat}, \& {Sch{\"a}fer}}]{IArev3}
{Kirk}, D., {Brown}, M.~L., {Hoekstra}, H., {et~al.} 2015, \ssr, 193, 139

\bibitem[{{Kitching} {et~al.}(2017){Kitching}, {Alsing}, {Heavens}, {Jimenez}, {McEwen}, \& {Verde}}]{Kitching2017}
{Kitching}, T.~D., {Alsing}, J., {Heavens}, A.~F., {et~al.} 2017, \mnras, 469, 2737

\bibitem[{{Kitching} \& {Deshpande}(2022)}]{TomAnu2022}
{Kitching}, T.~D. \& {Deshpande}, A.~C. 2022, The Open Journal of Astrophysics, 5, 6

\bibitem[{{Kitching} {et~al.}(2020){Kitching}, {Deshpande}, \& {Taylor}}]{Tom2020}
{Kitching}, T.~D., {Deshpande}, A.~C., \& {Taylor}, P.~L. 2020, The Open Journal of Astrophysics, 3, 14

\bibitem[{{Kitching} {et~al.}(2019){Kitching}, {Paykari}, {Hoekstra}, \& {Cropper}}]{Tom2019}
{Kitching}, T.~D., {Paykari}, P., {Hoekstra}, H., \& {Cropper}, M. 2019, The Open Journal of Astrophysics, 2, 5

\bibitem[{Kodama \& Sasaki(1984)}]{KS84}
Kodama, H. \& Sasaki, M. 1984, Prog. Theor. Phys. Suppl., 78, 1

\bibitem[{{K{\"o}hlinger} {et~al.}(2017){K{\"o}hlinger}, {Viola}, {Joachimi}, {Hoekstra}, {van Uitert}, {Hildebrandt}, {Choi}, {Erben}, {Heymans}, {Joudaki}, {Klaes}, {Kuijken}, {Merten}, {Miller}, {Schneider}, \& {Valentijn}}]{Kohl17}
{K{\"o}hlinger}, F., {Viola}, M., {Joachimi}, B., {et~al.} 2017, \mnras, 471, 4412

\bibitem[{{Komatsu} {et~al.}(2011){Komatsu}, {Smith}, {Dunkley}, {Bennett}, {Gold}, {Hinshaw}, {Jarosik}, {Larson}, {Nolta}, {Page}, {Spergel}, {Halpern}, {Hill}, {Kogut}, {Limon}, {Meyer}, {Odegard}, {Tucker}, {Weiland}, {Wollack}, \& {Wright}}]{Komatsu2011}
{Komatsu}, E., {Smith}, K.~M., {Dunkley}, J., {et~al.} 2011, \apjs, 192, 18

\bibitem[{{Koyama}(2016)}]{Kazuya2016}
{Koyama}, K. 2016, Reports on Progress in Physics, 79, 046902

\bibitem[{{Lamman} {et~al.}(2024){Lamman}, {Tsaprazi}, {Shi}, {{\v{S}}ar{\v{c}}evi{\'c}}, {Pyne}, {Legnani}, \& {Ferreira}}]{Lamman2024}
{Lamman}, C., {Tsaprazi}, E., {Shi}, J., {et~al.} 2024, The Open Journal of Astrophysics, 7, 14

\bibitem[{{Laureijs} {et~al.}(2011){Laureijs}, {Amiaux}, {Arduini}, {Augu{\`e}res}, {Brinchmann}, {Cole}, {Cropper}, {Dabin}, {Duvet}, {Ealet}, {Garilli}, {Gondoin}, {Guzzo}, {Hoar}, {Hoekstra}, {Holmes}, {Kitching}, {Maciaszek}, {Mellier}, {Pasian}, {Percival}, {Rhodes}, {Saavedra Criado}, {Sauvage}, {Scaramella}, {Valenziano}, {Warren}, {Bender}, {Castander}, {Cimatti}, {Le F{\`e}vre}, {Kurki-Suonio}, {Levi}, {Lilje}, {Meylan}, {Nichol}, {Pedersen}, {Popa}, {Rebolo Lopez}, {Rix}, {Rottgering}, {Zeilinger}, {Grupp}, {Hudelot}, {Massey}, {Meneghetti}, {Miller}, {Paltani}, {Paulin-Henriksson}, {Pires}, {Saxton}, {Schrabback}, {Seidel}, {Walsh}, {Aghanim}, {Amendola}, {Bartlett}, {Baccigalupi}, {Beaulieu}, {Benabed}, {Cuby}, {Elbaz}, {Fosalba}, {Gavazzi}, {Helmi}, {Hook}, {Irwin}, {Kneib}, {Kunz}, {Mannucci}, {Moscardini}, {Tao}, {Teyssier}, {Weller}, {Zamorani}, {Zapatero Osorio}, {Boulade}, {Foumond}, {Di Giorgio}, {Guttridge}, {James}, {Kemp}, {Martignac}, {Spencer}, {Walton}, {Bl{\"u}mchen}, {Bonoli},
  {Bortoletto}, {Cerna}, {Corcione}, {Fabron}, {Jahnke}, {Ligori}, {Madrid}, {Martin}, {Morgante}, {Pamplona}, {Prieto}, {Riva}, {Toledo}, {Trifoglio}, {Zerbi}, {Abdalla}, {Douspis}, {Grenet}, {Borgani}, {Bouwens}, {Courbin}, {Delouis}, {Dubath}, {Fontana}, {Frailis}, {Grazian}, {Koppenh{\"o}fer}, {Mansutti}, {Melchior}, {Mignoli}, {Mohr}, {Neissner}, {Noddle}, {Poncet}, {Scodeggio}, {Serrano}, {Shane}, {Starck}, {Surace}, {Taylor}, {Verdoes-Kleijn}, {Vuerli}, {Williams}, {Zacchei}, {Altieri}, {Escudero Sanz}, {Kohley}, {Oosterbroek}, {Astier}, {Bacon}, {Bardelli}, {Baugh}, {Bellagamba}, {Benoist}, {Bianchi}, {Biviano}, {Branchini}, {Carbone}, {Cardone}, {Clements}, {Colombi}, {Conselice}, {Cresci}, {Deacon}, {Dunlop}, {Fedeli}, {Fontanot}, {Franzetti}, {Giocoli}, {Garcia-Bellido}, {Gow}, {Heavens}, {Hewett}, {Heymans}, {Holland}, {Huang}, {Ilbert}, {Joachimi}, {Jennins}, {Kerins}, {Kiessling}, {Kirk}, {Kotak}, {Krause}, {Lahav}, {van Leeuwen}, {Lesgourgues}, {Lombardi}, {Magliocchetti}, {Maguire},
  {Majerotto}, {Maoli}, {Marulli}, {Maurogordato}, {McCracken}, {McLure}, {Melchiorri}, {Merson}, {Moresco}, {Nonino}, {Norberg}, {Peacock}, {Pello}, {Penny}, {Pettorino}, {Di Porto}, {Pozzetti}, {Quercellini}, {Radovich}, {Rassat}, {Roche}, {Ronayette}, {Rossetti}, {Sartoris}, {Schneider}, {Semboloni}, {Serjeant}, {Simpson}, {Skordis}, {Smadja}, {Smartt}, {Spano}, {Spiro}, {Sullivan}, {Tilquin}, {Trotta}, {Verde}, {Wang}, {Williger}, {Zhao}, {Zoubian}, \& {Zucca}}]{Laureijs11}
{Laureijs}, R., {Amiaux}, J., {Arduini}, S., {et~al.} 2011, ESA/SRE(2011)12, arXiv:1110.3193

\bibitem[{{Limber}(1953)}]{Limber1953}
{Limber}, D.~N. 1953, \apj, 117, 134

\bibitem[{{Linder}(2003)}]{Linder2003}
{Linder}, E.~V. 2003, \prl, 90, 091301

\bibitem[{{Linder}(2005)}]{Linder2005}
{Linder}, E.~V. 2005, \prd, 72, 043529

\bibitem[{{Linder} \& {Cahn}(2007)}]{LinderCahn2007}
{Linder}, E.~V. \& {Cahn}, R.~N. 2007, Astroparticle Physics, 28, 481

\bibitem[{{Linke} {et~al.}(2024){Linke}, {Unruh}, {Wittje}, {Schrabback}, {Grandis}, {Asgari}, {Dvornik}, {Hildebrandt}, {Hoekstra}, {Joachimi}, {Reischke}, {van den Busch}, {Wright}, {Schneider}, {Aghanim}, {Altieri}, {Amara}, {Andreon}, {Auricchio}, {Baccigalupi}, {Baldi}, {Bardelli}, {Bonino}, {Branchini}, {Brescia}, {Brinchmann}, {Camera}, {Capobianco}, {Carbone}, {Cardone}, {Carretero}, {Casas}, {Castander}, {Castellano}, {Cavuoti}, {Cimatti}, {Congedo}, {Conselice}, {Conversi}, {Copin}, {Courbin}, {Courtois}, {Da Silva}, {Degaudenzi}, {Dinis}, {Douspis}, {Dubath}, {Dupac}, {Dusini}, {Farina}, {Farrens}, {Ferriol}, {Fosalba}, {Frailis}, {Franceschi}, {Fumana}, {Galeotta}, {Gillis}, {Giocoli}, {Grazian}, {Grupp}, {Guzzo}, {Haugan}, {Holmes}, {Hook}, {Hormuth}, {Hornstrup}, {Hudelot}, {Jahnke}, {Keih{\"a}nen}, {Kermiche}, {Kiessling}, {Kilbinger}, {Kitching}, {Kubik}, {Kuijken}, {K{\"u}mmel}, {Kunz}, {Kurki-Suonio}, {Ligori}, {Lilje}, {Lindholm}, {Lloro}, {Maino}, {Maiorano}, {Mansutti}, {Marggraf},
  {Markovic}, {Martinet}, {Marulli}, {Massey}, {McCracken}, {Medinaceli}, {Mei}, {Mellier}, {Meneghetti}, {Merlin}, {Meylan}, {Moresco}, {Moscardini}, {Munari}, {Nakajima}, {Nichol}, {Niemi}, {Nightingale}, {Padilla}, {Paltani}, {Pasian}, {Pedersen}, {Pettorino}, {Pires}, {Polenta}, {Poncet}, {Popa}, {Raison}, {Rebolo}, {Renzi}, {Rhodes}, {Riccio}, {Romelli}, {Roncarelli}, {Saglia}, {Sakr}, {Sapone}, {Sartoris}, {Schirmer}, {Secroun}, {Seidel}, {Serrano}, {Sirignano}, {Sirri}, {Stanco}, {Starck}, {Tallada-Cresp{\'\i}}, {Taylor}, {Tereno}, {Toledo-Moreo}, {Torradeflot}, {Tutusaus}, {Valenziano}, {Vassallo}, {Verdoes Kleijn}, {Veropalumbo}, {Wang}, {Weller}, {Zamorani}, {Zucca}, {Burigana}, {Pezzotta}, {Porciani}, {Scottez}, {Viel}, \& {Le Brun}}]{Laila2024}
{Linke}, L., {Unruh}, S., {Wittje}, A., {et~al.} 2024, arXiv e-prints, arXiv:2407.09810

\bibitem[{{LoVerde} \& {Afshordi}(2008)}]{LA2008}
{LoVerde}, M. \& {Afshordi}, N. 2008, \prd, 78, 123506

\bibitem[{Ma \& Bertschinger(1995)}]{Ma95}
Ma, C.-P. \& Bertschinger, E. 1995, \apj, 455, 7

\bibitem[{{Maraio} {et~al.}(2023){Maraio}, {Hall}, \& {Taylor}}]{2023MNRAS.520.4836M}
{Maraio}, A., {Hall}, A., \& {Taylor}, A. 2023, \mnras, 520, 4836

\bibitem[{{Martinelli} \& {Casas}(2021)}]{MatteoSanti2021}
{Martinelli}, M. \& {Casas}, S. 2021, Universe, 7, 506

\bibitem[{{Martinelli} {et~al.}(2021){Martinelli}, {Tutusaus}, {Archidiacono}, {Camera}, {Cardone}, {Clesse}, {Casas}, {Casarini}, {Mota}, {Hoekstra}, {Carbone}, {Ili{\'c}}, {Kitching}, {Pettorino}, {Pourtsidou}, {Sakr}, {Sapone}, {Auricchio}, {Balestra}, {Boucaud}, {Branchini}, {Brescia}, {Capobianco}, {Carretero}, {Castellano}, {Cavuoti}, {Cimatti}, {Cledassou}, {Congedo}, {Conselice}, {Conversi}, {Corcione}, {Costille}, {Douspis}, {Dubath}, {Dusini}, {Fabbian}, {Fosalba}, {Frailis}, {Franceschi}, {Gillis}, {Giocoli}, {Grupp}, {Guzzo}, {Holmes}, {Hormuth}, {Jahnke}, {Kermiche}, {Kiessling}, {Kilbinger}, {Kunz}, {Kurki-Suonio}, {Ligori}, {Lilje}, {Lloro}, {Maiorano}, {Marggraf}, {Markovic}, {Massey}, {Meneghetti}, {Meylan}, {Morin}, {Moscardini}, {Niemi}, {Padilla}, {Paltani}, {Pasian}, {Pedersen}, {Pires}, {Polenta}, {Poncet}, {Popa}, {Raison}, {Rhodes}, {Roncarelli}, {Rossetti}, {Saglia}, {Schneider}, {Secroun}, {Serrano}, {Sirignano}, {Sirri}, {Starck}, {Sureau}, {Taylor}, {Tereno}, {Toledo-Moreo},
  {Valentijn}, {Valenziano}, {Vassallo}, {Wang}, {Welikala}, {Zacchei}, \& {Zoubian}}]{Pirates20}
{Martinelli}, M., {Tutusaus}, I., {Archidiacono}, M., {et~al.} 2021, \aap, 649, A100

\bibitem[{{McCarthy} {et~al.}(2017){McCarthy}, {Schaye}, {Bird}, \& {Le Brun}}]{McCarthy2017}
{McCarthy}, I.~G., {Schaye}, J., {Bird}, S., \& {Le Brun}, A. M.~C. 2017, \mnras, 465, 2936

\bibitem[{{Mead} {et~al.}(2021){Mead}, {Brieden}, {Tr{\"o}ster}, \& {Heymans}}]{HMCode2020}
{Mead}, A.~J., {Brieden}, S., {Tr{\"o}ster}, T., \& {Heymans}, C. 2021, \mnras, 502, 1401

\bibitem[{{Mead} {et~al.}(2016){Mead}, {Heymans}, {Lombriser}, {Peacock}, {Steele}, \& {Winther}}]{HMCode2016}
{Mead}, A.~J., {Heymans}, C., {Lombriser}, L., {et~al.} 2016, \mnras, 459, 1468

\bibitem[{{Miralda-Escude}(1991)}]{ME91}
{Miralda-Escude}, J. 1991, \apj, 370, 1

\bibitem[{{Miyatake} {et~al.}(2023){Miyatake}, {Sugiyama}, {Takada}, {Nishimichi}, {Li}, {Shirasaki}, {More}, {Kobayashi}, {Nishizawa}, {Rau}, {Zhang}, {Takahashi}, {Dalal}, {Mandelbaum}, {Strauss}, {Hamana}, {Oguri}, {Osato}, {Luo}, {Kannawadi}, {Hsieh}, {Armstrong}, {Bosch}, {Komiyama}, {Lupton}, {Lust}, {MacArthur}, {Miyazaki}, {Murayama}, {Okura}, {Price}, {Sunayama}, {Tait}, {Tanaka}, \& {Wang}}]{hscy3second}
{Miyatake}, H., {Sugiyama}, S., {Takada}, M., {et~al.} 2023, \prd, 108, 123517

\bibitem[{Moretti {et~al.}(2024)}]{Moretti24}
Moretti, C. {et~al.} 2024, in preparation

\bibitem[{Pandey {et~al.}(2022)Pandey, Krause, DeRose, MacCrann, Jain, Crocce, Blazek, Choi, Huang, To, Fang, Elvin-Poole, Prat, Porredon, Secco, Rodriguez-Monroy, Weaverdyck, Park, Raveri, Rozo, Rykoff, Bernstein, S\'anchez, Jarvis, Troxel, Zacharegkas, Chang, Alarcon, Alves, Amon, Andrade-Oliveira, Baxter, Bechtol, Becker, Camacho, Campos, Carnero~Rosell, Carrasco~Kind, Cawthon, Chen, Chintalapati, Davis, Di~Valentino, Diehl, Dodelson, Doux, Drlica-Wagner, Eckert, Eifler, Elsner, Everett, Farahi, Fert\'e, Fosalba, Friedrich, Gatti, Giannini, Gruen, Gruendl, Harrison, Hartley, Huff, Huterer, Kovacs, Leget, McCullough, Muir, Myles, Navarro-Alsina, Omori, Rollins, Roodman, Rosenfeld, Sevilla-Noarbe, Sheldon, Shin, Troja, Tutusaus, Varga, Wechsler, Yanny, Yin, Zhang, Zuntz, Abbott, Aguena, Allam, Annis, Bacon, Bertin, Brooks, Burke, Carretero, Conselice, Costanzi, da~Costa, Pereira, De~Vicente, Dietrich, Doel, Evrard, Ferrero, Flaugher, Frieman, Garc\'{\i}a-Bellido, Gaztanaga, Gerdes, Giannantonio, Gschwend,
  Gutierrez, Hinton, Hollowood, Honscheid, James, Jeltema, Kuehn, Kuropatkin, Lahav, Lima, Lin, Maia, Marshall, Melchior, Menanteau, Miller, Miquel, Mohr, Morgan, Palmese, Paz-Chinch\'on, Petravick, Pieres, Plazas~Malag\'on, Sanchez, Scarpine, Serrano, Smith, Soares-Santos, Suchyta, Tarle, Thomas, \& Weller}]{Pandey2022}
Pandey, S., Krause, E., DeRose, J., {et~al.} 2022, Phys. Rev. D, 106, 043520

\bibitem[{{Parkinson} {et~al.}(2012){Parkinson}, {Riemer-S{\o}rensen}, {Blake}, {Poole}, {Davis}, {Brough}, {Colless}, {Contreras}, {Couch}, {Croom}, {Croton}, {Drinkwater}, {Forster}, {Gilbank}, {Gladders}, {Glazebrook}, {Jelliffe}, {Jurek}, {Li}, {Madore}, {Martin}, {Pimbblet}, {Pracy}, {Sharp}, {Wisnioski}, {Woods}, {Wyder}, \& {Yee}}]{WiggleZ2012}
{Parkinson}, D., {Riemer-S{\o}rensen}, S., {Blake}, C., {et~al.} 2012, \prd, 86, 103518

\bibitem[{{Peebles}(1984)}]{Peebles1984}
{Peebles}, P.~J.~E. 1984, \apj, 284, 439

\bibitem[{{Perivolaropoulos} \& {Skara}(2022)}]{Peri2022}
{Perivolaropoulos}, L. \& {Skara}, F. 2022, \nar, 95, 101659

\bibitem[{{Perlmutter} {et~al.}(1999){Perlmutter}, {Aldering}, {Goldhaber}, {Knop}, {Nugent}, {Castro}, {Deustua}, {Fabbro}, {Goobar}, {Groom}, {Hook}, {Kim}, {Kim}, {Lee}, {Nunes}, {Pain}, {Pennypacker}, {Quimby}, {Lidman}, {Ellis}, {Irwin}, {McMahon}, {Ruiz-Lapuente}, {Walton}, {Schaefer}, {Boyle}, {Filippenko}, {Matheson}, {Fruchter}, {Panagia}, {Newberg}, {Couch}, \& {Project}}]{Perl1999}
{Perlmutter}, S., {Aldering}, G., {Goldhaber}, G., {et~al.} 1999, \apj, 517, 565

\bibitem[{{Planck Collaboration} {et~al.}(2020){Planck Collaboration}, {Aghanim}, {Akrami}, {Ashdown}, {Aumont}, {Baccigalupi}, {Ballardini}, {Banday}, {Barreiro}, {Bartolo}, {Basak}, {Battye}, {Benabed}, {Bernard}, {Bersanelli}, {Bielewicz}, {Bock}, {Bond}, {Borrill}, {Bouchet}, {Boulanger}, {Bucher}, {Burigana}, {Butler}, {Calabrese}, {Cardoso}, {Carron}, {Challinor}, {Chiang}, {Chluba}, {Colombo}, {Combet}, {Contreras}, {Crill}, {Cuttaia}, {de Bernardis}, {de Zotti}, {Delabrouille}, {Delouis}, {Di Valentino}, {Diego}, {Dor{\'e}}, {Douspis}, {Ducout}, {Dupac}, {Dusini}, {Efstathiou}, {Elsner}, {En{\ss}lin}, {Eriksen}, {Fantaye}, {Farhang}, {Fergusson}, {Fernandez-Cobos}, {Finelli}, {Forastieri}, {Frailis}, {Fraisse}, {Franceschi}, {Frolov}, {Galeotta}, {Galli}, {Ganga}, {G{\'e}nova-Santos}, {Gerbino}, {Ghosh}, {Gonz{\'a}lez-Nuevo}, {G{\'o}rski}, {Gratton}, {Gruppuso}, {Gudmundsson}, {Hamann}, {Handley}, {Hansen}, {Herranz}, {Hildebrandt}, {Hivon}, {Huang}, {Jaffe}, {Jones}, {Karakci}, {Keih{\"a}nen},
  {Keskitalo}, {Kiiveri}, {Kim}, {Kisner}, {Knox}, {Krachmalnicoff}, {Kunz}, {Kurki-Suonio}, {Lagache}, {Lamarre}, {Lasenby}, {Lattanzi}, {Lawrence}, {Le Jeune}, {Lemos}, {Lesgourgues}, {Levrier}, {Lewis}, {Liguori}, {Lilje}, {Lilley}, {Lindholm}, {L{\'o}pez-Caniego}, {Lubin}, {Ma}, {Mac{\'\i}as-P{\'e}rez}, {Maggio}, {Maino}, {Mandolesi}, {Mangilli}, {Marcos-Caballero}, {Maris}, {Martin}, {Martinelli}, {Mart{\'\i}nez-Gonz{\'a}lez}, {Matarrese}, {Mauri}, {McEwen}, {Meinhold}, {Melchiorri}, {Mennella}, {Migliaccio}, {Millea}, {Mitra}, {Miville-Desch{\^e}nes}, {Molinari}, {Montier}, {Morgante}, {Moss}, {Natoli}, {N{\o}rgaard-Nielsen}, {Pagano}, {Paoletti}, {Partridge}, {Patanchon}, {Peiris}, {Perrotta}, {Pettorino}, {Piacentini}, {Polastri}, {Polenta}, {Puget}, {Rachen}, {Reinecke}, {Remazeilles}, {Renzi}, {Rocha}, {Rosset}, {Roudier}, {Rubi{\~n}o-Mart{\'\i}n}, {Ruiz-Granados}, {Salvati}, {Sandri}, {Savelainen}, {Scott}, {Shellard}, {Sirignano}, {Sirri}, {Spencer}, {Sunyaev}, {Suur-Uski}, {Tauber}, {Tavagnacco},
  {Tenti}, {Toffolatti}, {Tomasi}, {Trombetti}, {Valenziano}, {Valiviita}, {Van Tent}, {Vibert}, {Vielva}, {Villa}, {Vittorio}, {Wandelt}, {Wehus}, {White}, {White}, {Zacchei}, \& {Zonca}}]{Planck2018}
{Planck Collaboration}, {Aghanim}, N., {Akrami}, Y., {et~al.} 2020, \aap, 641, A6

\bibitem[{{Pogosian} {et~al.}(2010){Pogosian}, {Silvestri}, {Koyama}, \& {Zhao}}]{Pogosian2010}
{Pogosian}, L., {Silvestri}, A., {Koyama}, K., \& {Zhao}, G.-B. 2010, \prd, 81, 104023

\bibitem[{{Porredon} {et~al.}(2022){Porredon}, {Crocce}, {Elvin-Poole}, {Cawthon}, {Giannini}, {De Vicente}, {Carnero Rosell}, {Ferrero}, {Krause}, {Fang}, {Prat}, {Rodriguez-Monroy}, {Pandey}, {Pocino}, {Castander}, {Choi}, {Amon}, {Tutusaus}, {Dodelson}, {Sevilla-Noarbe}, {Fosalba}, {Gaztanaga}, {Alarcon}, {Alves}, {Andrade-Oliveira}, {Baxter}, {Bechtol}, {Becker}, {Bernstein}, {Blazek}, {Camacho}, {Campos}, {Carrasco Kind}, {Chintalapati}, {Cordero}, {DeRose}, {Di Valentino}, {Doux}, {Eifler}, {Everett}, {Fert{\'e}}, {Friedrich}, {Gatti}, {Gruen}, {Harrison}, {Hartley}, {Herner}, {Huff}, {Huterer}, {Jain}, {Jarvis}, {Lee}, {Lemos}, {MacCrann}, {Mena-Fern{\'a}ndez}, {Muir}, {Myles}, {Park}, {Raveri}, {Rosenfeld}, {Ross}, {Rykoff}, {Samuroff}, {S{\'a}nchez}, {Sanchez}, {Sanchez}, {Sanchez Cid}, {Scolnic}, {Secco}, {Sheldon}, {Troja}, {Troxel}, {Weaverdyck}, {Yanny}, {Zuntz}, {Abbott}, {Aguena}, {Allam}, {Annis}, {Avila}, {Bacon}, {Bertin}, {Bhargava}, {Brooks}, {Buckley-Geer}, {Burke}, {Carretero},
  {Costanzi}, {da Costa}, {Pereira}, {Davis}, {Desai}, {Diehl}, {Dietrich}, {Doel}, {Drlica-Wagner}, {Eckert}, {Evrard}, {Flaugher}, {Frieman}, {Garc{\'\i}a-Bellido}, {Gerdes}, {Giannantonio}, {Gruendl}, {Gschwend}, {Gutierrez}, {Hinton}, {Hollowood}, {Honscheid}, {Hoyle}, {James}, {Kuehn}, {Kuropatkin}, {Lahav}, {Lidman}, {Lima}, {Lin}, {Maia}, {Marshall}, {Martini}, {Melchior}, {Menanteau}, {Miquel}, {Mohr}, {Morgan}, {Ogando}, {Palmese}, {Paz-Chinch{\'o}n}, {Petravick}, {Pieres}, {Plazas Malag{\'o}n}, {Romer}, {Santiago}, {Scarpine}, {Schubnell}, {Serrano}, {Smith}, {Soares-Santos}, {Suchyta}, {Tarle}, {Thomas}, {To}, {Varga}, {Weller}, \& {DES Collaboration}}]{Porredon2022}
{Porredon}, A., {Crocce}, M., {Elvin-Poole}, J., {et~al.} 2022, \prd, 106, 103530

\bibitem[{{Riess} {et~al.}(1998){Riess}, {Filippenko}, {Challis}, {Clocchiatti}, {Diercks}, {Garnavich}, {Gilliland}, {Hogan}, {Jha}, {Kirshner}, {Leibundgut}, {Phillips}, {Reiss}, {Schmidt}, {Schommer}, {Smith}, {Spyromilio}, {Stubbs}, {Suntzeff}, \& {Tonry}}]{Riess1998}
{Riess}, A.~G., {Filippenko}, A.~V., {Challis}, P., {et~al.} 1998, \aj, 116, 1009

\bibitem[{{Ross} {et~al.}(2025){Ross}, {Aguilar}, {Ahlen}, {Alam}, {Anand}, {Bailey}, {Bianchi}, {Brieden}, {Brooks}, {Burtin}, {Carnero Rosell}, {Chaussidon}, {Claybaugh}, {Cole}, {Dawson}, {de la Macorra}, {de Mattia}, {Dey}, {Dey}, {Doel}, {Fanning}, {Ferraro}, {Ereza}, {Font-Ribera}, {Forero-Romero}, {Gazta{\~n}aga}, {Gil-Mar{\'\i}n}, {Gontcho A Gontcho}, {Gonzalez-Morales}, {Guy}, {Hahn}, {Heydenreich}, {Honscheid}, {Howlett}, {Ishak}, {Karim}, {Kirkby}, {Kisner}, {Kong}, {Kremin}, {Krolewski}, {Lambert}, {Landriau}, {Lasker}, {Guillou}, {Levi}, {Manera}, {Martini}, {McDonald}, {Meisner}, {Miquel}, {Moon}, {Moustakas}, {Mu{\~n}oz-Guti{\'e}rrez}, {Myers}, {Nadathur}, {Napolitano}, {Newman}, {Nie}, {Niz}, {Palanque-Delabrouille}, {Percival}, {Poppett}, {Prada}, {Raichoor}, {Ravoux}, {Rezaie}, {Rosado-Marin}, {Rossi}, {Samushia}, {Sanchez}, {Schlafly}, {Schlegel}, {Seo}, {Smith}, {Sprayberry}, {Tarl{\'e}}, {Valcin}, {Vargas-Maga{\~n}a}, {Weaver}, {Wilson}, {Yu}, {Zarrouk}, {Zhao}, {Zhou}, \& {Zou}}]{ross25}
{Ross}, A.~J., {Aguilar}, J., {Ahlen}, S., {et~al.} 2025, JCAP, 01, 125

\bibitem[{{Samuroff} {et~al.}(2023){Samuroff}, {Mandelbaum}, {Blazek}, {Campos}, {MacCrann}, {Zacharegkas}, {Amon}, {Prat}, {Singh}, {Elvin-Poole}, {Ross}, {Alarcon}, {Baxter}, {Bechtol}, {Becker}, {Bernstein}, {Rosell}, {Kind}, {Cawthon}, {Chang}, {Chen}, {Choi}, {Crocce}, {Davis}, {DeRose}, {Dodelson}, {Doux}, {Drlica-Wagner}, {Eckert}, {Everett}, {Fert{\'e}}, {Gatti}, {Giannini}, {Gruen}, {Gruendl}, {Harrison}, {Herner}, {Huff}, {Jarvis}, {Kuropatkin}, {Leget}, {Lemos}, {McCullough}, {Myles}, {Navarro-Alsina}, {Pandey}, {Porredon}, {Raveri}, {Rodriguez-Monroy}, {Rollins}, {Roodman}, {Rossi}, {Rykoff}, {S{\'a}nchez}, {Secco}, {Sevilla-Noarbe}, {Sheldon}, {Shin}, {Troxel}, {Tutusaus}, {Weaverdyck}, {Yanny}, {Yin}, {Zhang}, {Zuntz}, {Aguena}, {Alves}, {Annis}, {Bacon}, {Bertin}, {Bocquet}, {Brooks}, {Burke}, {Carretero}, {Costanzi}, {da Costa}, {Pereira}, {De Vicente}, {Desai}, {Diehl}, {Dietrich}, {Doel}, {Ferrero}, {Flaugher}, {Frieman}, {Garc{\'\i}a-Bellido}, {Hinton}, {Hollowood}, {Honscheid}, {James},
  {Kuehn}, {Lahav}, {Marshall}, {Melchior}, {Mena-Fern{\'a}ndez}, {Menanteau}, {Miquel}, {Newman}, {Palmese}, {Pieres}, {Malag{\'o}n}, {Sanchez}, {Scarpine}, {Smith}, {Suchyta}, {Swanson}, {Tarle}, {To}, \& {DES Collaboration}}]{Samuroff2023}
{Samuroff}, S., {Mandelbaum}, R., {Blazek}, J., {et~al.} 2023, \mnras, 524, 2195

\bibitem[{{Saridakis} {et~al.}(2021){Saridakis}, {Lazkoz}, {Salzano}, {Vargas Moniz}, {Capozziello}, {Beltr{\'a}n Jim{\'e}nez}, {De Laurentis}, {Olmo}, {Akrami}, {Bahamonde}, {Bl{\'a}zquez-Salcedo}, {B{\"o}hmer}, {Bonvin}, {Bouhmadi-L{\'o}pez}, {Brax}, {Calcagni}, {Casadio}, {Cembranos}, {de la Cruz-Dombriz}, {Davis}, {Delhom}, {Di Valentino}, {Dialektopoulos}, {Elder}, {Mar{\'\i}a Ezquiaga}, {Frusciante}, {Garattini}, {Gergely}, {Giusti}, {Heisenberg}, {Hohmann}, {Iosifidis}, {Kazantzidis}, {Kleihaus}, {Koivisto}, {Kunz}, {Lobo}, {Martinelli}, {Mart{\'\i}n-Moruno}, {Mimoso}, {Mota}, {Peirone}, {Perivolaropoulos}, {Pettorino}, {Pfeifer}, {Pizzuti}, {Rubiera-Garcia}, {Levi Said}, {Sakellariadou}, {Saltas}, {Spurio Mancini}, {Voicu}, \& {Wojnar}}]{Cantata2021}
{Saridakis}, E.~N., {Lazkoz}, R., {Salzano}, V., {et~al.} 2021, arXiv e-prints, arXiv:2105.12582

\bibitem[{{Sawicki} \& {Bellini}(2015)}]{SB15}
{Sawicki}, I. \& {Bellini}, E. 2015, \prd, 92, 084061

\bibitem[{{Schmidt} {et~al.}(2009{\natexlab{a}}){Schmidt}, {Rozo}, {Dodelson}, {Hui}, \& {Sheldon}}]{Schmidt2009a}
{Schmidt}, F., {Rozo}, E., {Dodelson}, S., {Hui}, L., \& {Sheldon}, E. 2009{\natexlab{a}}, \apj, 702, 593

\bibitem[{{Schmidt} {et~al.}(2009{\natexlab{b}}){Schmidt}, {Rozo}, {Dodelson}, {Hui}, \& {Sheldon}}]{Schmidt2009b}
{Schmidt}, F., {Rozo}, E., {Dodelson}, S., {Hui}, L., \& {Sheldon}, E. 2009{\natexlab{b}}, \prl, 103, 051301

\bibitem[{{Schneider} {et~al.}(2010){Schneider}, {Eifler}, \& {Krause}}]{SEK10}
{Schneider}, P., {Eifler}, T., \& {Krause}, E. 2010, \aap, 520, A116

\bibitem[{{Schneider} {et~al.}(2002){Schneider}, {van Waerbeke}, {Kilbinger}, \& {Mellier}}]{Schneider2002}
{Schneider}, P., {van Waerbeke}, L., {Kilbinger}, M., \& {Mellier}, Y. 2002, \aap, 396, 1

\bibitem[{{Schrabback} {et~al.}(2010){Schrabback}, {Hartlap}, {Joachimi}, {Kilbinger}, {Simon}, {Benabed}, {Brada{\v{c}}}, {Eifler}, {Erben}, {Fassnacht}, {High}, {Hilbert}, {Hildebrandt}, {Hoekstra}, {Kuijken}, {Marshall}, {Mellier}, {Morganson}, {Schneider}, {Semboloni}, {van Waerbeke}, \& {Velander}}]{schrabback10}
{Schrabback}, T., {Hartlap}, J., {Joachimi}, B., {et~al.} 2010, \aap, 516, A63

\bibitem[{{Scolnic} {et~al.}(2022){Scolnic}, {Brout}, {Carr}, {Riess}, {Davis}, {Dwomoh}, {Jones}, {Ali}, {Charvu}, {Chen}, {Peterson}, {Popovic}, {Rose}, {Wood}, {Brown}, {Chambers}, {Coulter}, {Dettman}, {Dimitriadis}, {Filippenko}, {Foley}, {Jha}, {Kilpatrick}, {Kirshner}, {Pan}, {Rest}, {Rojas-Bravo}, {Siebert}, {Stahl}, \& {Zheng}}]{Pantheon+}
{Scolnic}, D., {Brout}, D., {Carr}, A., {et~al.} 2022, \apj, 938, 113

\bibitem[{{Secco} {et~al.}(2022){Secco}, {Samuroff}, {Krause}, {Jain}, {Blazek}, {Raveri}, {Campos}, {Amon}, {Chen}, {Doux}, {Choi}, {Gruen}, {Bernstein}, {Chang}, {DeRose}, {Myles}, {Fert{\'e}}, {Lemos}, {Huterer}, {Prat}, {Troxel}, {MacCrann}, {Liddle}, {Kacprzak}, {Fang}, {S{\'a}nchez}, {Pandey}, {Dodelson}, {Chintalapati}, {Hoffmann}, {Alarcon}, {Alves}, {Andrade-Oliveira}, {Baxter}, {Bechtol}, {Becker}, {Brandao-Souza}, {Camacho}, {Carnero Rosell}, {Carrasco Kind}, {Cawthon}, {Cordero}, {Crocce}, {Davis}, {Di Valentino}, {Drlica-Wagner}, {Eckert}, {Eifler}, {Elidaiana}, {Elsner}, {Elvin-Poole}, {Everett}, {Fosalba}, {Friedrich}, {Gatti}, {Giannini}, {Gruendl}, {Harrison}, {Hartley}, {Herner}, {Huang}, {Huff}, {Jarvis}, {Jeffrey}, {Kuropatkin}, {Leget}, {Muir}, {Mccullough}, {Navarro Alsina}, {Omori}, {Park}, {Porredon}, {Rollins}, {Roodman}, {Rosenfeld}, {Ross}, {Rykoff}, {Sanchez}, {Sevilla-Noarbe}, {Sheldon}, {Shin}, {Troja}, {Tutusaus}, {Varga}, {Weaverdyck}, {Wechsler}, {Yanny}, {Yin}, {Zhang},
  {Zuntz}, {Abbott}, {Aguena}, {Allam}, {Annis}, {Bacon}, {Bertin}, {Bhargava}, {Bridle}, {Brooks}, {Buckley-Geer}, {Burke}, {Carretero}, {Costanzi}, {da Costa}, {De Vicente}, {Diehl}, {Dietrich}, {Doel}, {Ferrero}, {Flaugher}, {Frieman}, {Garc{\'\i}a-Bellido}, {Gaztanaga}, {Gerdes}, {Giannantonio}, {Gschwend}, {Gutierrez}, {Hinton}, {Hollowood}, {Honscheid}, {Hoyle}, {James}, {Jeltema}, {Kuehn}, {Lahav}, {Lima}, {Lin}, {Maia}, {Marshall}, {Martini}, {Melchior}, {Menanteau}, {Miquel}, {Mohr}, {Morgan}, {Ogando}, {Palmese}, {Paz-Chinch{\'o}n}, {Petravick}, {Pieres}, {Plazas Malag{\'o}n}, {Rodriguez-Monroy}, {Romer}, {Sanchez}, {Scarpine}, {Schubnell}, {Scolnic}, {Serrano}, {Smith}, {Soares-Santos}, {Suchyta}, {Swanson}, {Tarle}, {Thomas}, {To}, \& {DES Collaboration}}]{DESY3Secco2022}
{Secco}, L.~F., {Samuroff}, S., {Krause}, E., {et~al.} 2022, \prd, 105, 023515

\bibitem[{Skordis(2008)}]{Skordis2008}
Skordis, C. 2008, Phys. Rev. D, 77, 123502

\bibitem[{{Spergel} {et~al.}(2003){Spergel}, {Verde}, {Peiris}, {Komatsu}, {Nolta}, {Bennett}, {Halpern}, {Hinshaw}, {Jarosik}, {Kogut}, {Limon}, {Meyer}, {Page}, {Tucker}, {Weiland}, {Wollack}, \& {Wright}}]{wmap1}
{Spergel}, D.~N., {Verde}, L., {Peiris}, H.~V., {et~al.} 2003, \apjs, 148, 175

\bibitem[{{Sugiyama} {et~al.}(2023){Sugiyama}, {Miyatake}, {More}, {Li}, {Shirasaki}, {Takada}, {Kobayashi}, {Takahashi}, {Nishimichi}, {Nishizawa}, {Rau}, {Zhang}, {Dalal}, {Mandelbaum}, {Strauss}, {Hamana}, {Oguri}, {Osato}, {Kannawadi}, {Hsieh}, {Luo}, {Armstrong}, {Bosch}, {Komiyama}, {Lupton}, {Lust}, {Miyazaki}, {Murayama}, {Okura}, {Price}, {Tait}, {Tanaka}, \& {Wang}}]{hscy3first}
{Sugiyama}, S., {Miyatake}, H., {More}, S., {et~al.} 2023, \prd, 108, 123521

\bibitem[{{Sugiyama} {et~al.}(2022){Sugiyama}, {Takada}, {Miyatake}, {Nishimichi}, {Shirasaki}, {Kobayashi}, {Mandelbaum}, {More}, {Takahashi}, {Osato}, {Oguri}, {Coupon}, {Hikage}, {Hsieh}, {Komiyama}, {Leauthaud}, {Li}, {Luo}, {Lupton}, {Murayama}, {Nishizawa}, {Park}, {Price}, {Simet}, {Speagle}, {Strauss}, \& {Tanaka}}]{Sugy2022}
{Sugiyama}, S., {Takada}, M., {Miyatake}, H., {et~al.} 2022, \prd, 105, 123537

\bibitem[{{Takahashi} {et~al.}(2012){Takahashi}, {Sato}, {Nishimichi}, {Taruya}, \& {Oguri}}]{HaloFit}
{Takahashi}, R., {Sato}, M., {Nishimichi}, T., {Taruya}, A., \& {Oguri}, M. 2012, \apj, 761, 152

\bibitem[{{Tanidis} \& {Camera}(2019)}]{Kostas2019}
{Tanidis}, K. \& {Camera}, S. 2019, \mnras, 489, 3385

\bibitem[{{Taylor} {et~al.}(2021{\natexlab{a}}){Taylor}, {Bernardeau}, \& {Huff}}]{xcut}
{Taylor}, P.~L., {Bernardeau}, F., \& {Huff}, E. 2021{\natexlab{a}}, \prd, 103, 043531

\bibitem[{{Taylor} {et~al.}(2018){Taylor}, {Bernardeau}, \& {Kitching}}]{kcut1}
{Taylor}, P.~L., {Bernardeau}, F., \& {Kitching}, T.~D. 2018, \prd, 98, 083514

\bibitem[{{Taylor} {et~al.}(2021{\natexlab{b}}){Taylor}, {Kitching}, {Cardone}, {Fert{\'e}}, {Huff}, {Bernardeau}, {Rhodes}, {Deshpande}, {Tutusaus}, {Pourtsidou}, {Camera}, {Carbone}, {Casas}, {Martinelli}, {Pettorino}, {Sakr}, {Sapone}, {Yankelevich}, {Auricchio}, {Balestra}, {Bodendorf}, {Bonino}, {Boucaud}, {Branchini}, {Brescia}, {Capobianco}, {Carretero}, {Castellano}, {Cavuoti}, {Cimatti}, {Cledassou}, {Congedo}, {Conversi}, {Corcione}, {Cropper}, {Franceschi}, {Garilli}, {Gillis}, {Giocoli}, {Guzzo}, {Haugan}, {Holmes}, {Hormuth}, {Jahnke}, {Kermiche}, {Kilbinger}, {Kunz}, {Kurki-Suonio}, {Ligori}, {Lilje}, {Lloro}, {Marggraf}, {Markovic}, {Massey}, {Mei}, {Medinaceli}, {Meneghetti}, {Meylan}, {Moresco}, {Morin}, {Moscardini}, {Niemi}, {Padilla}, {Pasian}, {Paltani}, {Pedersen}, {Pires}, {Percival}, {Polenta}, {Poncet}, {Popa}, {Raison}, {Roncarelli}, {Rossetti}, {Saglia}, {Schneider}, {Secroun}, {Seidel}, {Serrano}, {Sirignano}, {Sirri}, {Sureau}, {Cresp{\'\i}}, {Tavagnacco}, {Taylor}, {Teplitz},
  {Tereno}, {Toledo-Moreo}, {Valentijn}, {Valenziano}, {Vassallo}, {Wang}, {Weller}, {Zacchei}, \& {Zoubian}}]{kcut2}
{Taylor}, P.~L., {Kitching}, T., {Cardone}, V.~F., {et~al.} 2021{\natexlab{b}}, The Open Journal of Astrophysics, 4, 6

\bibitem[{{Thiele} {et~al.}(2020){Thiele}, {Duncan}, \& {Alonso}}]{Thiele2020}
{Thiele}, L., {Duncan}, C. A.~J., \& {Alonso}, D. 2020, \mnras, 491, 1746

\bibitem[{{Troxel} \& {Ishak}(2015)}]{TI2015}
{Troxel}, M.~A. \& {Ishak}, M. 2015, \physrep, 558, 1

\bibitem[{{Troxel} {et~al.}(2018){Troxel}, {MacCrann}, {Zuntz}, {Eifler}, {Krause}, {Dodelson}, {Gruen}, {Blazek}, {Friedrich}, {Samuroff}, {Prat}, {Secco}, {Davis}, {Fert{\'e}}, {DeRose}, {Alarcon}, {Amara}, {Baxter}, {Becker}, {Bernstein}, {Bridle}, {Cawthon}, {Chang}, {Choi}, {De Vicente}, {Drlica-Wagner}, {Elvin-Poole}, {Frieman}, {Gatti}, {Hartley}, {Honscheid}, {Hoyle}, {Huff}, {Huterer}, {Jain}, {Jarvis}, {Kacprzak}, {Kirk}, {Kokron}, {Krawiec}, {Lahav}, {Liddle}, {Peacock}, {Rau}, {Refregier}, {Rollins}, {Rozo}, {Rykoff}, {S{\'a}nchez}, {Sevilla-Noarbe}, {Sheldon}, {Stebbins}, {Varga}, {Vielzeuf}, {Wang}, {Wechsler}, {Yanny}, {Abbott}, {Abdalla}, {Allam}, {Annis}, {Bechtol}, {Benoit-L{\'e}vy}, {Bertin}, {Brooks}, {Buckley-Geer}, {Burke}, {Carnero Rosell}, {Carrasco Kind}, {Carretero}, {Castander}, {Crocce}, {Cunha}, {D'Andrea}, {da Costa}, {DePoy}, {Desai}, {Diehl}, {Dietrich}, {Doel}, {Fernandez}, {Flaugher}, {Fosalba}, {Garc{\'\i}a-Bellido}, {Gaztanaga}, {Gerdes}, {Giannantonio}, {Goldstein},
  {Gruendl}, {Gschwend}, {Gutierrez}, {James}, {Jeltema}, {Johnson}, {Johnson}, {Kent}, {Kuehn}, {Kuhlmann}, {Kuropatkin}, {Li}, {Lima}, {Lin}, {Maia}, {March}, {Marshall}, {Martini}, {Melchior}, {Menanteau}, {Miquel}, {Mohr}, {Neilsen}, {Nichol}, {Nord}, {Petravick}, {Plazas}, {Romer}, {Roodman}, {Sako}, {Sanchez}, {Scarpine}, {Schindler}, {Schubnell}, {Smith}, {Smith}, {Soares-Santos}, {Sobreira}, {Suchyta}, {Swanson}, {Tarle}, {Thomas}, {Tucker}, {Vikram}, {Walker}, {Weller}, {Zhang}, \& {DES Collaboration}}]{Troxel2018_DES_Y1}
{Troxel}, M.~A., {MacCrann}, N., {Zuntz}, J., {et~al.} 2018, \prd, 98, 043528

\bibitem[{Tsujikawa(2007)}]{Tsujikawa07}
Tsujikawa, S. 2007, Phys. Rev. D, 76, 023514

\bibitem[{{Tutusaus} {et~al.}(2020){Tutusaus}, {Martinelli}, {Cardone}, {Camera}, {Yahia-Cherif}, {Casas}, {Blanchard}, {Kilbinger}, {Lacasa}, {Sakr}, {Ili{\'c}}, {Kunz}, {Carbone}, {Castander}, {Dournac}, {Fosalba}, {Kitching}, {Markovic}, {Mangilli}, {Pettorino}, {Sapone}, {Yankelevich}, {Auricchio}, {Bender}, {Bonino}, {Boucaud}, {Brescia}, {Capobianco}, {Carretero}, {Castellano}, {Cavuoti}, {Cledassou}, {Congedo}, {Conversi}, {Corcione}, {Costille}, {Crocce}, {Cropper}, {Dubath}, {Dusini}, {Fabbian}, {Frailis}, {Franceschi}, {Garilli}, {Grupp}, {Guzzo}, {Hoekstra}, {Hormuth}, {Israel}, {Jahnke}, {Kermiche}, {Kubik}, {Laureijs}, {Ligori}, {Lilje}, {Lloro}, {Maiorano}, {Marggraf}, {Massey}, {Mei}, {Merlin}, {Meylan}, {Moscardini}, {Ntelis}, {Padilla}, {Paltani}, {Pasian}, {Percival}, {Pires}, {Poncet}, {Raison}, {Rhodes}, {Roncarelli}, {Rossetti}, {Saglia}, {Schneider}, {Secroun}, {Serrano}, {Sirignano}, {Sirri}, {Starck}, {Sureau}, {Taylor}, {Tereno}, {Toledo-Moreo}, {Valenziano}, {Wang}, {Welikala},
  {Weller}, {Zacchei}, \& {Zoubian}}]{Isaac2020}
{Tutusaus}, I., {Martinelli}, M., {Cardone}, V.~F., {et~al.} 2020, \aap, 643, A70

\bibitem[{{van Uitert} {et~al.}(2018){van Uitert}, {Joachimi}, {Joudaki}, {Amon}, {Heymans}, {K{\"o}hlinger}, {Asgari}, {Blake}, {Choi}, {Erben}, {Farrow}, {Harnois-D{\'e}raps}, {Hildebrandt}, {Hoekstra}, {Kitching}, {Klaes}, {Kuijken}, {Merten}, {Miller}, {Nakajima}, {Schneider}, {Valentijn}, \& {Viola}}]{vanUitert2018}
{van Uitert}, E., {Joachimi}, B., {Joudaki}, S., {et~al.} 2018, \mnras, 476, 4662

\bibitem[{{Van Waerbeke} {et~al.}(2000){Van Waerbeke}, {Mellier}, {Erben}, {Cuillandre}, {Bernardeau}, {Maoli}, {Bertin}, {McCracken}, {Le F{\`e}vre}, {Fort}, {Dantel-Fort}, {Jain}, \& {Schneider}}]{VanWaerbeke2000}
{Van Waerbeke}, L., {Mellier}, Y., {Erben}, T., {et~al.} 2000, \aap, 358, 30

\bibitem[{{Vazsonyi} {et~al.}(2021){Vazsonyi}, {Taylor}, {Valogiannis}, {Ramachandra}, {Fert{\'e}}, \& {Rhodes}}]{Vaz21}
{Vazsonyi}, L., {Taylor}, P.~L., {Valogiannis}, G., {et~al.} 2021, \prd, 104, 083527

\bibitem[{Verde {et~al.}(2019)Verde, Treu, \& Riess}]{Verde2019}
Verde, L., Treu, T., \& Riess, A.~G. 2019, Nature Astron., 3, 891

\bibitem[{{Vincenzi} {et~al.}(2024){Vincenzi}, {Brout}, {Armstrong}, {Popovic}, {Taylor}, {Acevedo}, {Camilleri}, {Chen}, {Davis}, {Hinton}, {Kelsey}, {Kessler}, {Lee}, {Lidman}, {M{\"o}ller}, {Qu}, {Sako}, {Sanchez}, {Scolnic}, {Smith}, {Sullivan}, {Wiseman}, {Asorey}, {Bassett}, {Carollo}, {Carr}, {Foley}, {Frohmaier}, {Galbany}, {Glazebrook}, {Graur}, {Kovacs}, {Kuehn}, {Malik}, {Nichol}, {Rose}, {Tucker}, {Toy}, {Tucker}, {Yuan}, {Abbott}, {Aguena}, {Alves}, {Andrade-Oliveira}, {Annis}, {Bacon}, {Bechtol}, {Bernstein}, {Brooks}, {Burke}, {Carnero Rosell}, {Carretero}, {Castander}, {Conselice}, {da Costa}, {Pereira}, {Desai}, {Diehl}, {Doel}, {Ferrero}, {Flaugher}, {Friedel}, {Frieman}, {Garc{\'\i}a-Bellido}, {Gatti}, {Giannini}, {Gruen}, {Gruendl}, {Hollowood}, {Honscheid}, {Huterer}, {James}, {Kuropatkin}, {Lahav}, {Lee}, {Lin}, {Marshall}, {Mena-Fern{\'a}ndez}, {Menanteau}, {Miquel}, {Palmese}, {Pieres}, {Plazas Malag{\'o}n}, {Porredon}, {Romer}, {Roodman}, {Sanchez}, {Sanchez Cid}, {Schubnell},
  {Sevilla-Noarbe}, {Suchyta}, {Swanson}, {Tarle}, {To}, {Walker}, \& {Weaverdyck}}]{DESY3SNeIA}
{Vincenzi}, M., {Brout}, D., {Armstrong}, P., {et~al.} 2024, arXiv e-prints, arXiv:2401.02945

\bibitem[{{Wang} \& {Steinhardt}(1998)}]{WS98}
{Wang}, L. \& {Steinhardt}, P.~J. 1998, \apj, 508, 483

\bibitem[{Weinberg(1989)}]{Weinberg1989}
Weinberg, S. 1989, Rev. Mod. Phys., 61, 1

\bibitem[{Wittman {et~al.}(2000)Wittman, Tyson, Kirkman, Dell'Antonio, \& Bernstein}]{Wittman2000}
Wittman, D.~M., Tyson, J.~A., Kirkman, D., Dell'Antonio, I., \& Bernstein, G. 2000, Nature, 405, 143

\bibitem[{{Yoo} \& {Watanabe}(2012)}]{YooWata2012}
{Yoo}, J. \& {Watanabe}, Y. 2012, International Journal of Modern Physics D, 21, 1230002

\bibitem[{{Zhao} {et~al.}(2009){Zhao}, {Pogosian}, {Silvestri}, \& {Zylberberg}}]{Zhao2008bn}
{Zhao}, G.-B., {Pogosian}, L., {Silvestri}, A., \& {Zylberberg}, J. 2009, \prl, 103, 241301

\end{thebibliography}

\label{LastPage}
\end{document}